\documentstyle[seceq,epsf]{ptptex}

\begin{document}
\def\agt{\mathrel{\raise.3ex\hbox{$>$}\mkern-14mu\lower0.6ex\hbox{$\sim$}}}
\def\alt{\mathrel{\raise.3ex\hbox{$<$}\mkern-14mu\lower0.6ex\hbox{$\sim$}}}

\newcommand{\beq}{\begin{equation}}
\newcommand{\eeq}{\end{equation}}
\newcommand{\beqn}{\begin{eqnarray}}
\newcommand{\eeqn}{\end{eqnarray}}
\newcommand{\pa}{\partial}
\newcommand{\vp}{\varphi}
\newcommand{\varep}{\varepsilon}
\newcommand{\ep}{\epsilon}
\newcommand{\comp}{(M/R)_\infty} 


\title{Gravitational Waves from the Merger of 
Binary Neutron Stars in a Fully General Relativistic Simulation}

\author{Masaru Shibata$^{1}$ and K\=oji Ury\=u$^2$}

\inst{$^1$ Graduate School of Arts and Sciences, 
University of Tokyo, Komaba, Meguro, Tokyo 153-8902, Japan \\
$^2$ Department of Physics, University of Wisconsin-Milwaukee, 
P.O. Box413, Milwaukee, WI 53201, USA}

\maketitle

\begin{abstract}
~\\
We performed 3D numerical simulations of the merger of equal-mass 
binary neutron stars in full general relativity 
using a new large scale supercomputer. We take 
the typical grid size as $(505,505,253)$ for $(x,y,z)$ and 
the maximum grid size as $(633,633,317)$. These grid numbers enable us to put 
the outer boundaries of the computational domain near the local wave zone 
and hence to calculate gravitational waveforms of good accuracy 
(within $\sim 10\%$ error) for the first time. To model neutron stars, 
we adopt a $\Gamma$-law 
equation of state in the form $P=(\Gamma-1)\rho\varepsilon$, 
where $P$, $\rho$, $\varep$ and $\Gamma$ are the pressure, 
rest mass density, 
specific internal energy, and adiabatic constant. 
It is found that gravitational waves in the merger stage 
have characteristic features that reflect the formed objects. 
In the case that a massive, transient neutron star is formed, 
its quasi-periodic oscillations are excited for a long duration, 
and this property is reflected clearly by the quasi-periodic nature of 
waveforms and the energy luminosity. 
In the case of black hole formation, 
the waveform and energy luminosity are likely damped after a short merger 
stage. However, a quasi-periodic oscillation can still be seen for a 
certain duration, because an oscillating transient massive object is 
formed during the merger. 
This duration depends strongly on the initial compactness of neutron stars 
and is reflected in the Fourier spectrum of gravitational waves. 
To confirm our results and to calibrate the accuracy of
gravitational waveforms, 
we carried out a wide variety of test simulations, 
changing the resolution and size of the computational domain. 
\end{abstract}

\section{Introduction}

Binary neutron stars such as the Hulse-Taylor binary 
pulsar \cite{HT} adiabatically 
inspiral as a result of the radiation reaction of gravitational waves, 
and they eventually merge. 
The latest statistical study suggests that such mergers could occur 
approximately once per year within a
distance of about 30 Mpc in the most optimistic scenario.\cite{BNST} 
Even in the most conservative scenario, 
the event rate would be approximately
once per year within a distance of about 400 Mpc.\cite{BNST} 
This implies that the merger of binary neutron stars is 
one promising source for kilo-meter-size laser interferometric 
detectors, such as LIGO, TAMA, GEO600 and VIRGO.\cite{KIP} 

The waveform in the merger stage\footnote{Hereafter, we refer to 
the stage after which the hydrodynamic interaction 
between two neutron stars sets in as the `merger stage'.
The stage before the merger stage is referred to as the `early stage'.  
} of binary neutron stars 
is believed to have two characteristic frequencies. 
One is that of quasi-normal modes (QNMs) of a black hole 
that is likely formed in the final stage of the merger in most cases. 
Perturbation studies have revealed that the frequency of the 
fundamental quadrupole QNM, $f_{\rm QNM}$, is between 
$\sim 0.1(\pi M_{\rm BH})^{-1}$ and $\sim 0.5(\pi M_{\rm BH})^{-1}$, 
where $M_{\rm BH}$ is the gravitational mass of the black hole.\cite{Leaver} 
The frequency depends 
strongly on the angular momentum parameter of the black hole
$q=J/M_{\rm BH}^2$ where $J$ denotes the angular momentum, and 
is higher for larger $q$. 
(For example, for $q=0$, $f_{\rm QNM}=0.1(\pi M_{\rm BH})^{-1}$ 
and for $q \alt 1$, $f_{\rm QNM} \alt 0.5(\pi M_{\rm BH})^{-1}$.) 
Studies of quasiequilibrium configurations just before merger 
suggest that the value of $q$ for a black hole formed after a merger is 
$\sim 0.8-1$ \cite{USE} (see also Table I). 
Then, the frequency of dominant QNMs is \cite{Leaver}
\beq
f_{\rm QNM} \sim (0.25-0.5) 
(\pi M_{\rm BH})^{-1} \approx  (5 - 10)~{\rm kHz} 
\biggl({3M_{\odot} \over M_{\rm BH}}\biggr),
\eeq
where $M_{\odot}$ denotes the solar mass. 
This frequency will be rather high, so that it may be difficult to 
detect QNMs of black holes, 
even with advanced laser interferometric detectors such as 
LIGO II or resonant mass detectors, in the near future. 

The other characteristic frequency is associated with fundamental 
oscillation modes of a merged massive object formed transiently 
after the onset of a merger.
Numerical simulations in the frameworks of 
Newtonian,\cite{RS,C} post-Newtonian,\cite{ONS,RJ,FR,AP} 
semi-relativistic,\cite{ORT} and 
fully general relativistic (GR) gravity \cite{bina,binas} 
have indicated that the frequency is approximately between 
$\sim 2$ and $\sim3$ kHz, depending on 
the equations of state and the compactness of neutron stars. 
Therefore, it is also too high for 
first generation laser interferometric detectors to detect. 
However, in contrast to the frequency of QNMs 
of a formed black hole, it is not extremely high, and it may be in the 
frequency range for advanced laser interferometric 
detectors, such as LIGO II,\cite{Ale} 
or resonant mass detectors in the near future. 
This implies that 
from gravitational waves induced by such fundamental oscillations, 
it may be possible to obtain information 
on the dynamical merger stage and the nature of 
transient massive neutron stars, which will be useful in 
understanding the nature of neutron stars.

Numerical hydrodynamic simulations employing full general relativity 
provide the best approach for understanding the waveforms from 
the merger of binary neutron stars. Over the last few years, 
numerical methods for solving coupled 
Einstein equations and hydrodynamic equations have been 
developed,\cite{gr3d,bina,other} and now such simulations are feasible. 
The purpose of this paper is to investigate the 
waveforms in the merger stage using a fully GR 
simulation carried out on 
a large scale supercomputer which has become available 
recently. This is the first paper 
that investigates gravitational waveforms from the merger of 
binary neutron stars using full general relativity with good accuracy.

A minimum grid number for 
extracting gravitational waveforms in numerical relativity 
is estimated in the following manner. 
Since it is impossible to carry out a simulation 
from the late inspiraling stage to the 
merger stage, due to the limitation on 
computational resources, we usually start simulations from innermost orbits 
of binaries. Assuming that a binary is in a circular orbit with
a Kepler angular velocity, the wavelength of 
gravitational waves in this orbit is estimated as
\beq
\lambda_{\rm IO}
={\pi \over \Omega_{\rm IO}} 
\approx 70 M\biggl({0.125 \over (M\Omega_{\rm IO})^{2/3}}\biggr)^{3/2} 
\approx 70 M\biggl({a \over 8M}\biggr)^{3/2}, 
\label{eqlam}
\eeq
where $\Omega_{\rm IO}$, $M$, and $a$ denote the angular velocity of the 
innermost orbit, gravitational mass and orbital separation, respectively.
Since the system becomes more compact after the onset of the merger, 
the typical wavelength of 
gravitational waves is shorter in the merger stage. 
Thus, the wavelength given in
Eq.~(\ref{eqlam}) is the longest one throughout a simulation. 
This implies that the location of the 
outer boundaries in the computational domain, $L$, has to be 
larger than $\lambda_{\rm IO}$, i.e., 
\beq
L \geq \lambda_{\rm IO}.\label{con1}
\eeq 
On the other hand, the grid size $\Delta x$ has to be 
small enough to resolve 
neutron stars and a black hole that is likely formed after the 
merger in many cases. The shortest length scale in the simulation 
appears to be the radius of the formed black hole, $2M$.
Here we assume that the mass of the black hole 
is approximately equal to the initial total gravitational mass $M$.  
Requiring the radius to be covered by at least 
10 grid points, we stipulate a upper bound as 
\beq
\Delta x \leq 0.2 M .\label{con2}
\eeq
{}From the constraints (\ref{con1}) and (\ref{con2}), the 
required minimum grid number for covering one dimension is estimated as
\beq
N = 2 {L \over \Delta x} 
\geq 700 \biggl({0.125 \over (M\Omega_{\rm IO})^{2/3}}\biggr)^{3/2} 
\approx 700 \biggl({a \over 8M}\biggr)^{3/2},
\eeq
where the factor 2 appears because both plus and minus 
directions need to be covered as the computational domain. 

In previous works,\cite{bina,binas} we performed simulations 
using FACOM VPP300 in the data processing center of the 
National Astronomical Observatory of Japan (NAOJ). 
The available memory of this machine was $\sim 30$ GBytes, so that 
the maximum grid size we used was $(293,293,147)$ for $(x, y, z)$. 
With this grid size, 
$L$ was typically $\sim \lambda_{0}/3$, where $\lambda_{0}$ 
denotes the wavelength of gravitational waves for 
the $l=|m|=2$ mode at $t=0$. 
Thus, we were not able to calculate 
gravitational waves in the early stage accurately nor 
to investigate the convergence of waveforms while varying 
$L$ and $\Delta x$ for a wide range in the previous simulations. 
Since the wavelength of gravitational waves becomes 
short during the merger, $L$ becomes comparable to the 
wavelength $\lambda$, 
and thus the situation is improved in late stages of the merger. 
Even so, the outer boundaries were still located in the local wave zone 
(i.e., $L \alt \lambda$), and therefore 
the accuracy of extracted gravitational waves is suspect. 

Since April 2001, a more powerful vector-parallel supercomputer 
FACOM VPP5000 has been available at NAOJ. 
About $700$ GBytes memory is available on this machine. Thus, it is 
possible to use 2--3 times as large $N$ 
as that in previous simulations carried out on VPP300.
This allows us to take $L$ as $\alt\lambda_{0}$, 
keeping a grid spacing of $\Delta x \sim 0.2M$.
This implies that gravitational waves 
can be computed fairly accurately throughout the simulation.
In addition, with this new supercomputer, 
it is possible to investigate the convergence of waveforms 
as the size of the computational domain $L$ 
(and the location of the wave extraction at $\sim L$) 
is increased from $\ll \lambda_{0}$ to $\sim \lambda_{0}$. 
In the merger stage, the wavelength of gravitational waves 
becomes between $\sim \lambda_0/3$ and $\sim \lambda_0/2$, 
since a merged object is compact. 
As a result, the accuracy of the extracted gravitational waveforms 
is significantly improved, and it is feasible to draw scientifically 
meaningful conclusions regarding 
gravitational waveforms, luminosity and spectra. 

The paper is organized as follows. In \S 2, we review 
the basic equations, gauge conditions and methods for 
setting initial conditions that we currently adopt 
in fully GR simulations of binary neutron star mergers. 
In \S 3, we summarize the methods used for analysis of gravitational waves. 
In \S 4, numerical results are presented, paying particular attention 
to gravitational waveforms. Section 5 is devoted to a summary. 
Throughout this paper, we adopt geometrical 
units in which $G=c=1$, where $G$ and $c$
are the gravitational constant and the speed of light.  
Latin and Greek indices denote spatial components (1--3) 
and space-time components (0--3), respectively. 
$\delta_{ij}(=\delta^{ij})$ denotes the Kronecker delta. 
We use Cartesian coordinates $x^k=(x, y, z)$ 
as the spatial coordinates and $t=x^0$ denotes the time coordinate.

\section{Formulation}

\subsection{Basic equations}

We solve the Einstein and GR hydrodynamic equations 
without any approximation by numerical simulation. 
Our formulation for a numerical solution of these coupled equations 
is described in detail in Refs. \citen{SN,gw3p2} and \citen{gr3d},
and therefore, we here only briefly review the basic equations.

We write the line element in the form 
\beqn
ds^2=g_{\mu\nu}dx^{\mu}dx^{\nu} =(-\alpha^2+\beta_k\beta^k)dt^2
+2\beta_i dx^i dt+\gamma_{ij}dx^i dx^j ,
\eeqn
where $g_{\mu\nu}$, $\alpha$, 
$\beta^i~(\beta_i=\gamma_{ij}\beta^j)$, 
and $\gamma_{ij}$ are the 4D metric, 
lapse function, shift vector, and 3D spatial metric, respectively. 
Following Refs. \citen{SN,gw3p2} and \citen{gr3d}, we define the 
quantities as 
\beqn
&& \gamma={\rm det}(\gamma_{ij}) \equiv e^{12\phi},\\
&& \tilde \gamma_{ij} \equiv e^{-4\phi}\gamma_{ij}, \\
&& \tilde A_{ij} \equiv 
e^{-4\phi} \Bigl(K_{ij}-{1 \over 3} \gamma_{ij} K \Bigr),
\eeqn
where $K_{ij}$ is the extrinsic curvature, and $K$ its trace. 
With this definition, 
${\rm det}(\tilde \gamma_{ij})$ is required to be unity. 
In the numerical computations, 
we evolve $\phi$, $\tilde \gamma_{ij}$, $K$, and $\tilde A_{ij}$ in time, 
instead of $\gamma_{ij}$ and $K_{ij}$. 
We note that the indices of $\tilde A_{ij}$ ($\tilde A^{ij}$) are 
raised (lowered) in terms of $\tilde \gamma^{ij}$ ($\tilde \gamma_{ij}$). 
Hereafter, we use $D_i$ and $\tilde D_i$ 
as the covariant derivatives with respect to $\gamma_{ij}$ and 
$\tilde \gamma_{ij}$, respectively: We define $\Delta = D^i D_i$ and 
$\tilde \Delta = \tilde D^i \tilde D_i$. 

As the matter source of the Einstein equation, 
we adopt a perfect fluid, for which 
the energy-momentum tensor is written  
\beq
T_{\mu\nu}=(\rho+\rho\varep+P)u_{\mu}u_{\nu}+P g_{\mu\nu},
\eeq
where $\rho$, $\varep$, $P$, and $u_{\mu}$ are 
the baryon rest-mass density, specific internal energy density, 
pressure, and four-velocity, respectively. 
We give initial conditions using polytropic equations of state as
\beq
P=\kappa \rho^{\Gamma},~~~~~~~\Gamma=1 + {1 \over n},
\eeq
where $\kappa$ and $n$ are a polytropic constant and a polytropic index. 
In the numerical time evolution, 
we use a $\Gamma$-law equation of state of the form 
\beq
P=(\Gamma-1)\rho\varep.
\eeq
In the absence of shocks, the polytropic form of 
the equations of state is preserved. 
We set $n=0.8$ ($\Gamma = 2.25$) and 1 ($\Gamma=2$) 
as a reasonable qualitative 
approximation to moderately stiff equations of state
for neutron stars.\cite{ST} 

The hydrodynamic equations [continuity, Euler and energy (or entropy) 
equations] are written in the forms 
\beqn
&&\pa_t \rho_* + \pa_i (\rho_* v^i )=0,\label{eqrho}\\
&&\pa_t (\rho_* \hat u_k)+ \pa_i (\rho_* \hat u_k v^i ) 
=-\alpha e^{6\phi}\pa_k (P + P_{\rm art})
-\rho_* \biggl[w h \pa_k \alpha - \hat u_j\pa_k \beta^j \nonumber \\
&&\hskip 5cm
+{\alpha e^{-4\phi} \hat u_i \hat u_j \over 2 w h} 
\pa_k \tilde \gamma^{ij}
-{2\alpha h (w^2-1) \over w} \pa_k \phi \biggr],\label{euler}\\
&&\pa_t e_* + \pa_i (e_* v^i )=\dot e_{\rm art},\label{energy}
\eeqn
where $\pa_{\mu}=\pa/\pa x^{\mu}$, 
$\rho_*=\rho w e^{6\phi}$, $h=1+\varep+P/\rho$, $w=\alpha u^t$, 
$\hat u_k=h  u_k$, 
$e_*=(\rho\varep)^{1/\Gamma} w e^{6\phi}$, and 
\beq
v^i \equiv {u^i \over u^t} = -\beta^i + \gamma^{ij}{u_j \over u^t}=
-\beta^i + {\alpha \tilde \gamma^{ij} \hat u_j 
\over w h e^{4\phi}}. \label{eqvelo}
\eeq
Here, $P_{\rm art}$ and $\dot e_{\rm art}$ are artificial 
viscous terms.\cite{gr3d} 
In numerical simulation, we solve Eqs. (\ref{eqrho})--(\ref{energy}) 
to carry out the evolution of $\rho_*$, $\hat u_k$ and $e_*$. 

Once $\hat u_i$ is obtained, $w$ 
is determined from the normalization relation of the four-velocity, 
$u^{\mu}u_{\mu}=-1$, which can be written as 
\beq
w^2=1+e^{-4\phi} \tilde \gamma^{ij} \hat u_i \hat u_j
\biggl[1+ {\Gamma e_*^{\Gamma} \over \rho_* (w e^{6\phi})^{\Gamma-1}}
\biggr]^{-2}.  \label{eqforw}
\eeq

The Einstein equation is split into constraint 
and evolution equations. 
The Hamiltonian and momentum constraint equations are 
written as 
\beqn
&& R_k^{~k}- \tilde A_{ij} \tilde A^{ij}+
{2 \over 3} K^2=16\pi \rho_{\rm H},
\label{ham}\\
&& D_i \tilde A^i_{~j}-{2 \over 3}D_j K=8\pi J_j, \label{mom}
\eeqn
or, equivalently 
\beqn
&& \tilde \Delta \psi = {\psi \over 8}\tilde R_k^{~k} 
- 2\pi \rho_{\rm H} \psi^5 
-{\psi^5 \over 8} \Bigl(\tilde A_{ij} \tilde A^{ij}
-{2 \over 3}K^2\Bigr), \label{hameq} \\
&& \tilde D_i (\psi^6  \tilde A^i_{~j}) - {2 \over 3} \psi^6 
\tilde D_j K = 8\pi J_j \psi^6, \label{momeq}
\eeqn
where $\psi\equiv e^{\phi}$, 
$\rho_{\rm H} \equiv T^{\mu\nu}n_{\mu}n_{\nu}$ and 
$J_i \equiv -T^{\mu\nu} n_{\mu}\gamma_{\nu i}$ with 
$n_{\mu}=(-\alpha, 0)$. Here, 
$R_{ij}$ denotes the Ricci tensor with respect to $\gamma_{ij}$, 
and $R_k^{~k}$ the Ricci scalar. These constraint equations are 
solved only at $t=0$ to set initial conditions (see \S 2.3). 

Following Refs. \citen{SN,gw3p2} and \citen{gr3d}, 
we write the evolution equations for the geometric variables as 
\beqn
&&(\pa_t - \beta^l \pa_l) \tilde \gamma_{ij} 
=-2\alpha \tilde A_{ij} 
+\tilde \gamma_{ik} \beta^k_{~,j}+\tilde \gamma_{jk} \beta^k_{~,i}
-{2 \over 3}\tilde \gamma_{ij} \beta^k_{~,k}, \label{heq} \\
&&(\pa_t - \beta^l \pa_l) \tilde A_{ij} 
= e^{ -4\phi } \biggl[ \alpha \Bigl(R_{ij}
-{1 \over 3}e^{4\phi}\tilde \gamma_{ij} R_k^{~k} \Bigr) 
-\Bigl( D_i D_j \alpha - {1 \over 3}e^{4\phi}
\tilde \gamma_{ij} \Delta \alpha \Bigr)
\biggr] \nonumber \\
&& \hskip 2.5cm +\alpha (K \tilde A_{ij} 
- 2 \tilde A_{ik} \tilde A_j^{~k}) 
+\beta^k_{~,i} \tilde A_{kj}+\beta^k_{~,j} 
\tilde A_{ki}
-{2 \over 3} \beta^k_{~,k} \tilde A_{ij} \nonumber \\
&& \hskip 2.5cm-8\pi\alpha \Bigl( 
e^{-4\phi} S_{ij}-{1 \over 3} \tilde \gamma_{ij} S_k^{~k}
\Bigr), \label{aijeq} \\
&&(\pa_t - \beta^l \pa_l) \phi = {1 \over 6}\Bigl( 
-\alpha K + \beta^k_{~,k} \Bigr), \label{peq} \\
&&(\pa_t - \beta^l \pa_l) K 
=\alpha \Bigl[ \tilde A_{ij} \tilde A^{ij}+{1 \over 3}K^2
\Bigr] 
-\Delta \alpha +4\pi \alpha (\rho_{\rm H}+ S_k^{~k}), 
\label{keq}
\eeqn
where $``~{,i}~''$ denotes the partial derivative and 
$S_{ij} \equiv T^{\mu\nu}\gamma_{\mu i}\gamma_{\nu j}$. 
We solve these evolution equations to carry out the evolution of 
$\tilde \gamma_{ij}$, $\tilde A_{ij}$, $\phi$ and $K$. 

For the computation of $R_{ij}$, we decompose 
\beq
R_{ij}=\tilde R_{ij}+R^{\phi}_{ij},
\eeq
where $\tilde R_{ij}$ is the Ricci tensor with respect to 
$\tilde \gamma_{ij}$, and 
\beqn
R^{\phi}_{ij}=-2\tilde D_i \tilde D_j \phi 
- 2  \tilde \gamma_{ij}\tilde \Delta \phi 
+ 4 \tilde D_i \phi \tilde D_j \phi 
- 4 \tilde \gamma_{ij} \tilde D_k \phi \tilde D^k \phi. 
\eeqn
$\tilde R_{ij}$ is written as 
\beqn
\tilde R_{ij}={1 \over 2}\biggl[
\delta^{kl}(-h_{ij,kl}+h_{ik,lj}+h_{jk,li})
+2\pa_k( f^{kl} \tilde \Gamma_{l,ij} ) 
-2 \tilde \Gamma^l_{kj}\tilde \Gamma^k_{il}\biggr],\label{eqij}
\eeqn
where we split $\tilde \gamma_{ij}$ and $\tilde \gamma^{ij}$  
as $\delta_{ij}+h_{ij}$ and $\delta^{ij}+f^{ij}$, respectively. 
$\tilde \Gamma^k_{ij}$ is the Christoffel symbol with respect 
to $\tilde \gamma_{ij}$, and 
$\tilde \Gamma_{k,ij}=\tilde \gamma_{kl} \tilde \Gamma^l_{ij}$. 
Because of the definition det$(\tilde \gamma_{ij})=1$, 
we have $\tilde \Gamma^k_{ki}=0$. 

In addition to the Laplacian of $h_{ij}$, 
$\tilde R_{ij}$ involves terms linear in $h_{ij}$ as 
$\delta^{kl} h_{ik,lj}$ and $\delta^{kl} h_{jk,li}$. 
To perform numerical simulation stably, 
we introduce an auxiliary variable 
$F_i=\delta^{jl}\pa_l \tilde \gamma_{ij}$,\cite{SN} 
which evolves according to the evolution equation 
\beqn
(\pa_t - \beta^l \pa_l)F_i& =&-16\pi \alpha J_i+2\alpha 
\Bigl\{ f^{kj} \tilde A_{ik,j}
+f^{kj}_{~~,j} \tilde A_{ik} 
-{1 \over 2} \tilde A^{jl} h_{jl,i} 
+6\phi_{,k} \tilde A^k_{~i}-{2\over 3}K_{,i} \Bigr\} 
\nonumber \\
&+&\delta^{jk} \Bigl\{ -2\alpha_{,k} \tilde A_{ij} 
+ \beta^l_{~,k}h_{ij,l} 
+\Bigl(\tilde \gamma_{il}\beta^l_{~,j}+\tilde \gamma_{jl}\beta^l_{~,i}
-{2\over 3}\tilde \gamma_{ij} \beta^l_{~,l}\Bigr)_{,k}\Bigr\}. 
\label{eqF}
\eeqn
In the numerical simulations, 
we evaluate $\delta^{kl}\tilde \gamma_{ik,lj}$ in Eq. (\ref{eqij}) 
as $F_{i,j}$. 

In the evolution equation for $\tilde A_{ij}$, 
we compute $R_k^{~k}$ directly as 
$R_{ij}\gamma^{ij}=e^{-4\phi}(\tilde R_{ij}
+R^{\phi}_{ij})\tilde \gamma^{ij}$ to suppress 
violation of the condition $\tilde A_{i}^{~i}=0$. 
We do not substitute the Hamiltonian constraint equation 
in this case because the Hamiltonian 
constraint is not exactly satisfied in numerical simulations, and 
its violation could systematically 
accumulate to eventually
violate the condition $\tilde A_i^{~i}=0$ significantly. 

In addition to the above formulation for the basic equations, we
adopt a transformation for computation of $\tilde R_k^{~k}$. 
In computing $\tilde R_k^{~k}=\tilde \gamma^{ij} \tilde R_{ij}$, 
there exist two linear terms in $h_{ij}$, 
$\tilde \gamma^{ij}\delta^{kl}h_{ij,kl}$ and $\tilde \gamma^{ij}F_{i,j}$.
Although they are linear in $h_{ij}$, 
these terms should be small outside the strong field zone, 
because they are composed of the trace part of $h_{ij}$ 
($\delta^{ij}h_{ij}$) and a vector component of $h_{ij}$ ($F_i$),
respectively. The magnitude of the term associated with $F_i$
is controlled by the choice of the 
spatial gauge condition, and in our choice of the gauge (see \S3),
it is maintained to be small outside the strong field zone.\cite{gw3p2} 
As shown in (\ref{eq22}), the other term is 
essentially the nonlinear term and should be
small. However, in numerical calculations, 
the nonlinearity is not guaranteed in the form
$\tilde \gamma^{ij}\delta^{kl}h_{ij,kl}$. Thus, 
to guarantee the nonlinear nature of this term, 
we rewrite it as 
\beq
\tilde \gamma^{ij}\delta^{kl}h_{ij,kl}=
-\delta^{kl}h_{ij,k}f^{ij}_{~~,l}. \label{eq22}
\eeq
As a result of these procedures, $\tilde R_k^{~k}$ is
maintained to be small outside the strong field zone.

\subsection{Gauge conditions}

We adopt the approximate maximal slice (AMS) condition 
$K \approx 0$ as the time slicing condition, 
and an approximate minimal distortion (AMD) gauge condition 
as the spatial gauge condition.\cite{gw3p2,gr3d} 

The equation imposing maximal slicing, $K=0=\pa_t K$, is 
set up from the right-hand side of Eq. (\ref{keq}) as 
\beq
\Delta \alpha =
\alpha \biggl( \tilde A_{ij} \tilde A^{ij} +{1 \over 3}K^2\biggr)
+4\pi \alpha (\rho_{\rm H} + S_k^{~k})\equiv 4\pi S_{\alpha}. 
\label{maxeq}
\eeq
In Eq. (\ref{maxeq}), we leave $K$ on the right-hand side,  
because it is not possible to guarantee it to be exactly zero 
in numerical simulations. 

To impose the AMS condition, we solve the following 
parabolic-type equation for $\ln \alpha$ 
at each timestep, until an approximate convergence is achieved:  
\beqn
\pa_{\tau} \ln \alpha && = 
\Delta \ln \alpha + (D_k \ln \alpha) (D^k \ln \alpha) 
-4\pi (\rho_{\rm H} + S_k^{~k})
- \tilde A_{ij} \tilde A^{ij} - {1 \over 3}K^2
+f_{\alpha} K \rho_*^{1/2}.\nonumber \\
~\label{eqalp2}
\eeqn
Here $\tau$ is a control parameter, and 
$f_{\alpha}$ is a constant for which we take to be of $O(1)$. 
In solving Eq. (\ref{eqalp2}) at each timestep, we 
substitute a trial function of $\alpha$, which is extrapolated as 
$\alpha(x^i)=2\alpha_{-1}(x^i)-\alpha_{-2}(x^i)$, where 
$\alpha_{-1}$ and $\alpha_{-2}$ are the lapse functions at 
previous timesteps, and then carry out iteration to achieve 
convergence. We typically carry out this iteration for 
about 30 steps. 

The last term of Eq. (\ref{eqalp2}) is introduced 
as a driver to achieve $K=0$: 
Assuming that convergence is achieved and that 
the right-hand side of Eq. (\ref{eqalp2}) becomes zero, 
the evolution equation for $K$ can be written 
\beq
(\pa_t - \beta^l \pa_l) K =
-f_{\alpha} \alpha K \rho_*^{1/2}. \label{Keq}
\eeq
Thus, if $K$ is zero initially and 
convergence is completely achieved, the 
maximal slicing condition $K=0$ is preserved. 
Even when the convergence is incomplete 
and $K$ deviates from zero, 
the right-hand side of Eq. (\ref{Keq}) 
causes $|K|$ to approach to zero on the local dynamical 
timescale $\sim \rho_*^{-1/2}$.\footnote{
In the last term of Eq. (\ref{eqalp2}), 
another function in place of $\rho_*$, e.g., 
$S_{\alpha}$, might be a better choice as a driver 
to achieve $K=0$. Other methods may be tried in the future.}
Hence, the condition $K=0$ is expected to be satisfied approximately.

To impose the AMD gauge condition, we solve the 
simple elliptic-type equations 
\beq
\Delta_{\rm flat} P_i = S_i, ~~~~~
\Delta_{\rm flat} \eta= -S_i x^i, \label{eqeta}
\eeq
where $\Delta_{\rm flat}$ is the Laplacian in the flat 3D space, and 
\beq
S_i\equiv 16\pi\alpha J_i 
+2\tilde A_{ ij} (\tilde D^j \alpha - 6\alpha \tilde D^j \phi)
+{4 \over 3}\alpha \tilde D_i K.\label{eqeta2}
\eeq
The equations for $P_i$ and $\eta$ are solved under the outer boundary 
conditions as 
\beqn
&&P_x = {C_{xx} x \over r^3}+{C_{xy} y \over r^3}+O(r^{-4}),~~~
P_y = {C_{yx} x \over r^3}+{C_{yy} y \over r^3}+O(r^{-4}),\nonumber \\
&&~~~~~P_z = {C_{zz} z \over r^3}+O(r^{-4}),~~~ 
\eta = {C_{\eta}  \over r}+O(r^{-3}),\nonumber 
\eeqn
where $C_{xx}, C_{xy}, C_{yx}, C_{yy}, C_{zz}$, and $C_{\eta}$ are 
constants computed from the volume integrations of 
$S_i x^j$ and $S_i x^i$. Here, to derive these boundary conditions, 
we assume that the system is composed of two identical neutron stars. 

{}From $P_i$ and $\eta$, we determine $\beta^i$ as 
\beq
\beta^j=\delta^{ji}\biggl[
{7 \over 8}P_i - {1 \over 8}(\eta_{,i}+P_{k,i} x^k) \biggr]. \label{eqeta3}
\eeq
Hence, $\beta^i$ satisfies an elliptic-type equation
of the form 
\beq
\delta_{ij} \Delta_{\rm flat} \beta^i + {1 \over 3} \beta^k_{~,kj}=S_j. 
\eeq

As described in Ref. \citen{gw3p2}, if the action 
\beq
I=\int d^3x (\pa_t {\tilde \gamma_{ij}}) 
(\pa_t {\tilde \gamma_{kl}})
\tilde \gamma^{ik} \tilde \gamma^{jl}
\eeq
is minimized with respect to $\beta^i$, 
we obtain the equation of a minimal distortion gauge 
condition \cite{SY}\footnote{
Definition of the minimal distortion gauge in this paper is 
slightly different from the original version in Ref. \citen{SY}.}
for $\beta^i$ as
\beq
\tilde \gamma_{jk}  \tilde \Delta \beta^k
+{1 \over 3} \tilde D_j \tilde D_i \beta^i + \tilde R_{jk} \beta^k
=S_j. \label{eqMD}
\eeq
Thus, the equation for $\beta^i$ in 
the AMD gauge condition is obtained by ignoring the coupling terms 
between $\beta^i$ and $h_{ij}$ in Eq. (\ref{eqMD}). 
Since the ignored terms are expected to be small,\cite{gw3p2} 
we can expect that $I$ 
is approximately minimized in the AMD gauge condition. 
By ignoring these coupling terms, the equation for $\beta^i$ 
is significantly simplified, enabling us to 
impose the spatial gauge condition with a relatively cheap
computational cost. 

The other benefit of the AMD gauge condition is that 
$F_i$ is guaranteed to be small everywhere except in 
the strong field region just around a highly 
relativistic object.\cite{gw3p2} 
This implies that the transverse condition 
$\delta^{ij} \pa_i \tilde \gamma_{jk} = 0$ approximately holds 
for $\tilde \gamma_{ij}$  in the wave zone, helping 
extraction of gravitational waves near the 
outer boundaries of the computational domain. 

One drawback of the AMD gauge condition together with the maximum slicing 
is that the resolution quickly becomes bad in high density regions 
whenever the matter source collapses to a black hole. 
To improve the resolution, we modify the AMD gauge condition around 
a black hole formation region, 
subtracting the radial part of $\beta^k$ with a method 
described in Refs. \citen{gw3p2,bina} and \citen{binas}. 

\subsection{Solution for initial conditions}

Even just before the merger, binary neutron stars that are 
not extremely compact are in a quasiequilibrium state because 
the timescale of the gravitational radiation reaction at Newtonian order 
$\sim 5/\{64\Omega(M_{\rm N}\Omega)^{5/3}\}$ \cite{ST} 
(where $M_{\rm N}$ and $\Omega$ denote the Newtonian total mass of system 
and the orbital angular velocity of binary neutron stars) 
is several times longer than the orbital period. Thus, 
for a realistic simulation of the merger, 
we should prepare a quasiequilibrium state as the initial condition. 
In simulations carried out to this time,
we have constructed such initial data sets in the following manner. 

First, we assume the existence of a helical Killing vector, 
\beq
\ell^{\mu}=\biggl({\pa \over \pa t}\biggr)^{\mu}
+\Omega \biggl({\pa \over \pa \varphi}\biggr)^{\mu}. 
\eeq
Since the emission of gravitational waves violates the 
helical symmetry, this assumption does not strictly hold 
in reality. However, as mentioned 
above, the emission timescale of gravitational waves 
is several times 
longer than the orbital period, even just before the merger 
({\it cf}. Table I), so that this assumption is acceptable 
for computing an approximate quasiequilibrium state. 

In this paper, we consider irrotational binary neutron stars with 
polytropic equations of state. 
The assumption of an irrotational velocity field 
with the existence of the helical 
Killing vector yields the first integral of the Euler equation,\cite{irre}
\beqn
{h \over u^t} +h u_k V^k ={\rm const},\label{feuler}
\eeqn
where $V^k=v^k-\ell^k$. 
Because of the irrotational velocity field, $u_i$ can be written as 
\beq
u_i=h^{-1}D_i \Phi, 
\eeq
where $\Phi$ denotes the scalar velocity potential that 
satisfies the elliptic-type PDE \cite{irre}
\beq
D_i(\rho \alpha h^{-1} D^i \Phi)
-D_i[\rho \alpha  h^{-1} (\ell^i+\beta^i)]=0. \label{fcont}
\eeq
The details of numerical methods for obtaining 
$\Phi$ are given in Ref. \citen{UE}. 

Initial conditions for geometric variables are obtained by solving the 
constraint equations (\ref{ham}) and (\ref{mom}) and equations for 
gauge conditions. Currently, we restrict our attention only to 
initial conditions in which 
$h_{ij}=0=\pa_t h_{ij}$ and $K=0$. 

Using the conformal factor $\psi \equiv e^{\phi}$, 
$\hat A_{ij}=\psi^6 \tilde A_{ij}$ and 
$\hat A^{ij}=\psi^6 \tilde A^{ij}$, the 
Hamiltonian and momentum constraint equations can be rewritten 
in the forms
\beqn
&& \Delta_{\rm flat} \psi = -2\pi \rho_{\rm H} \psi^5 -{1 \over 8}
\hat A_{ij} \hat A^{ij}\psi^{-7} \equiv S_{\psi}, \label{ham2} \\
&& \hat A^{~j}_{i~,j} = 8\pi J_i \psi^6.\label{mom2}
\eeqn
The momentum constraint equation is further rewritten as 
elliptic-type equations 
using two methods. In one method, we use the York decomposition 
\cite{York} as 
\beq
\hat A_{ij}=W_{i,j}+W_{j,i}-{2 \over 3}\delta_{ij} \delta^{kl}
W_{k,l}.\label{hataij}
\eeq
Then, denoting $W_i$ as \cite{gw3p2}
\beq
W_i={7 \over 8}B_i - {1 \over 8}(\chi_{,i}+B_{k,i} x^k), 
\eeq
where $\chi$ and $B_i$ are auxiliary functions, 
Eq. (\ref{mom2}) can be decomposed into 
two simple elliptic-type equations, 
\beq
\Delta_{\rm flat} B_i = 8\pi J_i \psi^6,~~~~~
\Delta_{\rm flat} \chi= -8\pi J_i x^i \psi^6.
\eeq
Since $J_i \psi^6~(=\rho_* \hat u_i)$ is nonzero only in 
the strong field region,  the solution of the momentum constraint 
equation can be accurately obtained even when outer boundary conditions 
are imposed near the strong field zone. 
This method is in particular useful in the case that $J_i\psi^6$ is 
{\it a priori} given. 

In the second method, we write $\tilde A_{ij}$ in terms of $\beta^i$ 
as \cite{WM}
\beq
\tilde A_{ij}={1 \over 2\alpha}\biggl(
\delta_{ik}\beta^k_{~,j}+\delta_{jk}\beta^k_{~,i}
-{2 \over 3}\delta_{ij}\beta^k_{~,k}\biggr), \label{aijequation}
\eeq
where we use $h_{ij}=0=\pa_t h_{ij}$ at $t=0$. 
Then, the equation for $\beta^i$ is 
\beq
\delta_{ij} \Delta_{\rm flat} \beta^j + {1 \over 3} \pa_i \pa_j \beta^j 
+\pa_j \ln\biggl( {\psi^6\over \alpha} \biggr) 
\biggl(\pa_i \beta^j + \delta_{ik}\delta^{jl}\pa_l \beta^k 
-{2 \over 3}\delta^j_{~i}\pa_k \beta^k \biggr)
=16\pi \alpha J_i. \label{ibetaeq}
\eeq
Thus, $\tilde A_{ij}$ can be obtained from either Eq. 
(\ref{hataij}) or Eq. (\ref{aijequation}).  
Quasiequilibrium states 
are obtained by solving equations for $\beta^i$ instead of 
those for $W_i$,\cite{UE,GBM}  because we do not know $J_i\psi^6$ 
{\it a priori} in this case, and also because we need to 
obtain $\beta^i$ to solve the hydrostatic equations. 
Thus, to obtain a quasiequilibrium state, 
we solve Eqs. (\ref{feuler}), (\ref{fcont}), (\ref{ham2}), 
(\ref{ibetaeq}), and (\ref{alpsi}) (see below) iteratively until 
sufficient convergence is achieved. 

In the following, modifying $u_i$ for a fixed density configuration, 
we slightly reduce the angular momentum and/or add an approaching 
velocity to 
quasiequilibrium states at $t=0$ to slightly accelerate the merger. 
Whenever we add such perturbations, we recompute the equations 
for $W_i$ and $\psi$ for the perturbed values of $u_i$ 
to guarantee that the constraint equations are satisfied at $t=0$. 
Then, to adjust the gauge conditions, 
we also recompute Eq. (\ref{ibetaeq}) and then Eq. (\ref{alpsi})  
for given $\rho_*, u_i, K_{ij}$ and $\psi$. 

In addition to the constraint equations, we solve an 
elliptic-type equation for $\alpha$ 
to impose $K=0=\pa_t K$ initially. 
In the conformally flat 3D space, this equation is written 
\beq
\Delta_{\rm flat} (\alpha\psi) = 2\pi \alpha \psi^5 (\rho_{\rm H} + 2 S_k^{~k})
+{7 \over 8}\alpha \psi^{-7}\hat A_{ij} \hat A^{ij}\equiv S_{\alpha\psi} .
\label{alpsi}
\eeq

Equations for $\psi$, $\alpha\psi$, 
$B_i$ and $\chi$ are solved under outer boundary conditions as 
\beqn
&&\psi = 1 + {C_{\psi}  \over r}+O(r^{-3}),~~~
\alpha\psi = 1 + {C_{\alpha\psi}  \over r}+O(r^{-3}),~~~
B_x = {C'_{xx} x + C'_{xy} y \over r^3}+O(r^{-4}),\nonumber \\
&& B_y = {C'_{yx} x + C'_{yy} y \over r^3}+O(r^{-4}),~~~
B_z = {C'_{zz} z \over r^3}+O(r^{-4}),~~~
\chi = {C_{\chi}  \over r}+O(r^{-3}),\nonumber 
\eeqn
where $C_{\psi}, C_{\alpha\psi}$, 
$C'_{xx}, C'_{xy}, C'_{yx}, C'_{yy}, C'_{zz}$, and $C_{\chi}$ are 
constants that can be computed from the volume integrations of 
$S_{\psi}$, $S_{\alpha\psi}$, $J_i x^j$ and $J_i x^i$. 
Here, to derive these boundary conditions, 
we assume that the system is composed of two identical neutron stars. 

\subsection{Definitions of quantities}

In numerical simulations, we often refer to the total baryon rest-mass, 
ADM mass and angular momentum of the system, which are given by 
\beqn
M_* &&\equiv \int d^3x \rho_*, \\
M_{\rm ADM} &&\equiv -{1 \over 2\pi} 
\oint_{r\rightarrow\infty} D^i \psi dS_i \nonumber \\
&&=\int d^3x \biggl[ \rho_{\rm H} e^{5\phi} +{e^{5\phi} \over 16\pi}
\biggl(\tilde A_{ij} \tilde A^{ij}-{2 \over 3}K^2 -\tilde R_k^{~k} 
e^{-4\phi}\biggr)\biggr], \label{eqm00}\\
J &&\equiv {1 \over 8\pi}\oint_{r\rightarrow\infty} 
\varphi^ i \tilde A_i^{~j} e^{6\phi} dS_j \nonumber \\
&&=\int d^3x e^{6\phi}\biggl[J_i \varphi^i  
+{1 \over 8\pi}\biggl( \tilde A_i^{~j} \pa_j \varphi^i 
-{1 \over 2}\tilde A_{ij}\varphi^k\pa_k \tilde \gamma^{ij}
+{2 \over 3}\varphi^j \pa_j K \biggr) \biggr],~~~
\label{eqj00}
\eeqn
where $dS_j=r^2 D_j r d(\cos\theta)d\varphi$ 
and $\varphi^j=-y(\pa_x)^j + x(\pa_y)^j$. 
Here, $M_*$ is a conserved quantity, and it uniquely specifies a model 
of stable binary neutron stars for a given $\Gamma$.
$M_{\rm ADM}$ and $J$ decrease 
as a result of the radiation reaction of gravitational waves. 

At $t=0$, the ADM mass and angular momentum 
are given in the conformal flat three space with $K=0$ as 
\beqn
&&M_{\rm ADM0}=\int d^3x \biggl( \rho_{\rm H} \psi^5 +{1 \over 16\pi\psi^7}
\hat A_{ij} \hat A^{ij}\biggr), \\
&&J_0=\int d^3x J_i \varphi^i \psi^6. \label{eqj0}
\eeqn
{}From these quantities, we define a nondimensional angular momentum 
parameter as $q \equiv J_0/M_{\rm ADM0}^2$. 

Instead of $M_*$, we specify a model of binary neutron stars 
using the compactness $\comp$, which is defined as the ratio of 
the ADM mass to the circumferential radius of 
a spherical neutron star in isolation (see Tables I and II). 
To specify a model, 
we also use the ratio of the total baryon rest-mass 
of the system to the maximum allowed mass of spherical 
neutron stars for a given equation of state $M_{*\rm max}^{\rm sph}$,
\beq
R_{\rm mass} \equiv {M_* \over M_{*\rm max}^{\rm sph}}.
\eeq

In addition to the above quantities, 
we use the relation between the baryon rest-mass and 
specific angular momentum that we define as 
\beq
M_*(j)=\int_{j' > j} d^3x' \rho_*(x'),
\eeq
where $j$ denotes the specific angular momentum 
$j\equiv h u_i \varphi^i$. Here, $M_*(j)/M_*$ denotes a baryon 
rest-mass fraction whose specific angular momentum 
is larger than a value $j$ (i.e., $M_*(j=0)/M_*=1$). 

Physical units enter the problem only through the polytropic 
constant $\kappa$, which can be chosen arbitrarily or else completely 
scaled out of the problem. Since $\kappa^{n/2}$ (in geometrical 
units $c=G=1$) has the dimension of length, dimensionless variables 
can be constructed as 
\beqn
&& \bar M_* = M_* \kappa^{-n/2}, 
~~~\bar M_{\rm ADM} = M_{\rm ADM} \kappa^{-n/2}, ~~~
\bar R = R \kappa^{-n/2},  \nonumber \\
&& \bar J = J \kappa^{-n}, ~~~
\bar \rho = \rho \kappa^{n}, ~~~{\rm and}~~~
\bar \Omega = \Omega \kappa^{n/2}.
\eeqn
In the following, we present only these dimensionless quantities. 

\subsection{Code tests and accuracy estimation of simulations}

Several test simulations, including the 
spherical collapse of dust, stability of spherical stars,
mode analysis of spherical stars, 
and evolution of rotating stars, have been carried out 
to check the accuracy of the numerical code. A list of these
test simulations and their results 
are given in Ref. \citen{gr3d}. (See also Ref. \citen{rot1}.)

For the simulations presented in this paper, 
we monitored the violation of the Hamiltonian constraint,
conservation of the baryon rest-mass, conservation of the ADM mass
[$M_{\rm ADM}+$(energy loss through gravitational radiation)
should be conserved] and conservation of angular momentum
[$J+$(angular momentum loss through gravitational radiation].
Simulations were stopped when the conservation of the 
ADM mass was significantly violated. 

As described in Ref. \citen{bina}, 
the violation of the Hamiltonian constraint
is locally measured by the equation as
\beq
\displaystyle
f_{\psi} \equiv 
{\Bigl|\tilde \Delta \psi - {\psi \over 8}\tilde R_k^{~k} 
+ 2\pi \rho_{\rm H} \psi^5 
+{\psi^5 \over 8} \Bigl(\tilde A_{ij} \tilde A^{ij}
-{2 \over 3}K^2\Bigr)\Bigr| \over
|\tilde \Delta \psi | + |{\psi \over 8}\tilde R_k^{~k}| 
+ |2\pi \rho_{\rm H} \psi^5| 
+{\psi^5 \over 8} \Bigl(|\tilde A_{ij} \tilde A^{ij}|+
{2 \over 3}K^2\Bigr)}. 
\eeq
We briefly discuss the magnitude of $f_{\psi}$ and
conservation of the mass and angular momentum in the Appendix. 

\section{Analysis of gravitational waves} 

As in Ref. \citen{gw3p2}, 
we analyze gravitational waveforms using two methods. 
In both cases, we extract gravitational waves 
near the outer boundaries of the computational domain. 

In one method, we compute the following quantities along the $z$-axis: 
\beq
\bar h_+ \equiv {z_{\rm obs} \over 2 M_{\rm ADM0}\comp}
(\tilde \gamma_{xx} - \tilde \gamma_{yy}), \hskip 5mm
\bar h_{\times} \equiv
{z_{\rm obs} \over M_{\rm ADM0}\comp} \tilde \gamma_{xy}. 
\label{hhh}
\eeq
Here $z_{\rm obs}$ denotes 
the location at which we extract gravitational waves. 
Since the gauge condition we adopt 
is approximately transverse and traceless in the wave zone, 
$\bar h_+$ and $\bar h_{\times}$ are expected to be 
appropriate measures of gravitational waves emitted. 
The amplitude of gravitational waves will be 
largest along the $z$-axis as a result of our choice of the orbital plane, 
so that $\bar h_{+,\times}$ can be used to measure 
the maximum amplitude of gravitational waves. 
We also note that $\bar h_{+}$ and $\bar h_{\times}$
are composed only of $|m|=2$ modes,
because other modes vanish along the $z$-axis.

The amplitude of gravitational waves 
for equal-mass binaries along the $z$-axis 
in the quadrupole formula is 
\beq
h_{\rm GW} = {M_{\rm N}^2 \over z_{\rm obs} a},
\eeq
where $M_{\rm N}$ is a Newtonian mass and $a$ an orbital separation. 
At the point of contact of two spherical stars in a circular orbit, 
we have 
\beq
h_{\rm GW} = {M_{\rm N} \comp \over z_{\rm obs}}. 
\eeq
Thus, by normalizing the amplitude by $M_{\rm ADM0}$ and $\comp$, as 
in Eq. (\ref{hhh}), the maximum values of $\bar h_{+}$ and $\bar h_{\times}$
are scaled to become approximately unity. 

To search for the dominant frequency of gravitational waves, we 
compute the Fourier spectra defined as
\beq
\bar h_{+,\times}(f)=\int^{t_f}_{t_i} e^{2\pi i f t} \bar h_{+,\times}dt.
\eeq
In the analysis, we choose $t_f$ to be the time at which we stop simulation, 
for simplicity. Before $t < z_{\rm obs}$, no waves 
propagate to $z_{\rm obs}$, so that we choose $t_i \approx z_{\rm obs}$.

Gravitational waves are also measured in terms of the gauge invariant 
Moncrief variables in a flat spacetime in the following 
manner.\cite{moncrief}\footnote{
The Moncrief formalism was originally derived for the 
Schwarzschild spacetime. We here apply his formalism 
in a flat spacetime.} 
First, we carry out a coordinate transformation for
the three-metric from Cartesian coordinates 
to spherical polar coordinates, and then split $\gamma_{ij}$ 
as $\eta_{ij}+\sum_{lm} \zeta_{ij}^{lm}$, where 
$\eta_{ij}$ is the flat metric in spherical polar coordinates 
and $\zeta_{ij}^{lm}$ is given by 
\beqn
\zeta_{ij}^{lm}=&& \left(
\begin{array}{lll}
\displaystyle 
H_{2lm} Y_{lm} & h_{1lm} Y_{lm,\theta}& h_{1lm} Y_{lm,\varphi}\\
\ast& r^2(K_{lm}Y_{lm}+G_{lm}W_{lm})&r^2G_{lm}X_{lm} \\
\ast& \ast&r^2\sin^2\theta(K_{lm}Y_{lm}-G_{lm}W_{lm}) \\
\end{array}
\right) \nonumber \\
&&+\left(\begin{array}{ccc}
0 &  -C_{lm} \pa_{\varphi} Y_{lm}/\sin\theta
& C_{lm} \pa_{\theta}Y_{lm}\sin\theta  \\
\ast & r^2D_{lm}X_{lm}/\sin\theta
             & -r^2 D_{lm}W_{lm}\sin\theta  \\
\ast & \ast & -r^2 D_{lm}X_{lm}\sin\theta \\
\end{array}
\right). 
\eeqn
Here, $\ast$ denotes the relation of symmetry. The quantities 
$H_{2lm}$, $h_{1lm}$, $K_{lm}$, $G_{lm}$, $C_{lm}$ and 
$D_{lm}$ are functions of $r$ and $t$, and are calculated 
by performing integrations over a two-sphere of given radius 
[see Ref. \citen{SN} for details]. Also, 
$Y_{lm}$ is the spherical harmonic function, and 
$W_{lm}$ and $X_{lm}$ are defined as 
\beqn
W_{lm} \equiv \Bigl[ (\pa_{\theta})^2-\cot\theta \pa_{\theta}
-{1 \over \sin^2\theta} (\pa_{\varphi})^2 \Bigl] Y_{lm},
\hskip 5mm
X_{lm} \equiv 2 \pa_{\varphi} \Bigl[ \pa_{\theta}-\cot\theta \Bigr] 
Y_{lm}.~~~~~ 
\eeqn
The gauge invariant variables of even and odd parities are then 
defined as 
\beqn
&&R_{lm}^{\rm E}(t,r) \equiv 
\sqrt{2(l-2)! \over (l+2)!}
\Bigl\{ 4k_{2lm}+l(l+1)k_{1lm} \Bigr\}, \\
&&R_{lm}^{\rm O}(t,r) \equiv \sqrt{2(l+2)! \over (l-2)!}
\biggl({C_{lm} \over r}+r \pa_r D_{lm}\biggr),
\eeqn
where
\beqn
k_{1lm}&& \equiv K_{lm}+l(l+1)G_{lm}+2r \pa_r G_{lm}-2{h_{1lm} \over r},\\
k_{2lm}&& \equiv {H_{2lm} \over 2} - {1 \over 2}{\pa \over \pa r}
\Bigl[r\{ K_{lm}+l(l+1)G_{lm} \} \Bigr].
\eeqn

Using the above variables, 
the energy luminosity and angular momentum flux 
of gravitational waves can be calculated as 
\beqn
&&{dE \over dt}={r^2 \over 32\pi}\sum_{l,m}\Bigl[
|\pa_t R_{lm}^{\rm E}|^2+|\pa_t R_{lm}^{\rm O}|^2 \Bigr],
\label{dedt} \\
&&{dJ \over dt}={r^2 \over 32\pi}\sum_{l,m}\Bigl[
 |m(\pa_t R_{lm}^{\rm E}) R_{lm}^{\rm E} |
+|m(\pa_t R_{lm}^{\rm O}) R_{lm}^{\rm O} | \Bigr]. 
\label{dJdt} 
\eeqn
The total radiated energy and angular momentum are calculated from
\beq
\Delta E = \int dt {dE \over dt}, \hskip 5mm
\Delta J = \int dt {dJ \over dt}.
\eeq
We have computed modes with $l=2$, 3 and 4, and found that 
even modes with $l=|m|=2$ are dominant. For this reason,
in the following, we focus only on these modes. 

\section{Numerical results}

\subsection{Products after merger}

In Table I, we list several quantities that characterize 
quasiequilibrium states of irrotational binary neutron stars 
used as initial conditions in the present simulations. 
All quantities are dimensionless and 
appropriately scaled with respect to $\kappa$. 
We choose binaries at innermost 
orbits for which the Lagrange points appear at the inner edge of 
neutron stars.\cite{USE}  Since the orbits of such binaries are stable, 
we reduce the angular momentum slightly to accelerate the merger. 
Also, we add an approaching velocity at $t=0$ for some cases 
(see discussion below). The values of 
$M_{\rm ADM}$ and $q$ listed in Table I are those of 
quasiequilibrium states before reducing the angular momentum 
and before adding the approaching velocity. 
Note that the frequency of gravitational waves for these 
quasiequilibria are given by 
\beq
f_{\rm QE}={\Omega_0 \over \pi} \approx 960~{\rm Hz}
\biggl({2.8M_{\odot} \over M_{\rm ADM0}}\biggr)
\biggl({C_0 \over 0.12}\biggr)^{3/2}, 
\label{fqe}
\eeq
where $\Omega_0$ denotes the angular velocity of 
quasiequilibrium states. Thus, the orbital period of 
the quasiequilibria, $P_{t=0}$, is
\beq
P_{t=0} \approx 2.08~{\rm msec}
\biggl({2.8M_{\odot} \over M_{\rm ADM0}}\biggr)^{-1}
\biggl({C_0 \over 0.12}\biggr)^{-3/2}.  
\eeq
Here, $C_0$ is a compactness parameter of orbits defined as 
\beq
C_0 \equiv (M_{\rm ADM0}\Omega_0)^{2/3} \sim {M_{\rm ADM0} \over a}, 
\eeq
where $a$ denotes the initial orbital separation. Products of the merger
which we found when we stopped simulations are also described in 
Table I, in the last column. 


\begin{table}[t]
\caption{
A list of several quantities for 
quasiequilibria of irrotational binary neutron stars with $\Gamma=2.25$ 
and $2$. The 
compactness of each star in isolation, $(M/R)_{\infty}$, 
the maximum density, $\bar \rho_{\rm max}=\kappa^n \rho_{\rm max}$, 
the total baryon rest-mass, $\bar M_* = \kappa^{-n/2} M_*$, the 
ADM mass at $t=0$, $\bar M_{\rm ADM0} = \kappa^{-n/2} M_{\rm ADM0}$, 
$q=J_0/M_{\rm ADM0}^2$, $\hat P_{t=0}\equiv P_{t=0}/M_{\rm ADM0}$, 
the orbital compactness, $C_0\equiv (M_{\rm ADM0}\Omega_0)^{2/3}$, 
the ratio of the emission timescale of gravitational waves to 
the orbital period at Newtonian order, 
$R_{\tau}=5(M_{\rm ADM0}\Omega_0)^{-5/3}/128\pi$, 
the ratio of the baryon rest-mass of each star to 
the maximum allowed mass for a spherical star, 
$R_{\rm mass}\equiv M_*/M_{*~\rm max}^{\rm sph}$, 
and products we found when we stopped simulations are listed. 
All quantities are normalized by 
$\kappa$ appropriately to be dimensionless: 
We can rescale the mass to a desirable value  
by appropriately choosing $\kappa$. 
Here, $M_{*~\rm max}^{\rm sph}$ denotes the maximum allowed mass 
of a spherical star ($\bar M_{*~\rm max}^{\rm sph} \approx 0.162$ 
at $\bar \rho_{\rm max} \approx 0.524$ for $\Gamma=2.25$ and 
$\bar M_{*~\rm max}^{\rm sph} \approx 0.180$ 
at $\bar \rho_{\rm max} \approx 0.32$ for $\Gamma=2$). 
BH and NS mean ``black hole'' and ``neutron star''. 
MA (marginal) implies that we were not able to determine the product.
}
\begin{center}
\begin{tabular}{|c|c|c|c|c|c|c|c|c|c|c|c|} \hline
\hspace{-2mm} Model \hspace{-2mm} &\hspace{-4mm} $\Gamma$ \hspace{-4mm} 
&\hspace{-3mm} $\comp$ \hspace{-3mm} &
$\bar \rho_{\rm max} $  &  $\bar M_*$  & 
\hspace{-2mm} $\bar M_{\rm ADM}$ \hspace{-2mm} &  $q$  & 
$\hat P_{\rm t=0}$  &  $C_0$  &  $R_{\tau}$  & 
\hspace{-2mm}$R_{\rm mass}$ \hspace{-2mm} &
\hspace{-2mm} Product \hspace{-4mm}\\ \hline\hline
(A)& 2.25 &0.12 & 0.139  & 0.186 & 0.173 & 1.02 &233& 0.0897 & 5.1  &1.15 
&NS \\ \hline
(B)& 2.25 &0.14 & 0.169  & 0.216 & 0.198 & 0.97 &182& 0.106 & 3.4  &1.33
&BH  \\ \hline
(C)& 2.25 &0.16 & 0.202  & 0.244 & 0.220 & 0.93 &145& 0.124 & 2.3  &1.51 
&BH  \\ \hline
(D)& 2.25 &0.17 & 0.220  & 0.257 & 0.231 & 0.92 &130& 0.132 & 2.0  &1.59 
&BH  \\ \hline
(E)& 2 &0.10 & 0.0695  & 0.224 &  0.211 & 1.07 &315&0.0735 & 8.5  &1.24
&NS\\ \hline
(F)& 2 &0.11 & 0.0798  & 0.242 &  0.227 & 1.03 &271&0.0813 & 6.6  &1.35
&MA\\ \hline
(G)& 2 &0.12 & 0.0910  & 0.261 & 0.243 & 1.00 &236& 0.0892 & 5.2  &1.44
&BH  \\ \hline
(H)& 2 &0.14 & 0.117  & 0.292 & 0.270 & 0.94 &184& 0.105 & 3.4  &1.62
&BH  \\ \hline
(I)& 2 &0.16 & 0.149  & 0.320  & 0.292 & 0.90 &146& 0.123 & 2.3  & 1.78
&BH  \\ \hline
\end{tabular}
\end{center}
\end{table}


\subsubsection{Set-up for computational grids}

We have performed the simulations using a fixed uniform grid 
and assuming reflection symmetry with respect to the $z=0$ plane. 
(i.e., the equatorial plane is chosen as the orbital plane.) 
The typical grid size is (505, 505, 253) for $(x, y, z)$. 
The largest grid size is (633, 633, 317). 
To investigate numerical effects associated 
with the location of outer boundaries, we performed simulations
choosing smaller grid sizes. 
The grid covers the region $-L \leq x, y \leq L$ and 
$0 \leq z \leq L$, where $L$ is a constant. 
The grid spacing is determined 
from the condition that the major diameter of each star is covered 
with 33 grid points initially. 
To investigate effects of numerical dissipation and diffusion, 
we also choose larger grid spacing in the calibration. 
With a (505, 505, 253) grid size, 
the computational memory required is about 120 GBytes, and 
the computational time for one model is typically about 100 CPU hours 
for about 10000 timesteps using 32 processors 
on FACOM VPP5000 in the data processing center of NAOJ. 

\begin{table}[t]
\caption{
List of the various parameter values in the simulation for 
models (A), (B), (C), (D), (E), (F), (G), (H) and (I): 
The model, the compactness, the $q$ parameter at $t=0$, the ratio of the 
approaching velocity to orbital velocity, the grid number, 
$L$ in units of $\lambda_0$, 
$L$ in units of $M_{\rm ADM0}$, and the 
grid number inside the initial gravitational mass length 
$M_{\rm ADM0}/\Delta x$ are listed. 
Here, $v_{\rm orb}\equiv (M_{\rm ADM0} \Omega_0)^{1/3}=C_0^{1/2}$.
}
\begin{center}
\begin{tabular}{|c|c|c|c|c|c|c|c|} \hline
Model & 
\hspace{0mm} $(M/R)_{\infty}$ \hspace{0mm} &
\hspace{0mm} $q$ \hspace{0mm} &
\hspace{0mm} $v_{\rm app}/v_{\rm orb}$ \hspace{0mm} &
\hspace{0mm} Grid number  \hspace{0mm} &
$L/\lambda_{0}$ & $L/M_{\rm ADM}$ & 
$M_{\rm ADM} / \Delta x$  \\ \hline\hline
(A-0)&0.12 & 1.00  & 0   & (293, 293, 147) & 0.322 & 37.5 &3.90\\ \hline
(A-1)&0.12 & 1.00  & 0.1 & (313, 313, 157) & 0.344 & 40.1 &3.90\\ \hline
(A-2)&0.12 & 1.00  & 0   & (505, 505, 253) & 0.556 & 64.7 &3.90\\ \hline
(A-3)&0.12 & 1.00  & 0   & (377, 377, 189) & 0.556 & 64.7 &2.91\\ \hline
(B-0)&0.14 & 0.95  & 0   & (293, 293, 147) & 0.343 & 31.1 &4.69\\ \hline
(B-1)&0.14 & 0.96  & 0.1 & (313, 313, 157) & 0.366 & 31.1 &4.69\\ \hline
(B-2)&0.14 & 0.95  & 0   & (505, 505, 253) & 0.592 & 53.7 &4.69\\ \hline
(B-3)&0.14 & 0.95  & 0   & (377, 377, 189) & 0.592 & 53.7 &3.50\\ \hline
(C-0)&0.16 & 0.91  & 0   & (293, 293, 147) & 0.364 & 26.3 &5.55\\ \hline
(C-2)&0.16 & 0.91  & 0   & (505, 505, 253) & 0.628 & 45.4 &5.55\\ \hline
(C-3)&0.16 & 0.91  & 0   & (633, 633, 317) & 0.788 & 56.9 &5.55\\ \hline
(D-0)&0.17 & 0.89  & 0   & (313, 313, 157) & 0.398 & 26.0 &6.00\\ \hline
(E-1)&0.10 & 1.06 &0.01& (505, 505, 253) & 0.492 & 77.6&3.25\\ \hline
(F-1)&0.11 & 1.02 &0.01& (505, 505, 253) & 0.514 & 69.5&3.62\\ \hline
(G-0)&0.12 & 0.99 &0.01& (249, 249, 125) & 0.263 & 30.9&4.01 \\ \hline
(H-1)&0.14 & 0.93 &0.02& (505, 505, 253) & 0.569 & 52.3&4.82 \\ \hline
(I-1)&0.16 & 0.89 &0.02& (505, 505, 253) & 0.606 & 44.1&5.71 \\ \hline
\end{tabular}
\end{center}
\end{table}

To perform an accurate simulation excluding spurious effects 
associated with imposing outer boundary conditions in a local zone, 
$L$ should be larger than the wavelength of gravitational waves, $\lambda$. 
Using almost the entire capacity of the supercomputer 
VPP-5000 at NAOJ, it is possible to extend $L$ as large as $\agt \lambda$. 
However, such a large simulation is limited by the 
computational time assigned to us.  Since the grid spacing should be 
smaller than 1/30 of the diameter of each star 
to avoid large numerical dissipation and diffusion (see below), 
the size of the computational domain along each axis, $L$, becomes 
about 0.5--0.6 $\lambda_0$ in this work. 
(Here, $\lambda_0=\pi/\Omega_0$.) This implies 
that gravitational waves in the early stage are not very accurately  
computed. On the other hand, the wavelength of quasi-periodic 
waves of the merged object excited during merger 
is much smaller than $\lambda_0$ and $L$, so that 
the waveforms in the merger stage can be computed accurately. 
This point is discussed in detail in \S 4.2. 

\subsubsection{Set-up for initial data}

As found in Ref. \citen{USE}, orbits for all irrotational binaries 
of equal mass with 
$\Gamma < 2.5$ are dynamically stable from infinite separation 
to the innermost orbit at which 
Lagrange points appear at the inner edges of neutron stars. 
Thus, the merger in reality should be triggered by the 
radiation reaction of gravitational waves. As argued above, 
however, the radiation reaction in early stages during 
which $\lambda > L$ is not 
accurately computed with our current computational resources. 
To induce prompt merger by destabilizing the orbital motion, we slightly 
decrease the angular momentum at $t=0$ from the equilibrium value 
(by less than $1\%$).\footnote{
As initial data sets for the simulation with $\Gamma=2.25$ 
and for the simulations in Ref. \citen{bina},
we adopted solutions of binary neutron stars 
in quasiequilibria which are computed using the numerical code used 
in Ref. \citen{USE}. 
Our latest computation for the quasiequilibrium 
computed with a revised code indicates 
2--3 \% smaller angular momentum than in the previous computation, while 
other quantities, such as the gravitational mass and angular velocity, 
do not vary much. In our previous code, 
the magnitude of the shift vector is slightly overestimated around the
strong field region.  As a result, the angular momentum is systematically
overestimated. 
We became aware of this fact after we finished most of the simulations 
for $\Gamma=2.25$. Thus, all the results presented in this paper 
were computed using the old initial data. We have adopted the new 
data sets only for $\Gamma=2$ cases. 
The relevant quantities for $\Gamma=2.25$
computed using the new code are as follows (compare with Table I).  

\begin{center}
\vspace{0.cm}
\begin{tabular}{|c|c|c|c|c|c|c|c|} \hline
Model & $(M/R)_{\infty}$ &
$\bar \rho_{\rm max} $ & $\bar M_*$ & $\bar M_{\rm ADM0}$ &
$q$ &  $P_{\rm t=0}/M_{\rm ADM0}$  & $C_0$  \\ \hline
(A)&0.12 & 0.139  & 0.186 & 0.173 & 1.00 & 232 & 0.0901 \\ \hline
(B)&0.14 & 0.169  & 0.216 & 0.198 & 0.95 & 181 & 0.106 \\ \hline
(C)&0.16 & 0.202  & 0.244 & 0.220 & 0.91 & 145 & 0.123 \\ \hline
(D)&0.17 & 0.221  & 0.257 & 0.231 & 0.89 & 131 & 0.132 \\ \hline
\end{tabular}
\vspace{0.cm}
\end{center}

The latest computation is expected to provide
more accurate initial data sets. Hence, 
reader might think it desirable 
to re-perform all simulations with the new initial data for
$\Gamma=2.25$. However, we think it not necessary to do this at present. 
Our present main purpose is to demonstrate that it is possible to extract 
gravitational waveforms within $\sim 10\%$ error
as well as to determine the fate of mergers and 
the dependence of this fate
on the compactness of neutron stars and equations of state. 
For these purposes, slight inaccuracy 
in the initial data sets is not important. 
By performing several test simulations, 
we have confirmed that qualitative properties are not modified 
using the new data sets for $\Gamma=2.25$ with a 
small depletion factor (0--2\%) and small grid numbers. 
Even the quantitative results agree well with those 
presented in this paper for a particular
choice of the depletion factor in this range.
Therefore we can conclude that the qualitative features in 
the merger stage ($t>P_{t=0}$) depend very weakly on the 
depletion factor, as long as the magnitude is of order 0.01.}

Using the quadrupole formula, 
the angular momentum loss in one orbital period 
$\Delta J_{\rm one}$ is estimated as $4\pi M_{\rm ADM0} C_0^2/5$,\cite{ST} 
and hence the ratio of $\Delta J_{\rm one}$ to the total angular momentum 
$J$ is estimated as
\beq
{\Delta J_{\rm one} \over J}
= {4\pi \over 5q}C_0^2 \approx 3.6\%\biggl({1 \over q}\biggr)
\biggl({C_0 \over 0.12}\biggr)^2. 
\eeq
Thus, the artificial decrease of the orbital angular momentum
by $1\%$ 
is much smaller than the loss through gravitational radiation 
in one orbital period. 

As we mentioned in Ref. \citen{bina}, we also performed
several test simulations with small grid number varying the
depletion factor in the range 0 -- 2\%, but we found that the 
qualitative results for the merger stage are not modified at all. 
The depletion only affects the evolution of the late
inspiraling stage before the hydrodynamic interaction of
two neutron stars sets in. 

To investigate the effect of the approaching velocity on 
the product and gravitational waveforms, as well as 
to accelerate the merger in some cases, 
we also add an approaching velocity $v_{\rm app}$ 
for some models by changing the component of the four-velocity as
\beq
u_x = (u_x)_{\rm original} - {v_{\rm app} \over 2}{x \over |x|},
\eeq
where we assume that the center of mass of two stars is 
located along the $x$-axis at $t=0$. 
For $\Gamma=2$, we add $v_{\rm app}$ 
as in Ref. \citen{bina}. Explicitly, 
we set $v_{\rm app}=(0.01-0.02)v_{\rm orb}$, where the 
orbital velocity $v_{\rm orb}$ is defined as 
$(M_{\rm ADM0} \Omega_0)^{1/3}=C_0^{1/2}$.
The order of magnitude of this approaching velocity is 
in approximate agreement (within a factor of 2--3) 
with that induced by the radiation reaction of 
gravitational waves.\cite{SU01} 
For $\Gamma=2.25$, $v_{\rm app}$ was basically chosen to be zero. 
To investigate its effect for comparison, 
we choose a large nonzero value for some models as 
$v_{\rm app}=0.1v_{\rm orb}$ (see \S 4.2.3). 

These treatments of the initial conditions 
could cause a certain systematic deviation from a realistic merger in 
the early stages of the simulation, i.e., before the
hydrodynamic interaction of the two neutron stars sets in. Obviously, 
a more realistic simulation taking into account the radiation reaction 
in early stages with a large computational domain or 
with improved wave extraction techniques \cite{Bishop} 
is desirable. However, as we see below, 
qualitative and even quantitative results for the merger stage 
are not affected strongly by these initial conditions, nor by 
the few percent artificial decrease of the angular momentum.

\begin{figure}[t]
\begin{center}
\epsfxsize=2.2in
\leavevmode
\epsffile{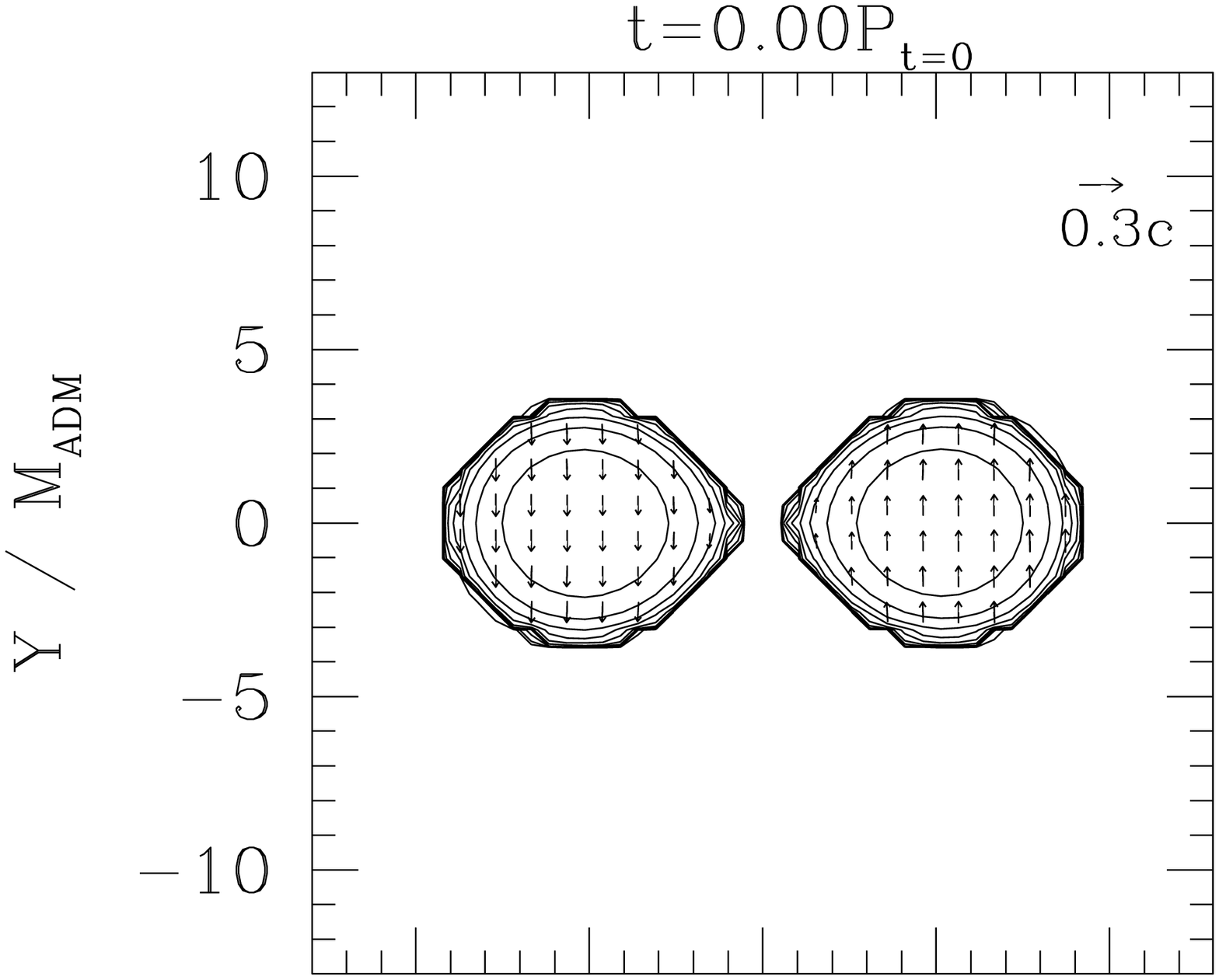}
\epsfxsize=2.2in
\leavevmode
\hspace{-1.6cm}\epsffile{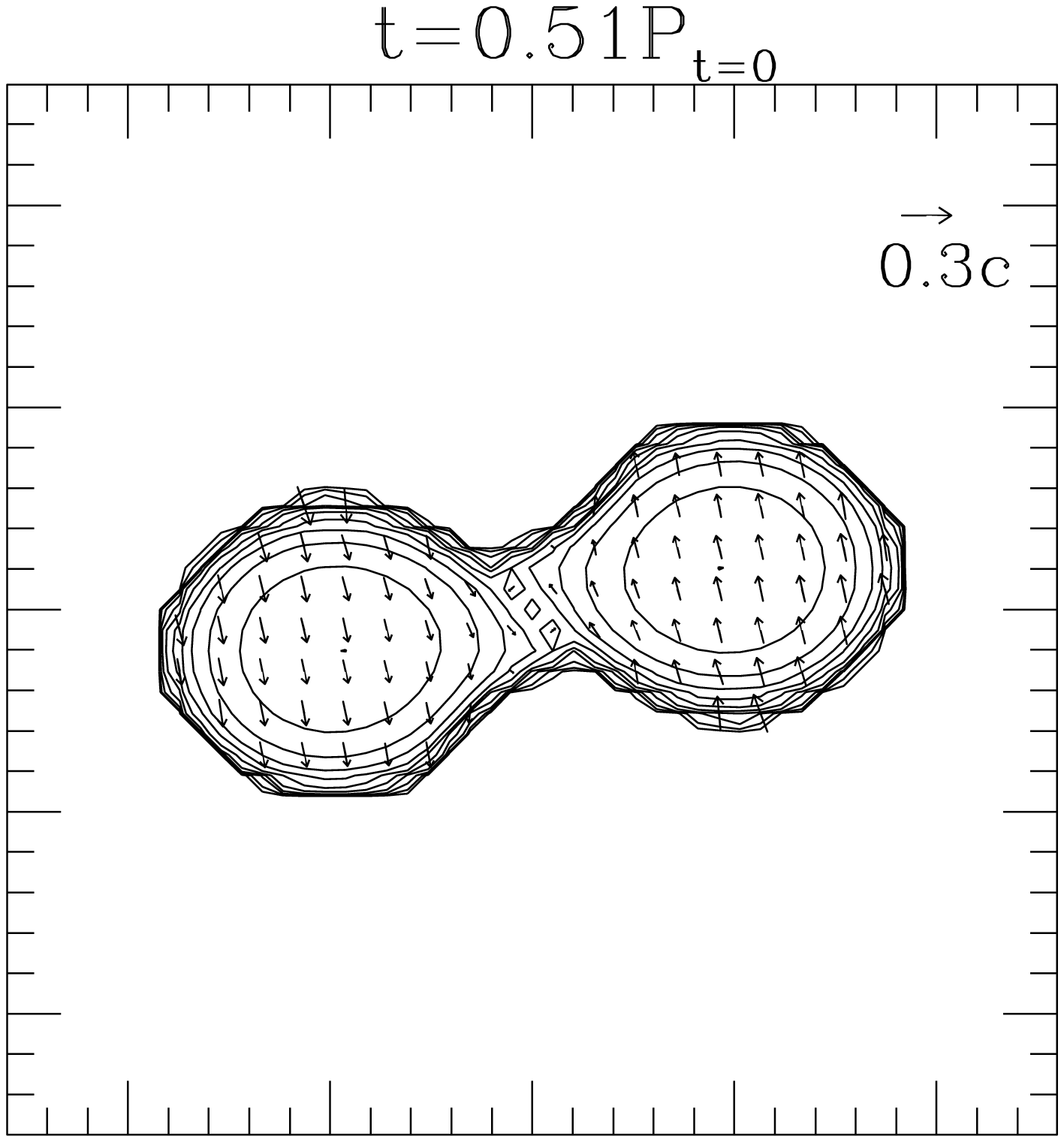} 
\epsfxsize=2.2in
\leavevmode
\hspace{-1.6cm}\epsffile{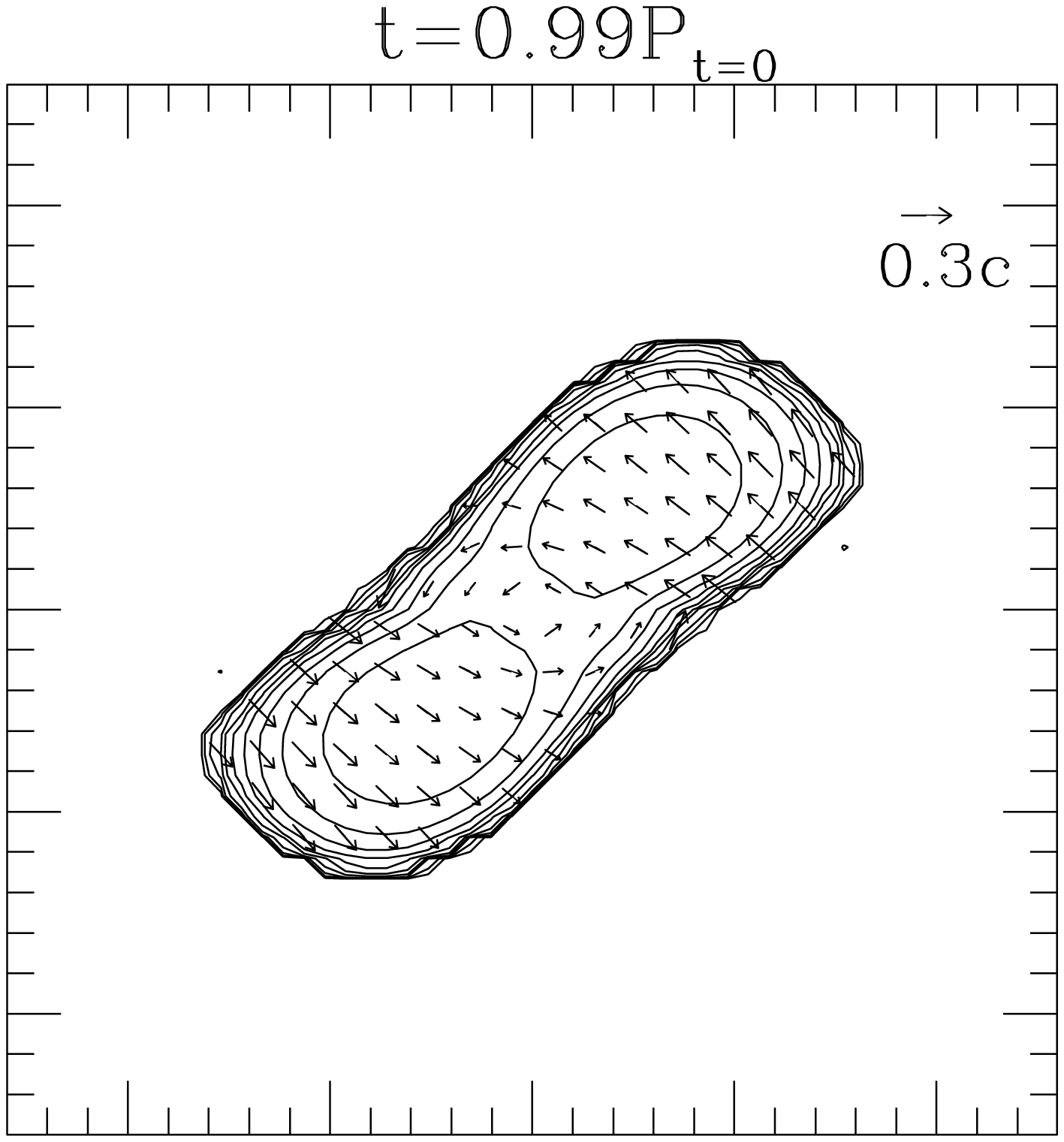} \\
\vspace{-1.1cm}
\epsfxsize=2.2in
\leavevmode
\epsffile{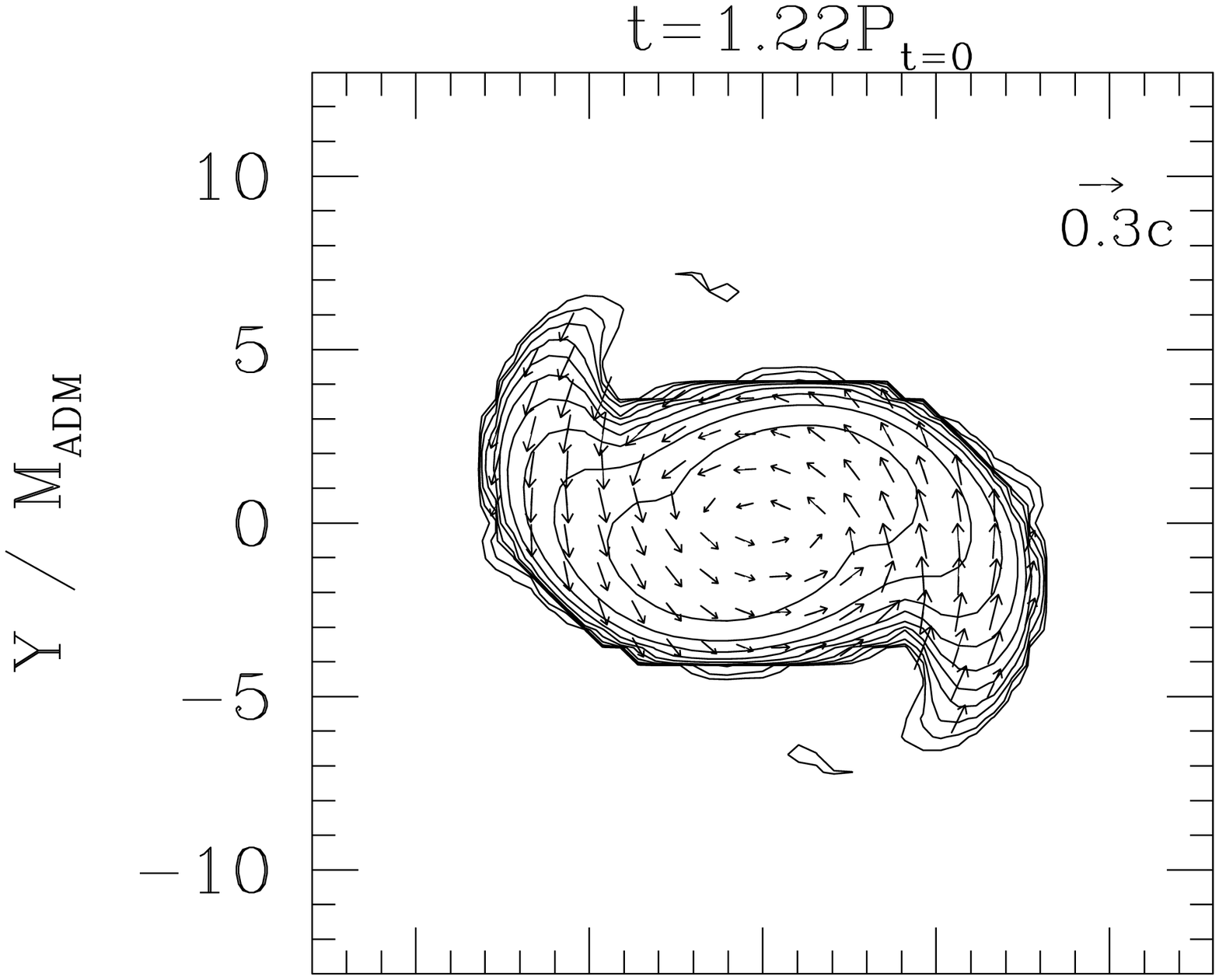} 
\epsfxsize=2.2in
\leavevmode
\hspace{-1.6cm}\epsffile{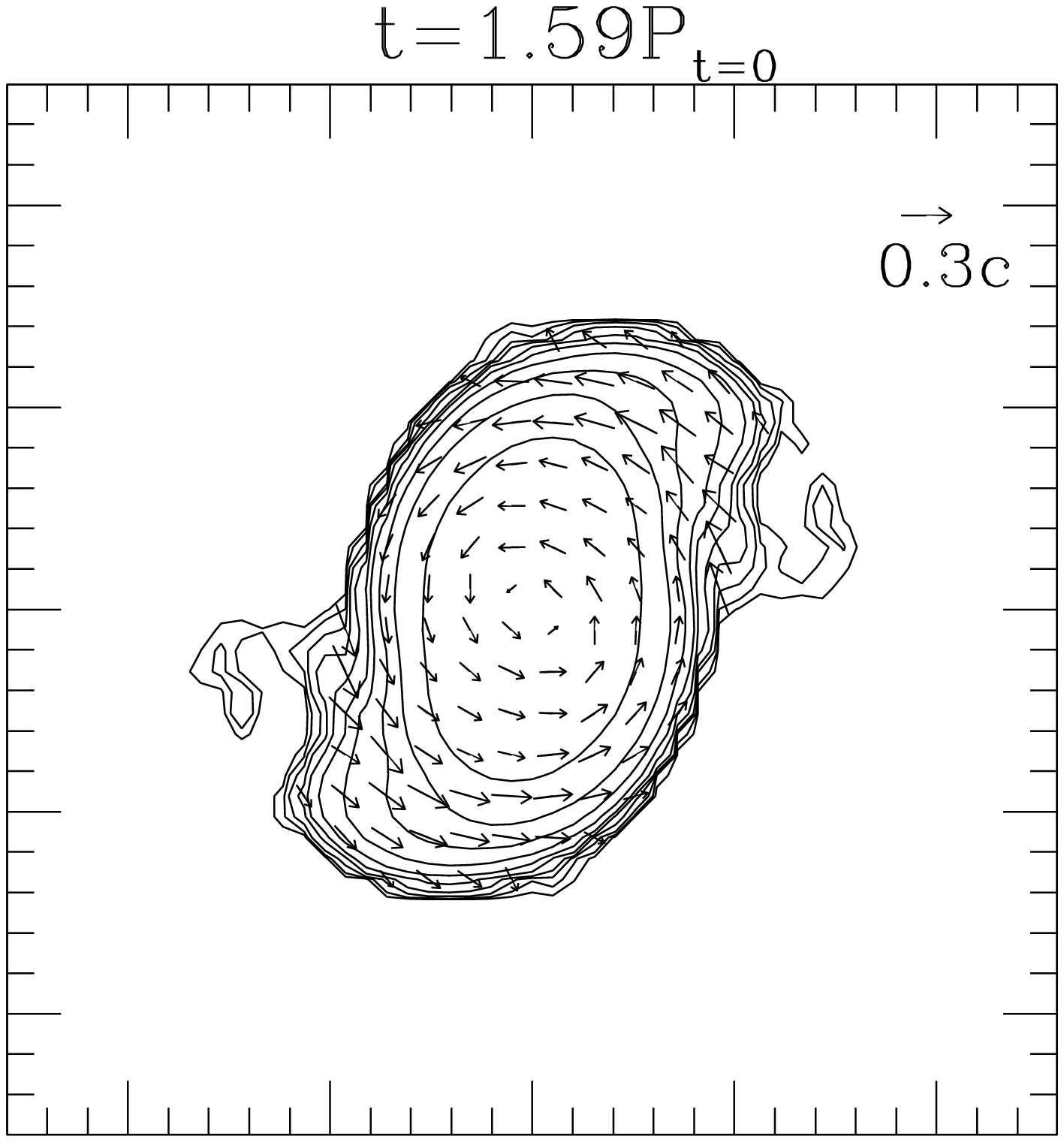}
\epsfxsize=2.2in
\leavevmode
\hspace{-1.6cm}\epsffile{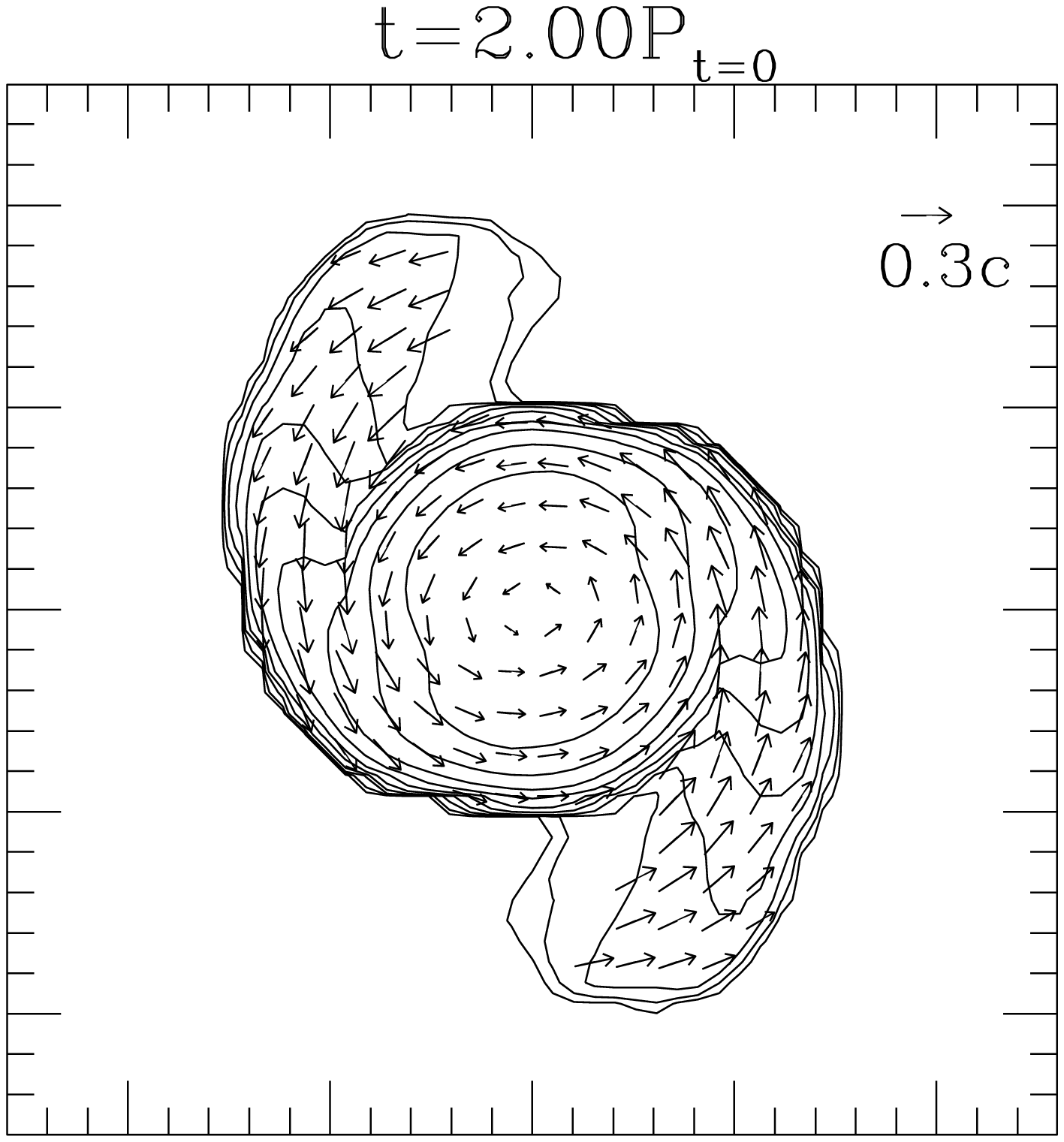}\\ 
\vspace{-1.1cm}
\epsfxsize=2.2in
\leavevmode
\epsffile{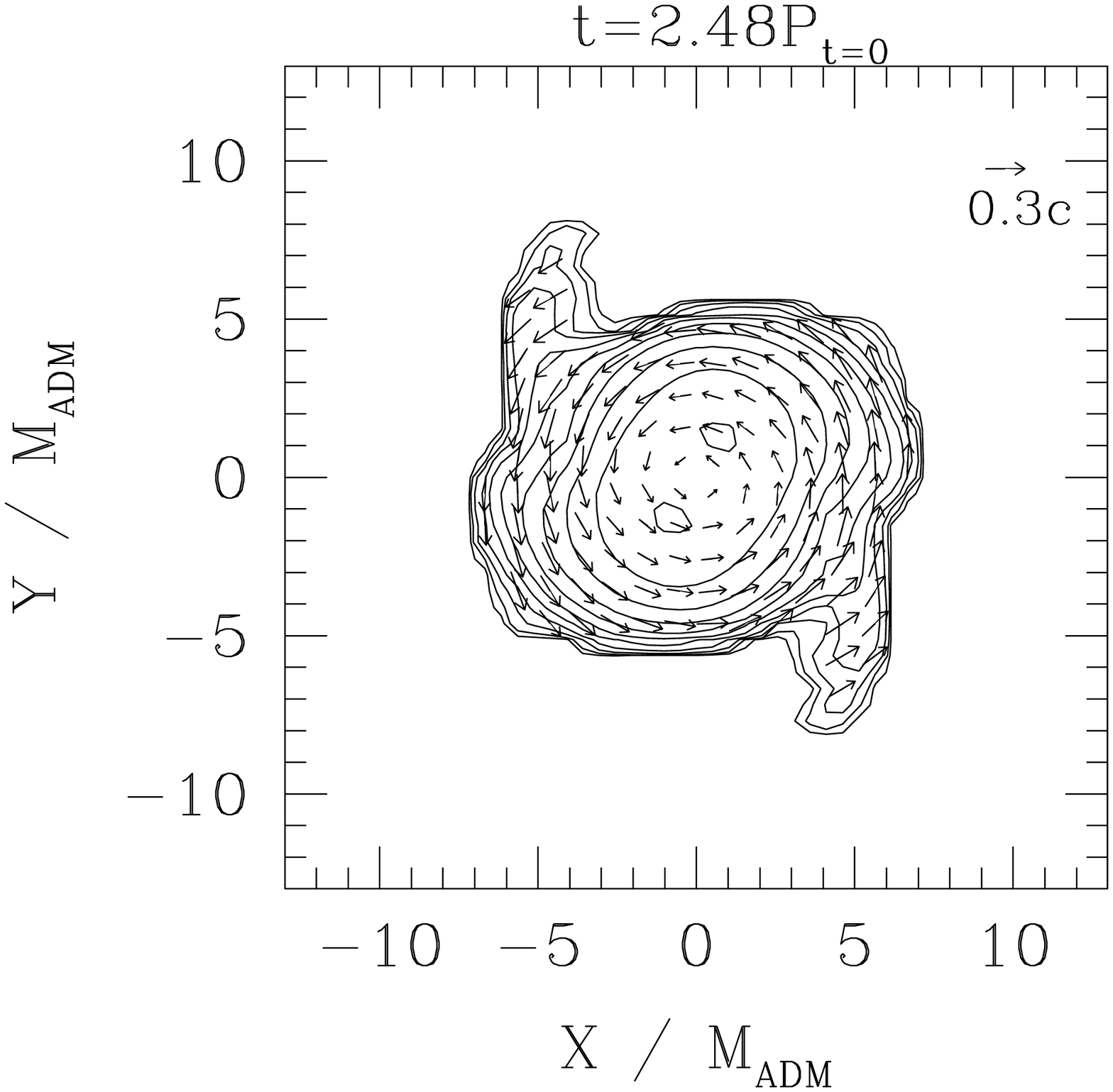} 
\epsfxsize=2.2in
\leavevmode
\hspace{-1.6cm}\epsffile{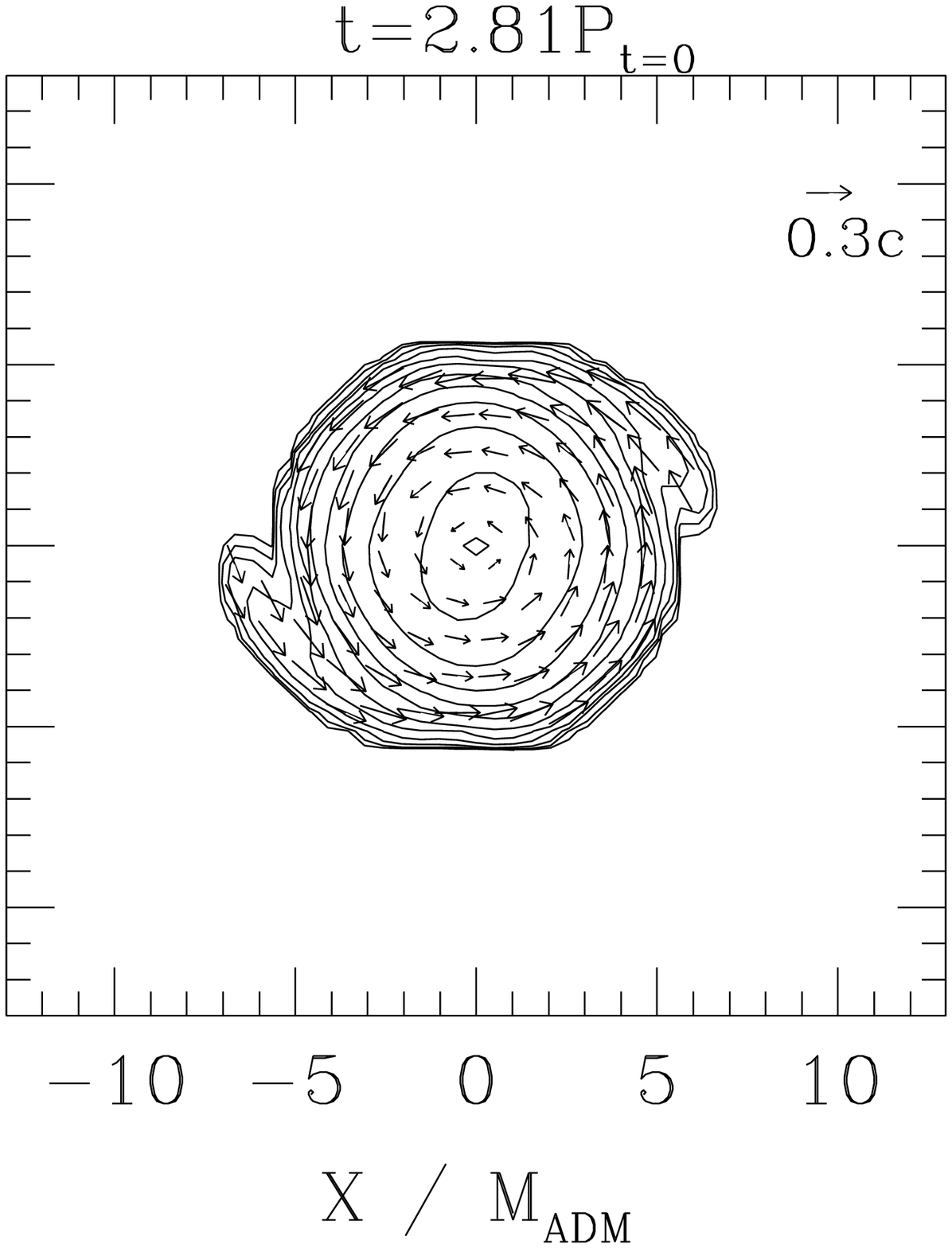}
\epsfxsize=2.2in
\leavevmode
\hspace{-1.6cm}\epsffile{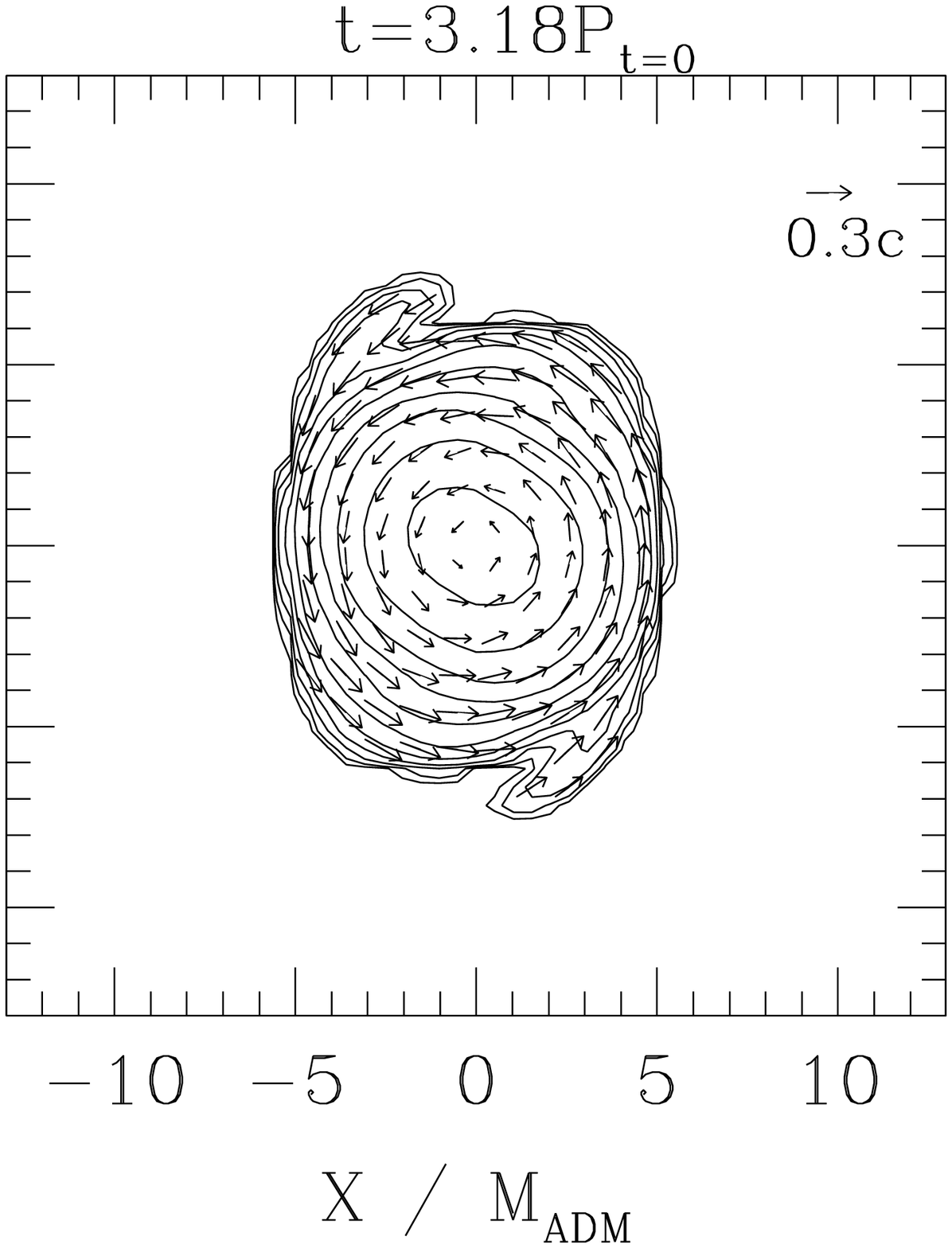} 
\end{center}
\vspace{-5mm}
\caption{
Snapshots of the density contours for $\rho$ 
in the equatorial plane for model (A-2). 
The contour curves are drawn for $\rho/\rho_{0}=10^{-0.3j}$, 
where $\rho_{0}$ is the maximum value of $\rho_0$ at 
$t=0$ (here $\rho_{0} = 0.139\kappa^{-n}$), for $j=0,1,2,\cdots,10$. 
The maximum density at each time step is found in Fig.~\ref{FIGALPRHO}. 
Vectors indicate the local velocity field $(v^x,v^y)$, and the scale 
is shown in the upper right-hand corner. 
$P_{\rm t=0}$ denotes the orbital period of the quasiequilibrium 
configuration given at $t=0$. 
The length scale is shown in units of $GM_{\rm ADM0}/c^2$, 
where $M_{\rm ADM0}$ is the gravitational mass at $t=0$.
Note that $t=3.18P_{t=0} \approx 740 M_{\rm ADM0}$. 
Each panel does not necessarily represent contours 
of a constant time interval.
\label{fig:FIG1} }
\end{figure}

\begin{figure}[t]
\begin{center}
\epsfxsize=2.2in
\leavevmode
\epsffile{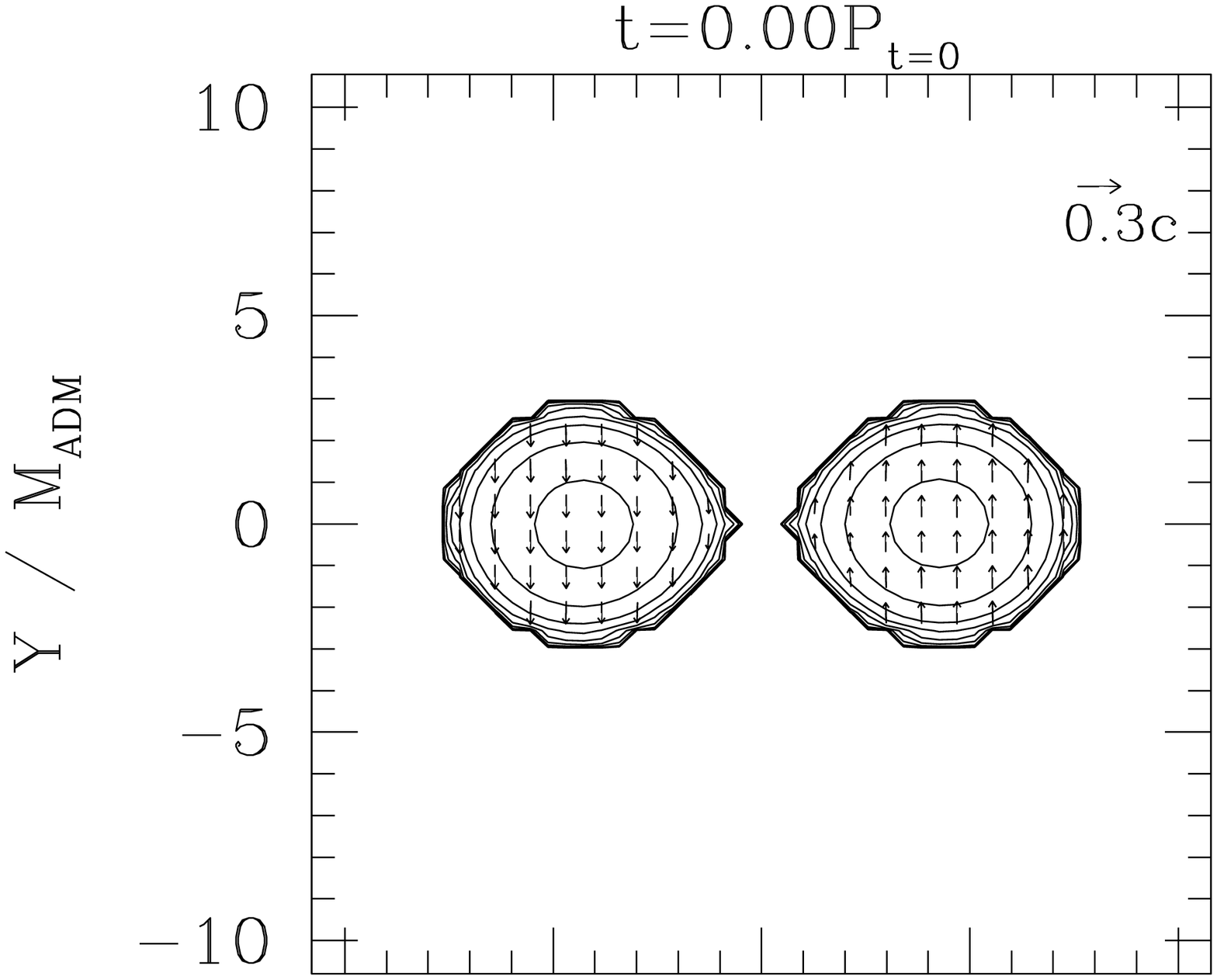}
\epsfxsize=2.2in
\leavevmode
\hspace{-1.6cm}\epsffile{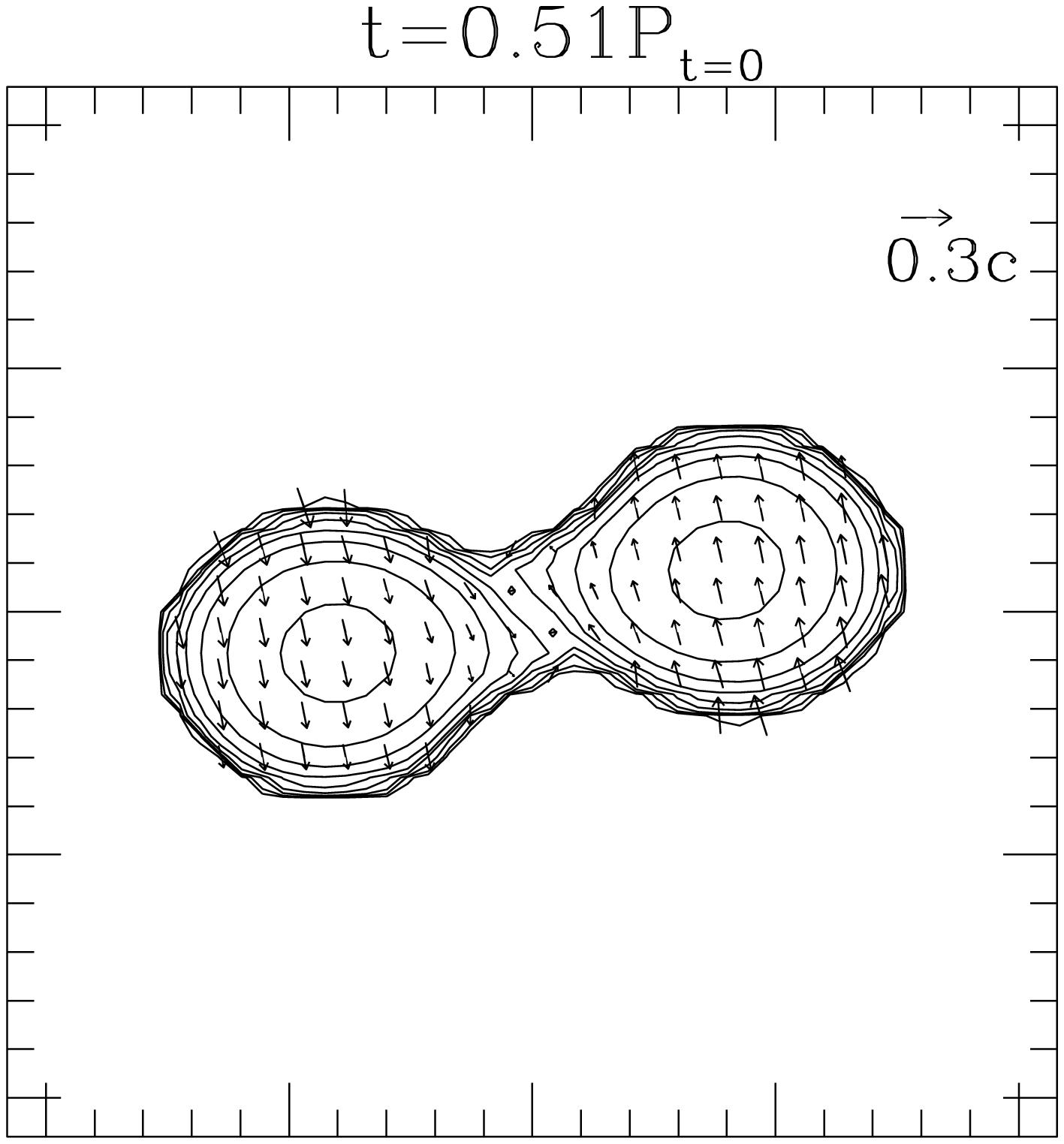} 
\epsfxsize=2.2in
\leavevmode
\hspace{-1.6cm}\epsffile{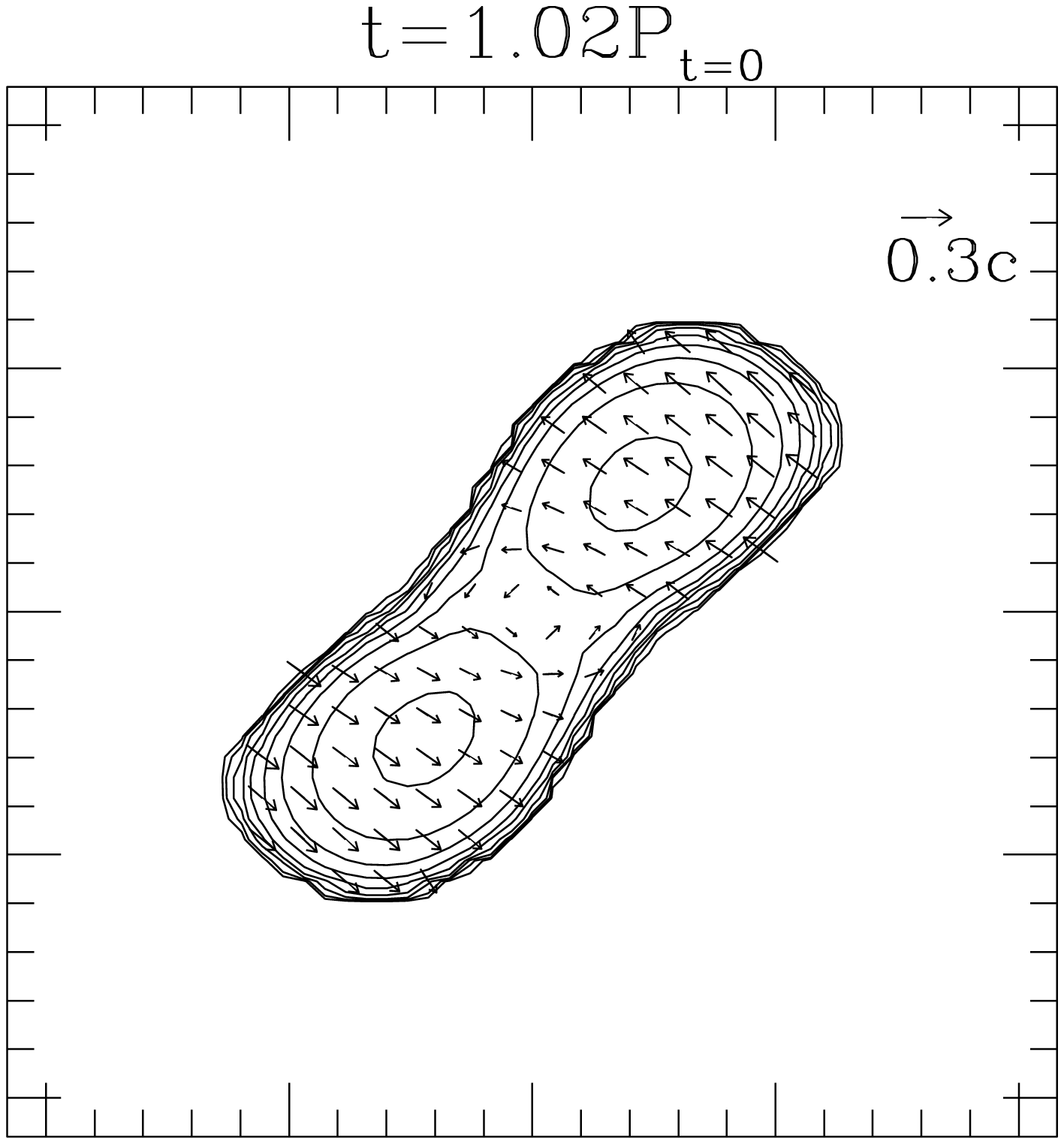} \\
\vspace{-1.1cm}
\epsfxsize=2.2in
\leavevmode
\epsffile{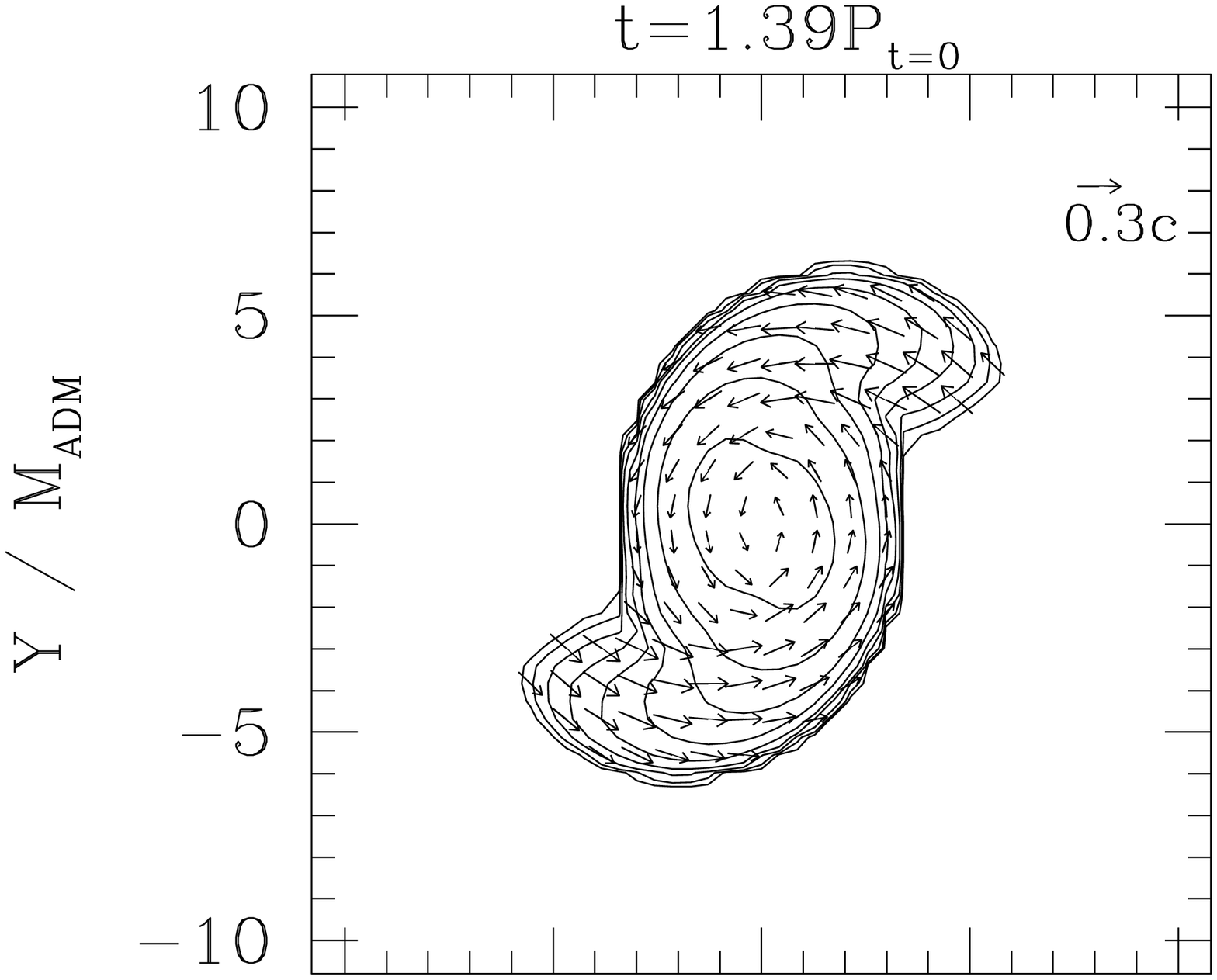} 
\epsfxsize=2.2in
\leavevmode
\hspace{-1.6cm}\epsffile{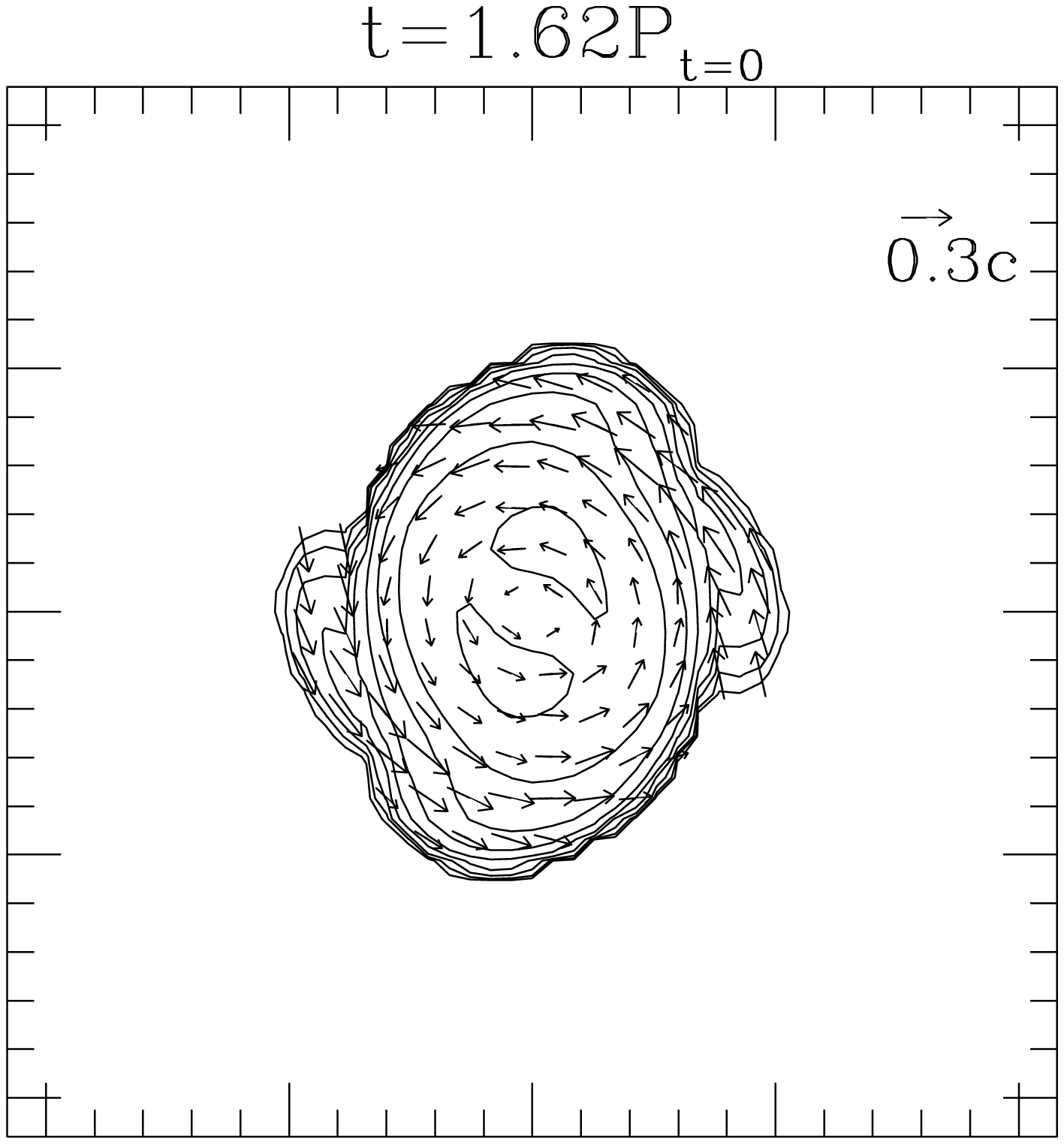}
\epsfxsize=2.2in
\leavevmode
\hspace{-1.6cm}\epsffile{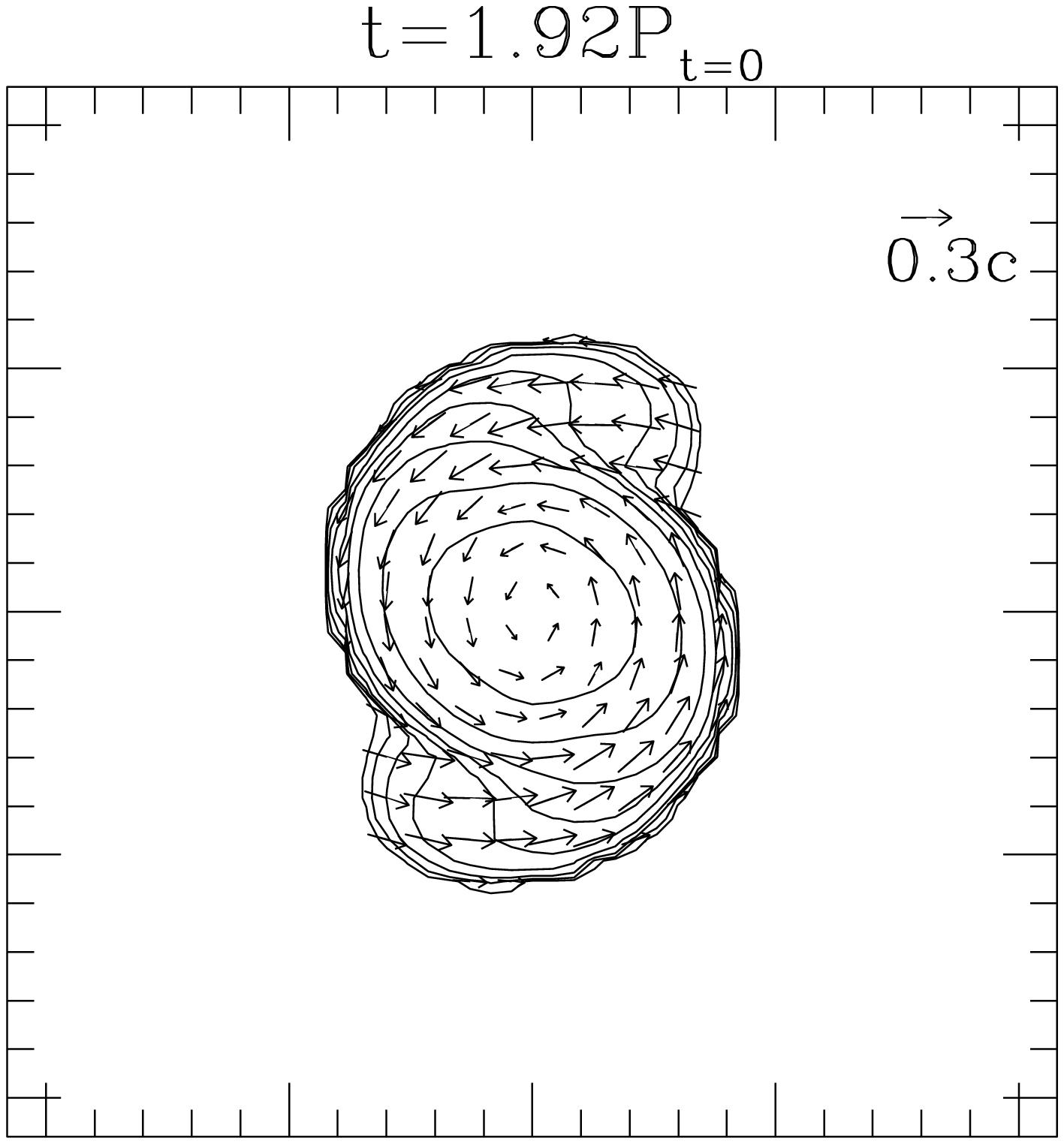}\\ 
\vspace{-1.1cm}
\epsfxsize=2.2in
\leavevmode
\epsffile{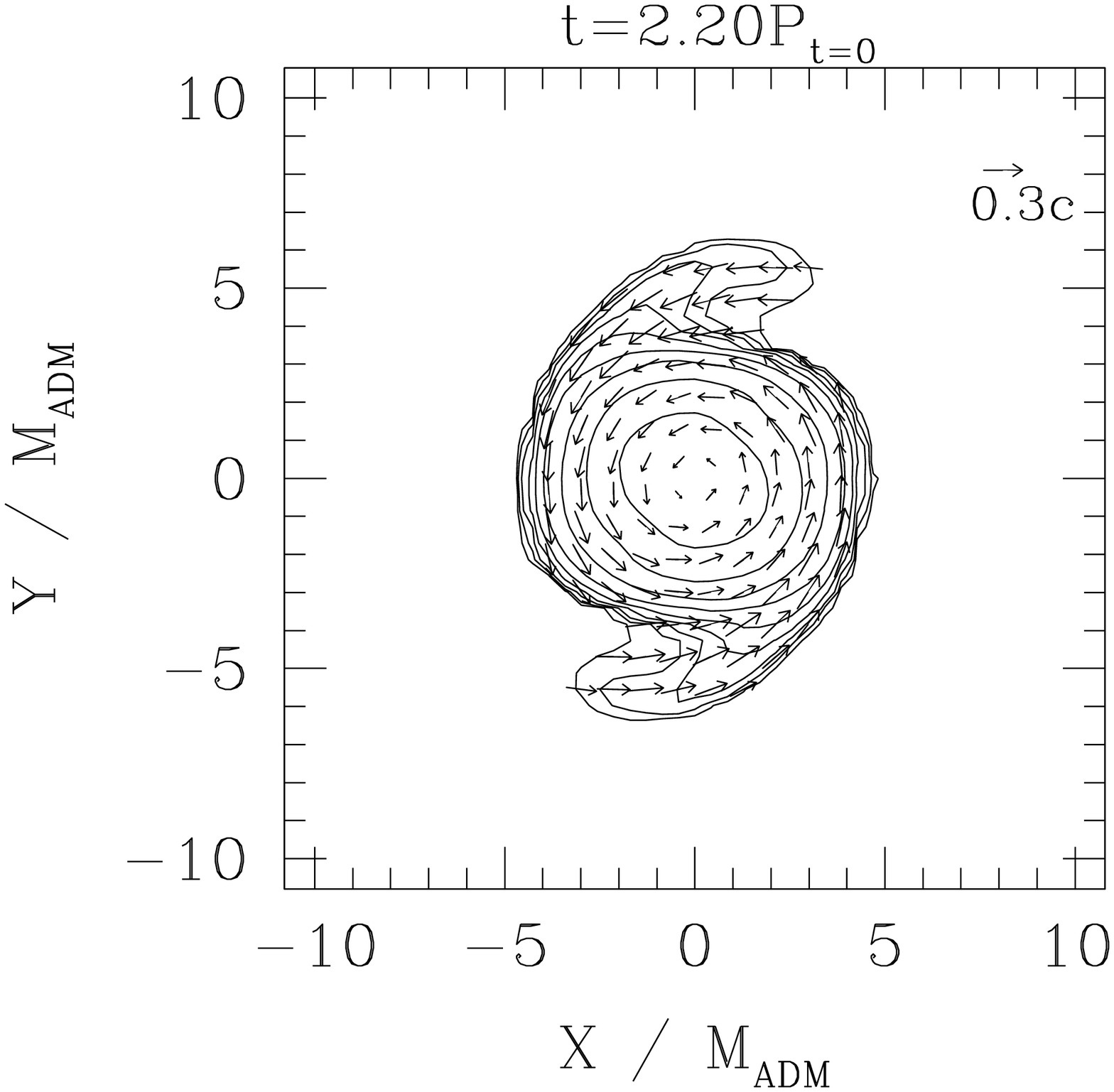} 
\epsfxsize=2.2in
\leavevmode
\hspace{-1.6cm}\epsffile{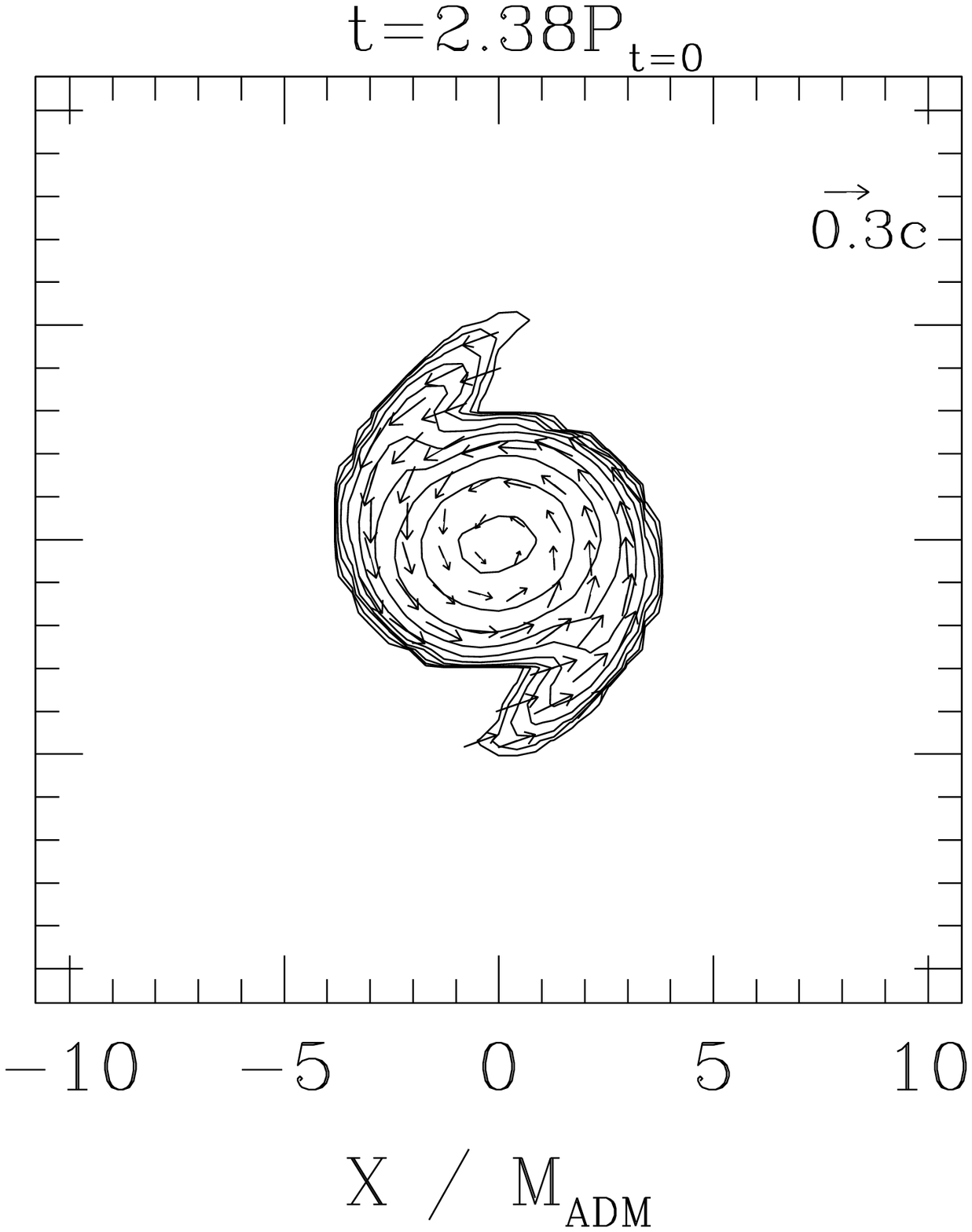}
\epsfxsize=2.2in
\leavevmode
\hspace{-1.6cm}\epsffile{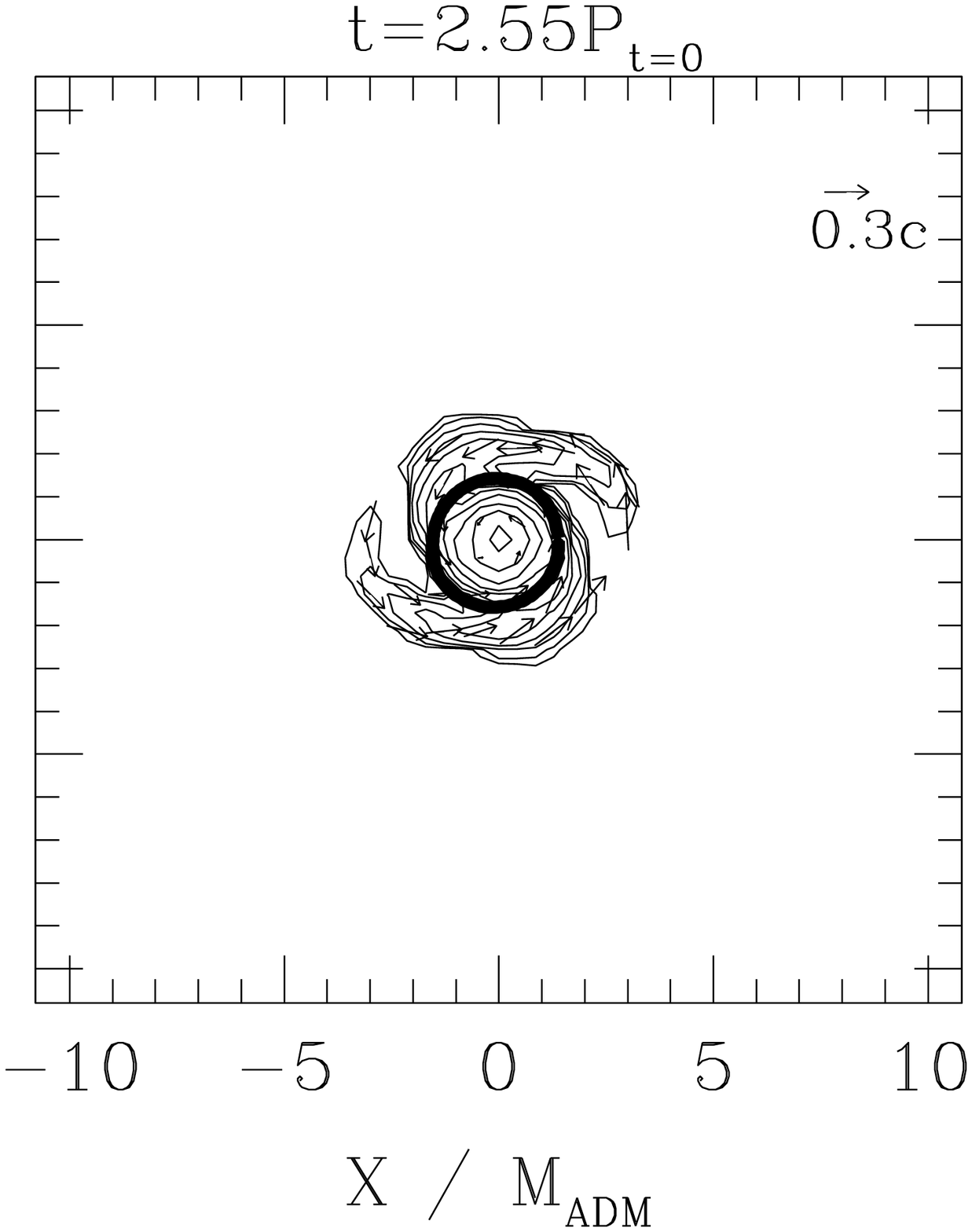} 
\end{center}
\vspace{-5mm}
\caption{The same as Fig.~\ref{fig:FIG1}, but 
for model (B-2). The contour curves are drawn for 
$\rho/\rho_{0}=10^{-0.3j}$ for $j=0,1,2,\cdots,10$. Here, 
$\rho_{0}=0.52 \kappa^{-n}$, which is the critical density of stability for 
spherical stars against gravitational collapse. 
The thick solid circle in the final panel indicates the apparent horizon. 
\label{fig:FIG2}}
\end{figure}

\begin{figure}[t]
\begin{center}
\epsfxsize=2.2in
\leavevmode
\epsffile{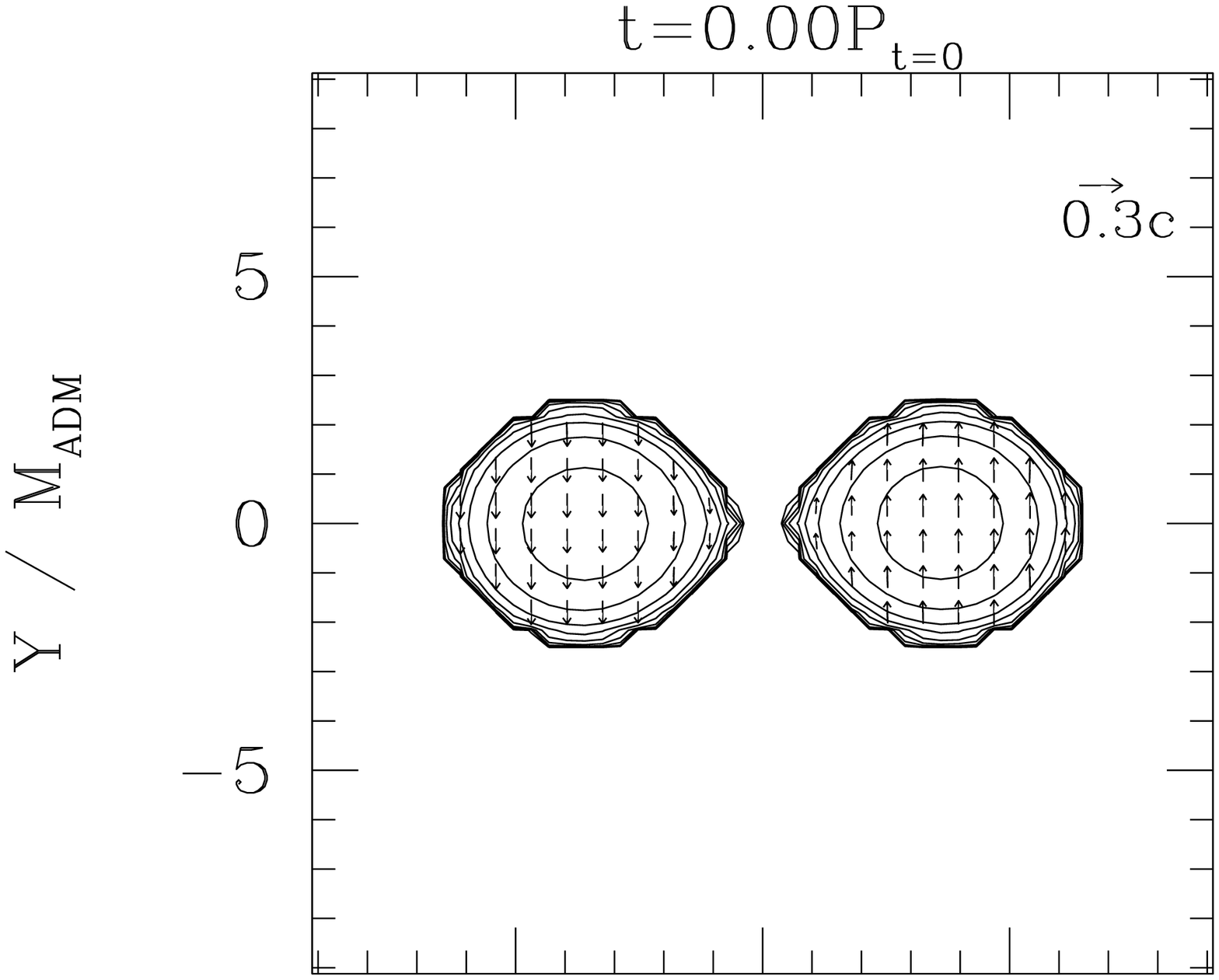}
\epsfxsize=2.2in
\leavevmode
\hspace{-1.6cm}\epsffile{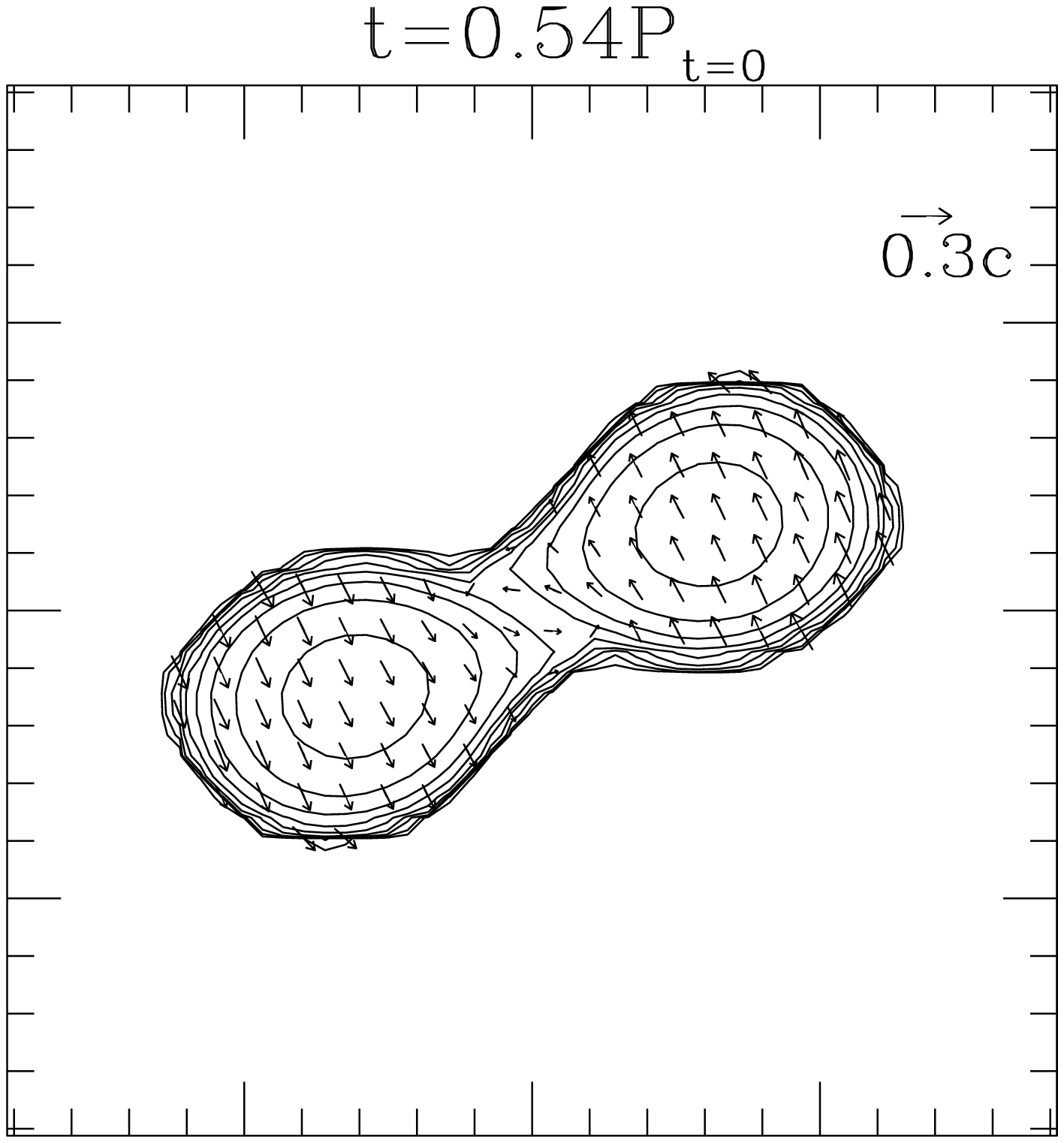} 
\epsfxsize=2.2in
\leavevmode
\hspace{-1.6cm}\epsffile{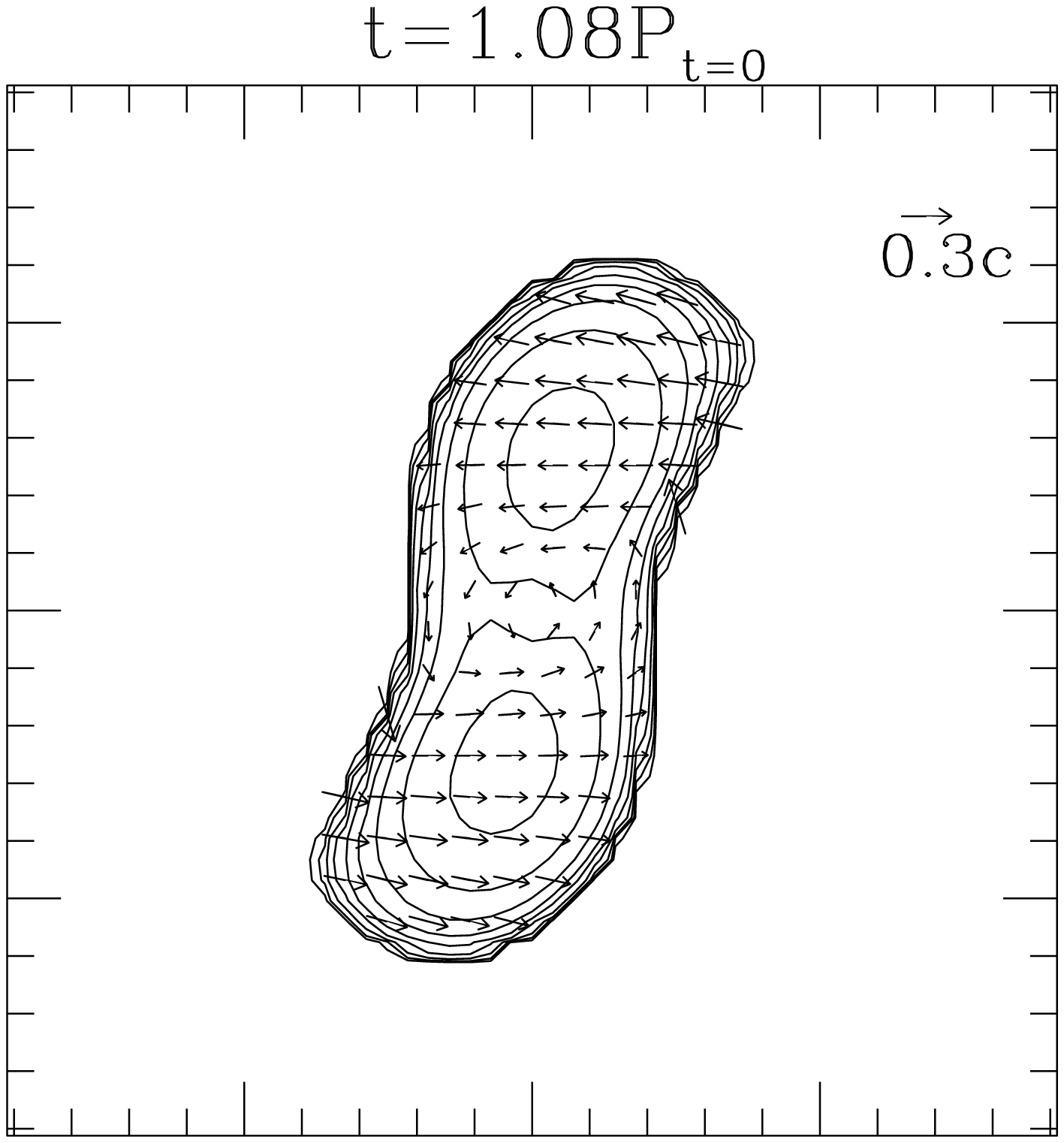} \\
\vspace{-1.1cm}
\epsfxsize=2.2in
\leavevmode
\epsffile{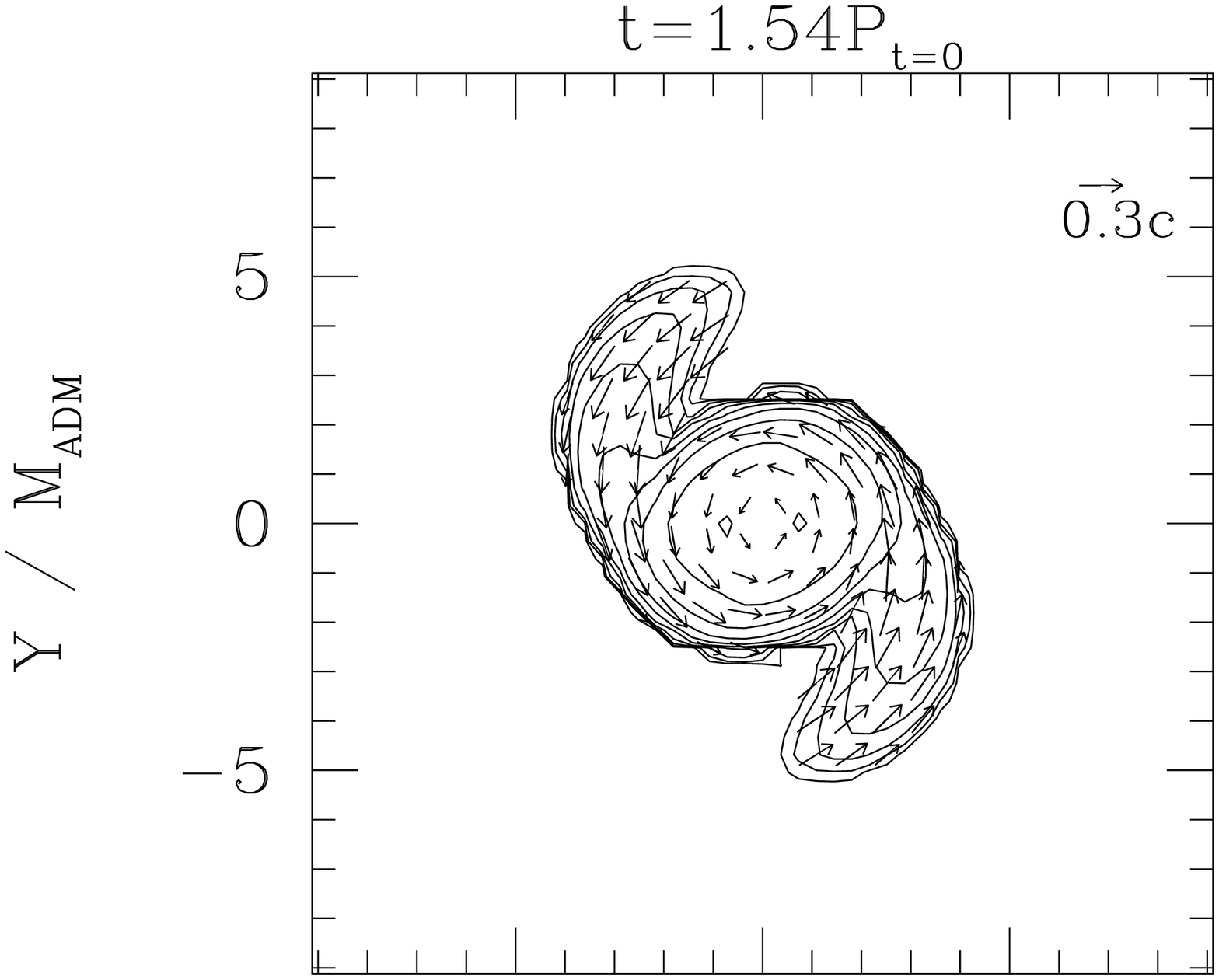} 
\epsfxsize=2.2in
\leavevmode
\hspace{-1.6cm}\epsffile{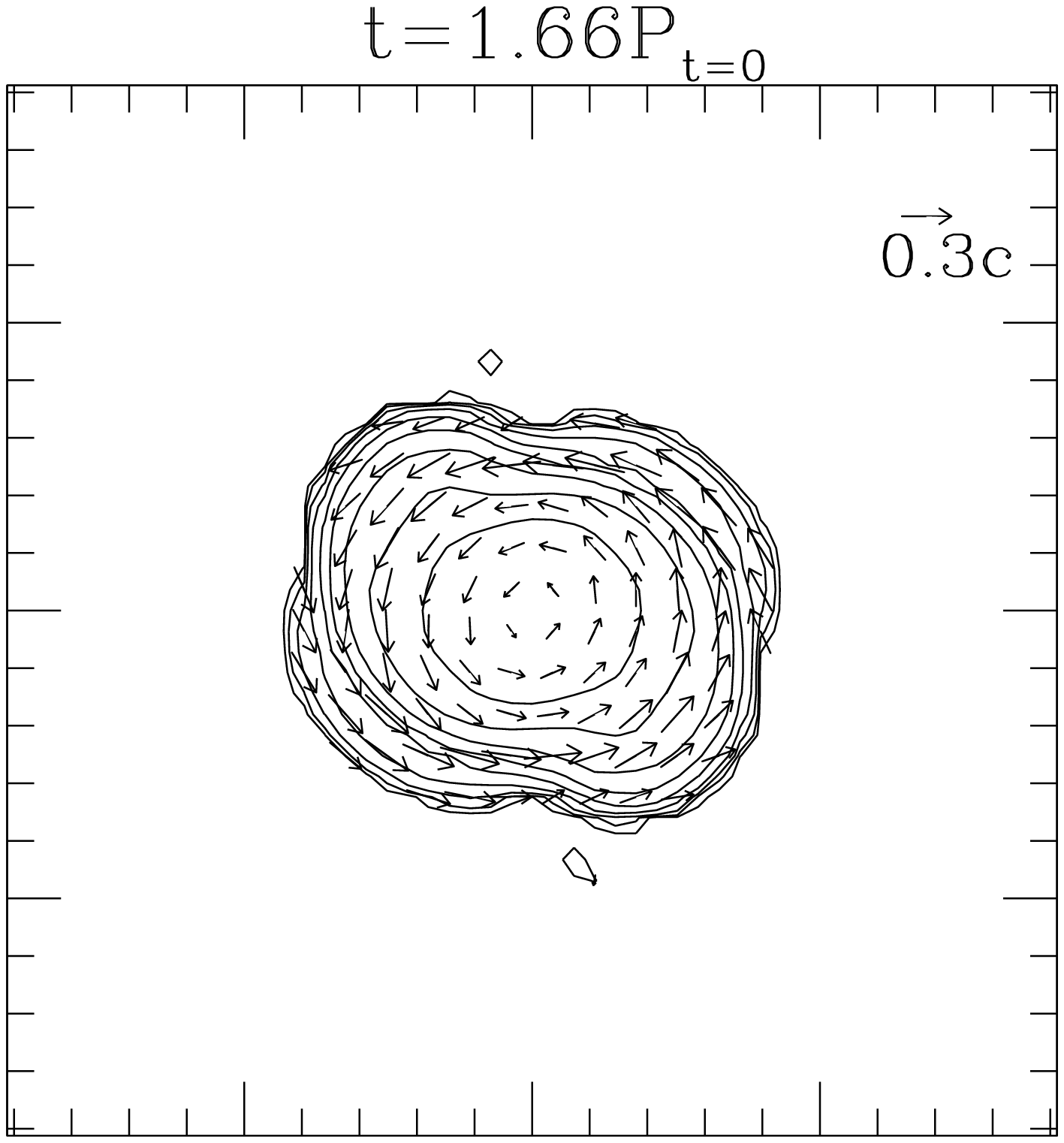}
\epsfxsize=2.2in
\leavevmode
\hspace{-1.6cm}\epsffile{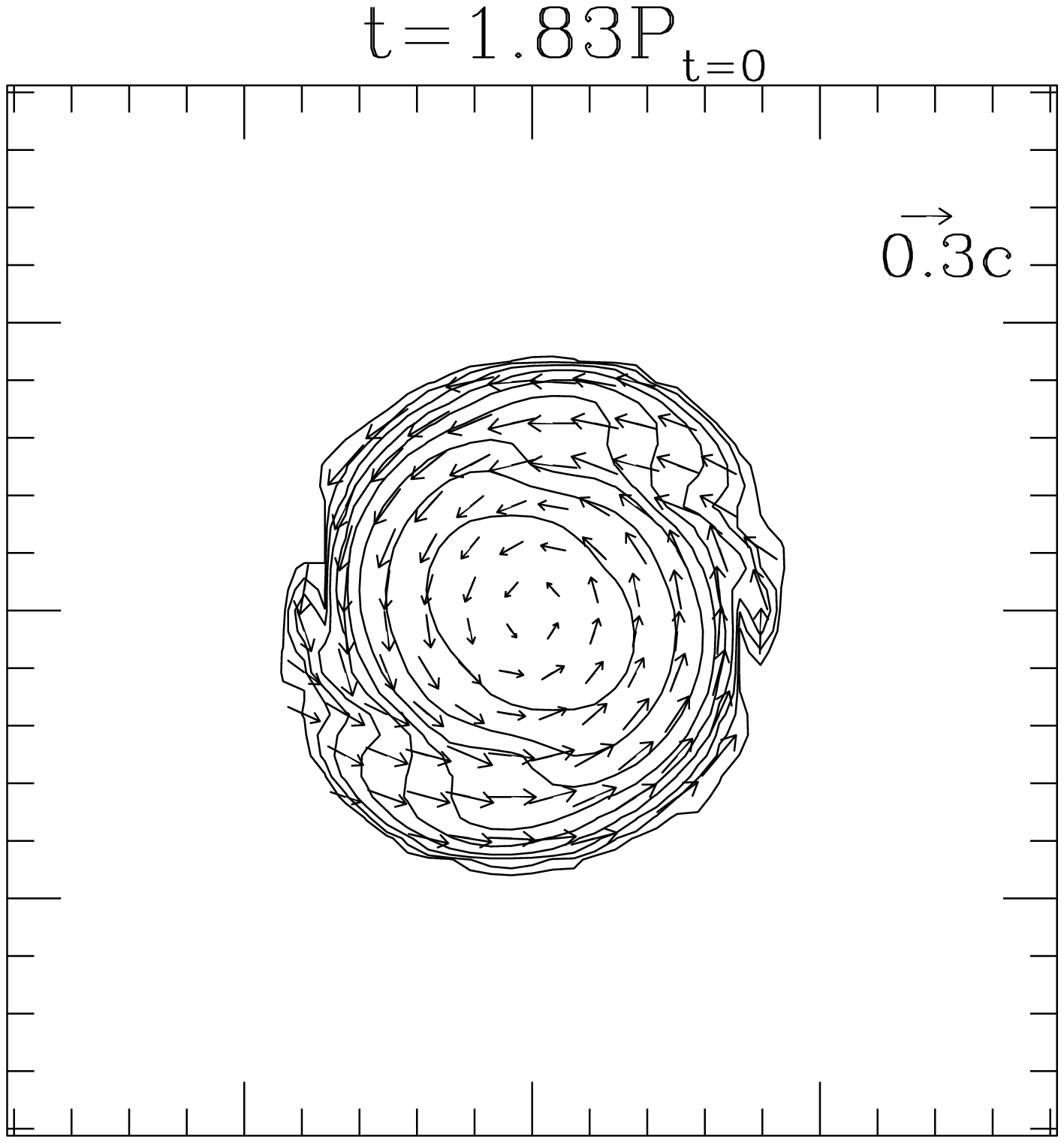}\\ 
\vspace{-1.1cm}
\epsfxsize=2.2in
\leavevmode
\epsffile{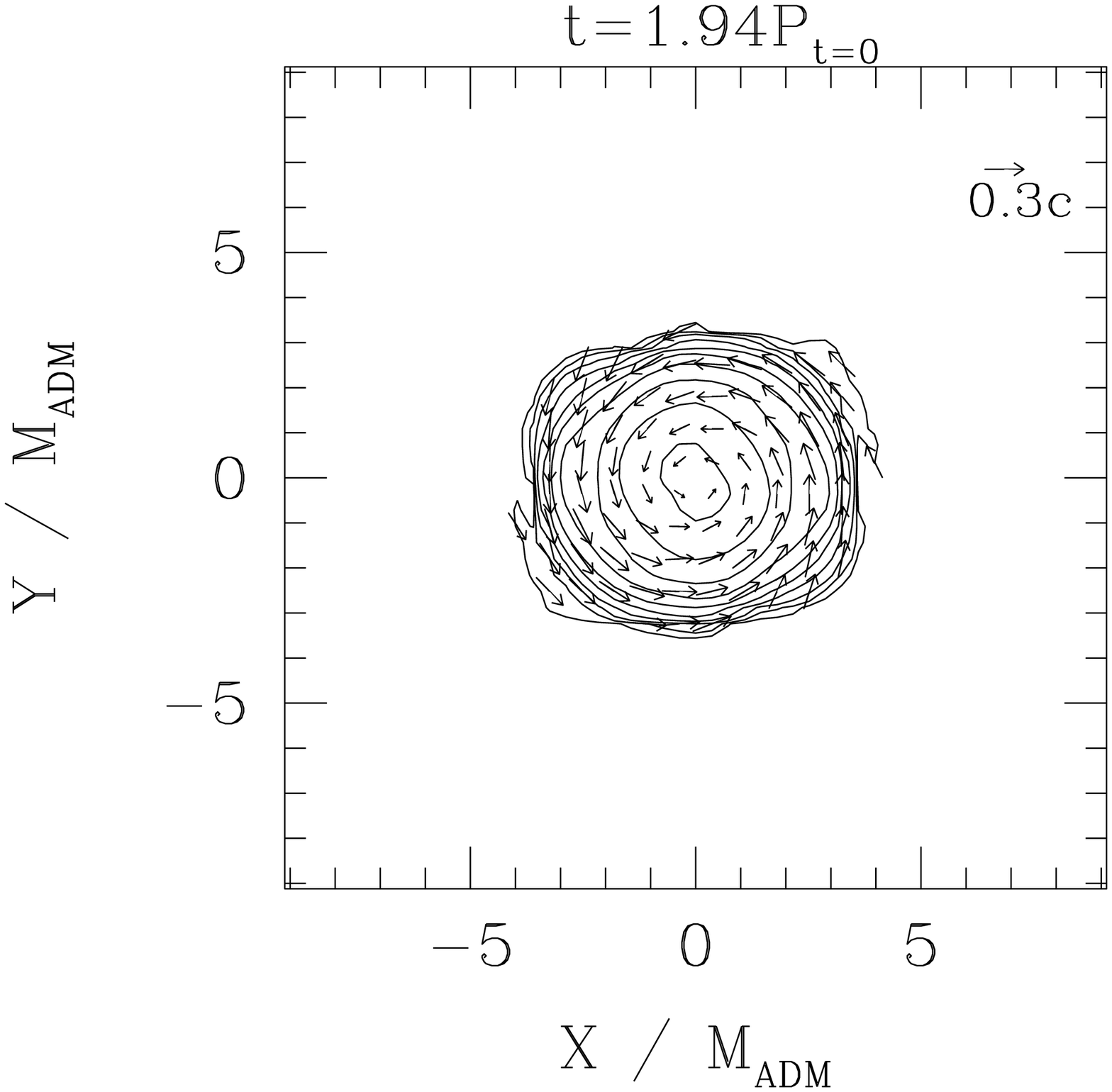} 
\epsfxsize=2.2in
\leavevmode
\hspace{-1.6cm}\epsffile{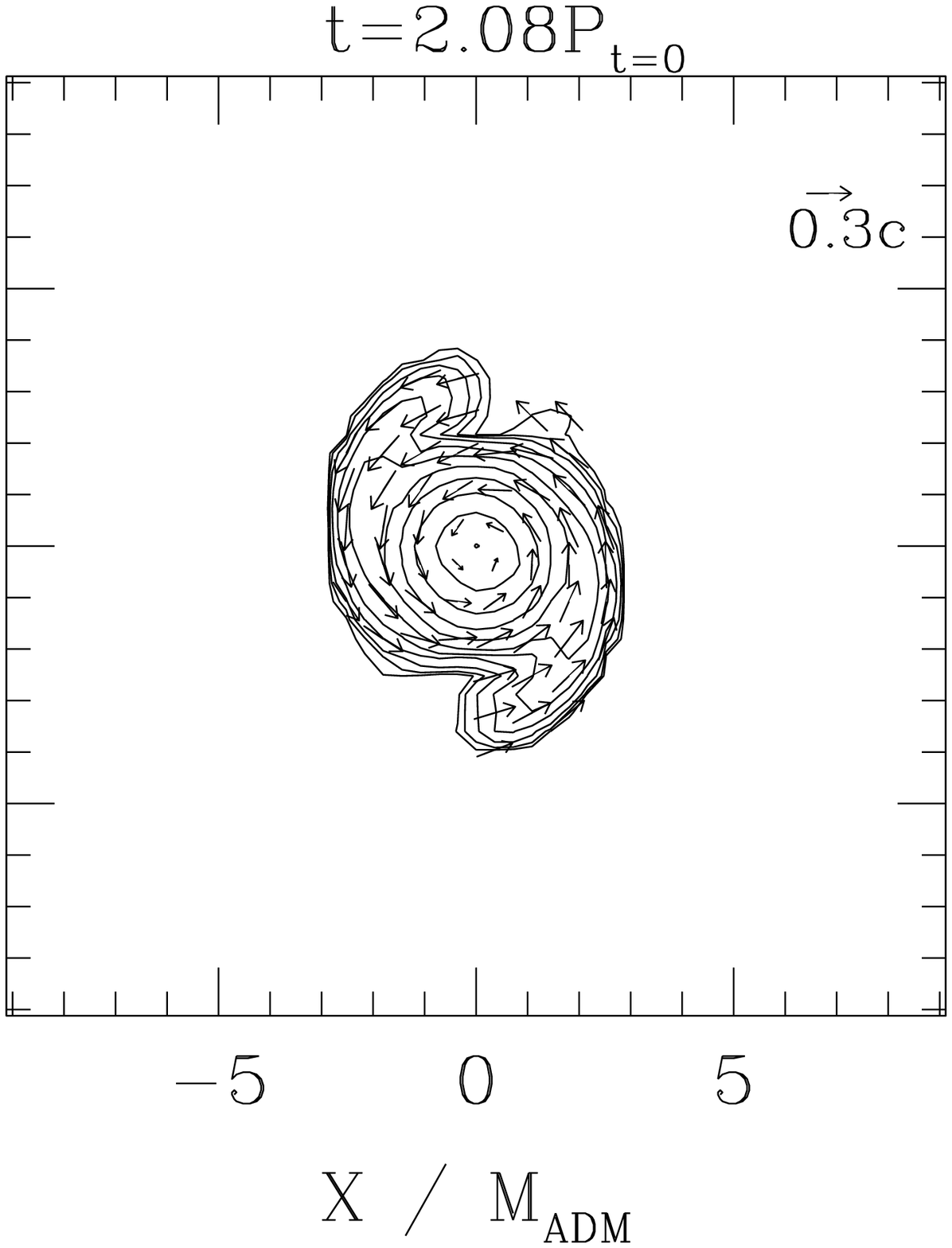}
\epsfxsize=2.2in
\leavevmode
\hspace{-1.6cm}\epsffile{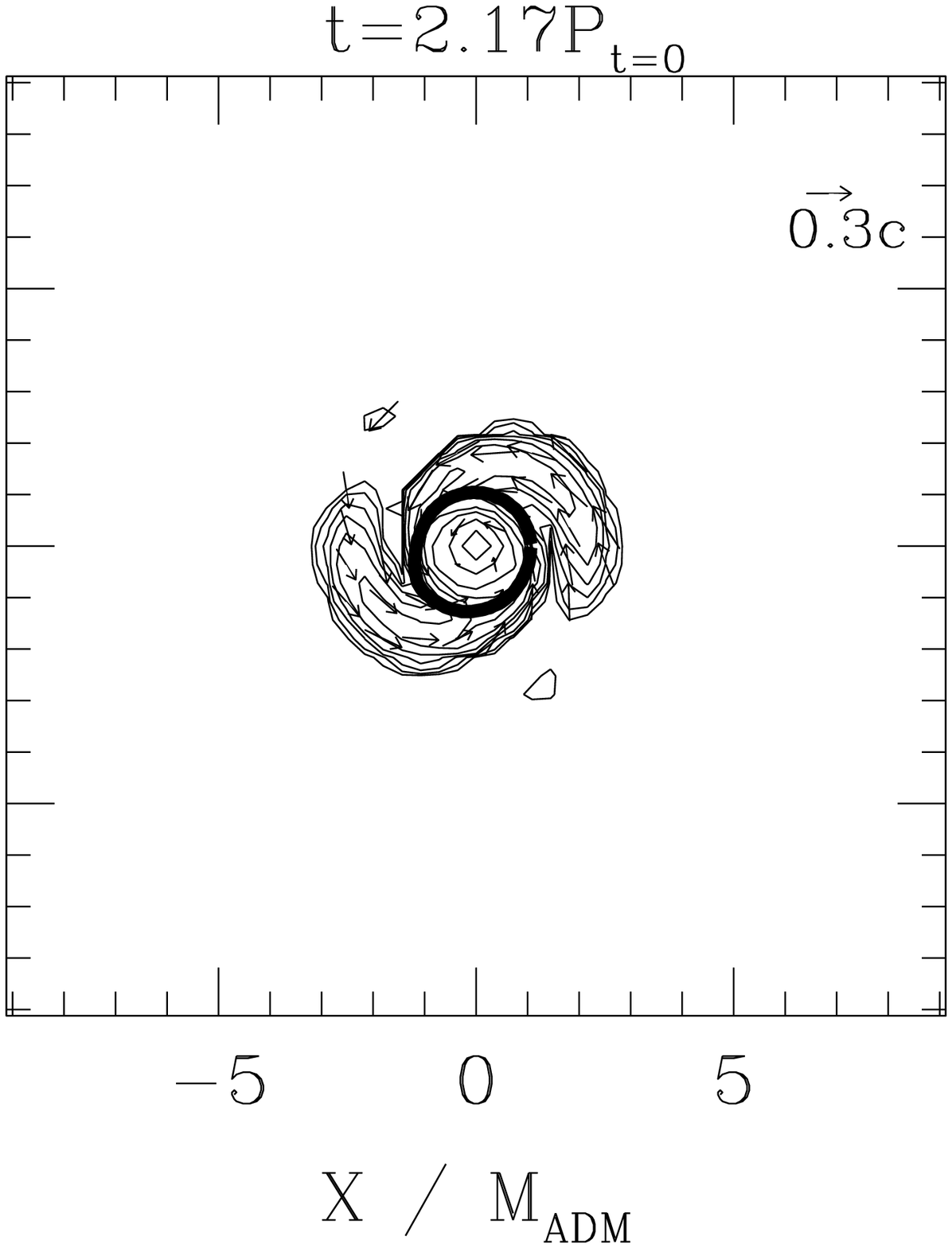}
\end{center}
\vspace{-5mm}
\caption{The same as Fig.~\ref{fig:FIG2}, but for model (C-3). 
\label{fig:FIG3}}
\end{figure}

\begin{figure}[t]
\begin{center}
\epsfxsize=2.8in
\leavevmode
\epsffile{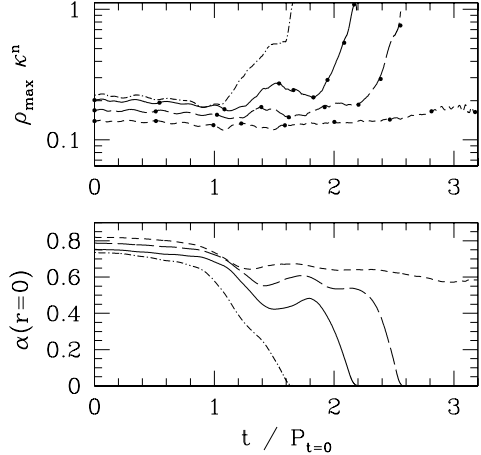}
\end{center}
\vspace{-8mm}
\caption{Evolution of the maximum value of $\rho$ and 
$\alpha$ at $r=0$ for model (A-2) 
(dashed curves), (B-2) (long-dashed curves), (C-3) (solid curves) 
and (D-0) (dotted-dashed curves). 
Solid circles denote the time slices which we choose to generate the 
contour plots displayed in Figs.~\ref{fig:FIG1}--\ref{fig:FIG3}.
Note that the merger starts at $t \sim P_{t=0}$ for each model
and that apparent horizons are formed when $\alpha(r=0)$ becomes 
$\sim 0.04$--0.05. 
\label{FIGALPRHO}}
\end{figure}

\subsubsection{Characteristics of merger for $\Gamma=2.25$}

In Figs.~1--3, we display the density contour curves 
and velocity vectors for $\rho$ and $v^i$ 
at selected timesteps for simulations of 
models (A-2), (B-2) and (C-3).\footnote{
Animations for these simulations can be seen 
at \\ http://www.esa.c.u-tokyo.ac.jp/\~{}shibata/anim.html.}
Time evolutions of the maximum density for $\rho$ and $\alpha$ at $r=0$ 
for models (A-2), (B-2), (C-3) and (D-0) 
are also shown in Fig.~4. The maximum density and central value of 
the lapse function at timesteps selected in Figs.~1--3 are 
plotted in Fig.~4. 

In the merger for model (A), the product we found when we stopped 
the simulation at $t \approx 3.3P_{t=0}$, 
where $P_{t=0}=2\pi/\Omega_0$, is a massive neutron star. 
However, the merged object does not promptly settle down to 
an axisymmetric rotating neutron star. Soon after the onset of the 
merger, the merged object develops spiral arms (see fourth panel 
of Fig.~1), because it is rapidly rotating, 
and hence appears to be dynamically unstable with respect to formation of 
bar and spiral arms. The spiral arms do not spread widely outward, but 
wind about the central core (see fifth panel). After the winding, 
spiral arms develop again (see sixth panel), and subsequently 
wind about the core after a short time (see seventh panel). After 
repeating this process several times, the merged object  
gradually settles into an ellipsoidal rotating star (see eighth and 
ninth panels). During these oscillations, gravitational waves 
of fairly large amplitude are excited over a long timescale. This 
gravitational radiation
eventually triggers the collapse of this massive neutron star 
to a black hole (see discussion in \S 4.2).

\begin{figure}[t]
\begin{center}
\epsfxsize=2.8in
\leavevmode
\epsffile{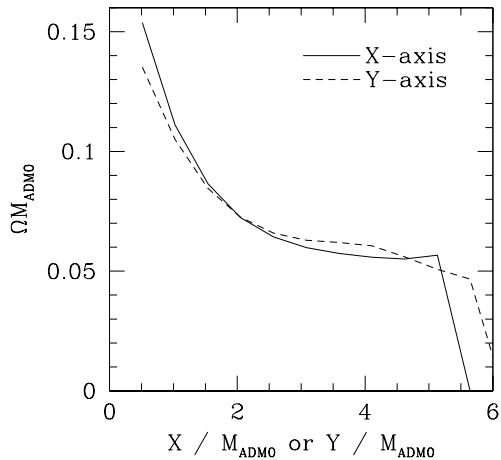}
\end{center}
\vspace{-5mm}
\caption{The angular velocity in units 
of $M_{\rm ADM0}^{-1}$ along $x$ and $y$ axes at $t/P_{t=0}=3.18$ 
for model (A-2).
\label{fig:FIGOME}}
\end{figure}

The total baryon rest-mass of the binary for model (A) 
is $\approx 16\%$ larger than the 
maximum allowed value of a spherical star in isolation. 
Even with such a large mass, the merged object does not collapse to 
a black hole within a couple of dynamical timescales. 
As indicated in Fig.~1, the merger proceeds very gradually, 
because the approaching velocity at the point of the
contact of two stars is not 
very large. Consequently, 
the shock heating does not appear to play an important role
in merging. Indeed, the function $P/\rho^{\Gamma}$, 
which represents the change of the entropy, increases 
by at most 10$\%$ from the initial value (i.e., $\kappa$) 
around the central region of the merged object. 
This implies that the rotational centrifugal force plays an 
important role for supporting the self-gravity of such 
a massive neutron star. To illustrate the importance of 
the rotation in supporting the large mass, 
the angular velocity along $x$ and $y$-axes, defined as 
$|v^y/x|$ and $|v^x/y|$, are displayed 
in Fig.~\ref{fig:FIGOME}. These plots show that the formed 
massive neutron star is 
differentially rotating, and the magnitude of the 
angular velocity is of order of the Kepler velocity, i.e., 
$\Omega M_{\rm ADM}(R/M_{\rm ADM})^{3/2} =O(1)$, where $R$ 
is the characteristic radius of the merged object 
($\approx 5M_{\rm ADM}$).

\begin{figure}[t]
\begin{center}
\epsfxsize=2.8in
\leavevmode
\epsffile{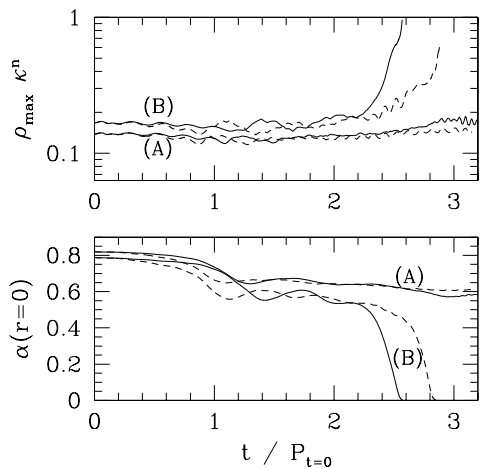}
\end{center}
\vspace{-8mm}
\caption{Evolution of the maximum values of $\rho$ and 
$\alpha$ at $r=0$ for models (A-2) (solid curves), 
(A-3) (dashed curves), (B-2) (solid curves) and (B-3) (dashed curves). 
\label{FIGALPRHO2}}
\end{figure}

For models (B), (C) and (D), in which $(M/R)_{\infty} \geq 0.14$, 
a black hole is formed in the present simulations. 
For all the cases, we were able to locate apparent horizons 
in the last stage of the
simulations using our apparent horizon finder.\cite{AH} 
However, the formation process is 
slightly different for each of these models ({\it cf}. Fig.~\ref{FIGALPRHO}). 
For model (D), a black hole is formed soon
after the first contact of the two neutron stars at $t \sim P_{t=0}$ 
(see Fig.~4). 
For model (C), the formation timescale is slightly 
longer. In this case, the merged object first experiences a bounce
(see fifth and sixth panels in Fig.~3, and Fig.~4), 
and then collapses into a black hole. 
For model (B), the formation timescale to a black hole is even longer. 
In this case, the merged object quasi-radially oscillates 
for several times after the first contact. Indeed, 
the lapse function at $r=0$ does not monotonically 
approach to zero, as shown in Fig.~\ref{FIGALPRHO}. 
The collapse toward a black hole seems to 
set in after the angular momentum is dissipated through 
gravitational radiation. (Indeed, the angular momentum decreases 
by a large amount of $O(10\%)$ from the late inspiral to 
the merger stages; see \S 4.2 and Table III.)
As we show in \S 4.2, the difference 
in the formation process of black holes is 
significantly reflected by the waveform of gravitational waves 
({\it cf}. Fig.~\ref{gw225}) and 
the Fourier spectra ({\it cf}. Fig.~\ref{spectrum}). 

To investigate the effect of resolution on numerical results, 
we compare the time evolution of the maximum density and 
the central lapse function for models (A-2) and (A-3) and 
for models (B-2) and (B-3) in Fig.~\ref{FIGALPRHO2}. In these comparisons, 
we fix the location of the outer boundaries, 
while the grid spacing for (A-3) and (B-3) is 
1.34 times larger than that for (A-2) and (B-2), respectively
(see Table II). Figure~\ref{FIGALPRHO2} indicates 
that convergence is achieved fairly well. It also shows that 
with lower resolution, the numerical dissipation of 
the angular momentum is larger, so that the merger happens earlier. 
Also, high density peaks are captured less accurately, because 
of larger numerical diffusion. 
This makes the formation timescale of a black hole longer. 
It is interesting to note that, if we use much lower resolution, 
model (B) might not result in black hole formation
in several rotational periods of the merger object, 
because of a large numerical diffusion. This result indicates that 
to determine the criterion of black hole formation 
and the formation timescale accurately, 
we need to perform better-resolved numerical simulations. 


A noteworthy result 
for models (B), (C) and (D) is that 
the baryon rest-mass fraction outside
the apparent horizon at its formation is less than $1\%$ 
of the total rest-mass. This implies that the fraction of the baryon 
rest-mass in the disk around the black hole is very small. 
We expect that 
the mass fraction of the disk is much less than
1\% of the total baryon rest-mass in these cases. 
In the following, we describe the reason for these results.

\begin{figure}[t]
\begin{center}
\epsfxsize=2.6in
\leavevmode
\epsffile{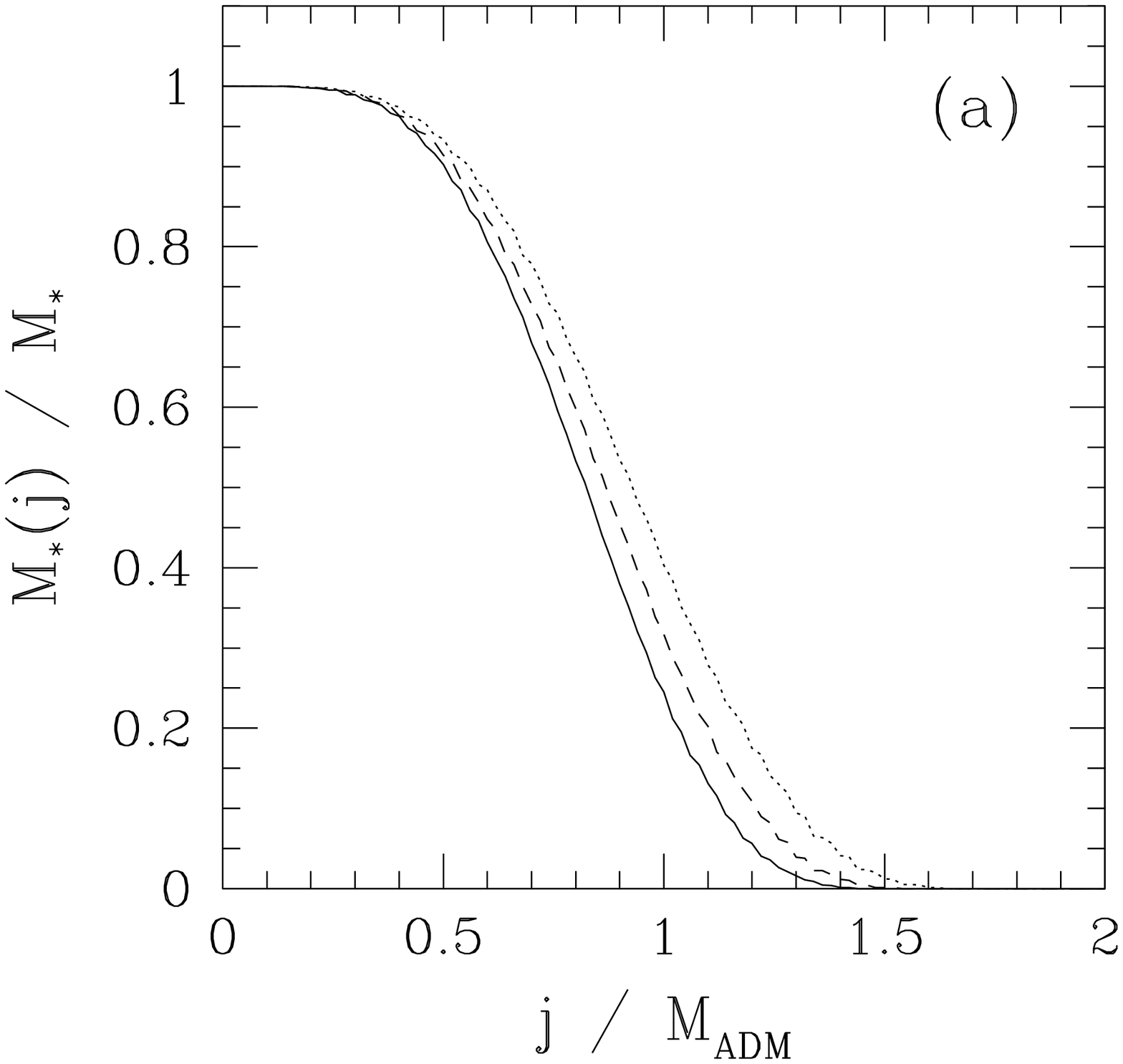}
\epsfxsize=2.6in
\leavevmode
\epsffile{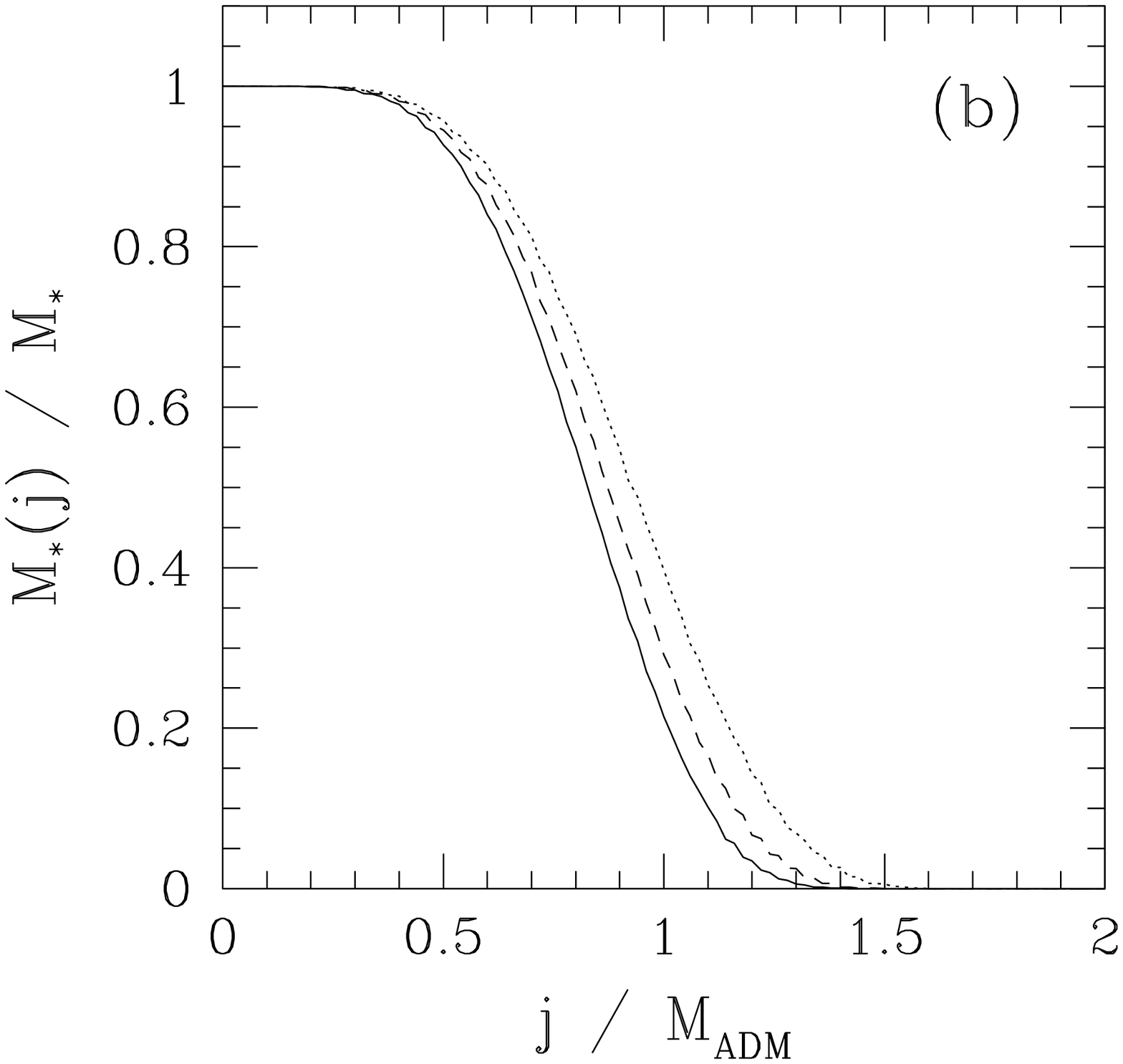}
\end{center}
\vspace{-7mm}
\caption{\footnotesize (a) $M_*(j)/M_*$ as a function of $j/M_{\rm ADM0}$ 
for models (A) (dotted curve), (B) (dashed curve) and (C) (solid curve). 
(b) The same as (a), but for models (G) (dotted curve), (H) (dashed curve) 
and (I) (solid curve).
\label{FIGANG}}
\end{figure}

In Fig.~\ref{FIGANG}(a), we plot $M_*(j)/M_*$ as a function of 
$j/M_{\rm ADM0}$ for models (A), (B) and (C) at $t=0$. 
It is found that there is no fluid element 
for which $j/M_{\rm ADM0} > 1.6$ for any model. 
(We also plot this curve for $\Gamma=2$ in Fig.~\ref{FIGANG}(b). This 
indicates that the spectrum of the specific angular momentum depends 
very weakly on $\Gamma$. Thus, the following conclusion
holds irrespective of the value of $\Gamma$
and the compactness of the neutron stars.) 
This small value of the specific angular momentum of
all the fluid elements at $t=0$ plays an important role 
for determining the final outcome. 

As we found in the present simulations [as well as 
in Refs. \citen{bina} and \citen{binas}] 
a quite large fraction of the fluid 
elements are eventually swallowed into the black hole whenever 
it is formed [i.e., for (B)--(D) as well as (G)--(I) (see Table I)]. 
Gravitational radiation carries energy from the system, 
but it is likely $\alt 1\%$ of $M_{\rm ADM0}$ 
(see \S 4.2 and Table III). Thus, the ADM 
mass of the formed black holes is approximately equal to the 
initial value of the system. 
In contrast to the much smaller value for the energy,
the angular momentum is dissipated 
by gravitational waves by $\sim 10\%$ of the initial value
before the formation of black holes (see Table III). 
These facts imply that $q=J/M_{\rm ADM}^2$ of formed black holes 
is not equal to the initial value, $\sim 0.9$--$0.95$
but to $\sim 0.8$--$0.85$ 
for models (B)--(D). 
The specific angular momentum of a test particle in the 
innermost stable circular orbit around a Kerr black hole of mass 
$M_{\rm ADM}$ and $q=0.8$ (0.9) is $\approx 2.4M_{\rm ADM}$ 
$(2.1M_{\rm ADM})$. Fluid elements with the specific
angular momentum less than this value have to be
swallowed into black holes. Therefore, Fig.~\ref{FIGANG} shows that 
{\em no fluid element for irrotational binary 
neutron stars just before the merger has large specific 
angular momentum large enough to form a disk around the formed black hole.}
For a disk formation, a certain transport mechanism of the angular 
momentum, such as a hydrodynamic interaction, is necessary. 
Since the black holes are formed in the dynamical timescale of 
the system, this mechanism has to be very effective 
to increase the specific angular momentum of the fluid elements
by more than 40 \% 
in such short timescale. However, such a rapid process is unlikely to work, 
as indicated by the present simulations. 

\begin{figure}[t]
\begin{center}
\epsfxsize=2.8in
\leavevmode
\epsffile{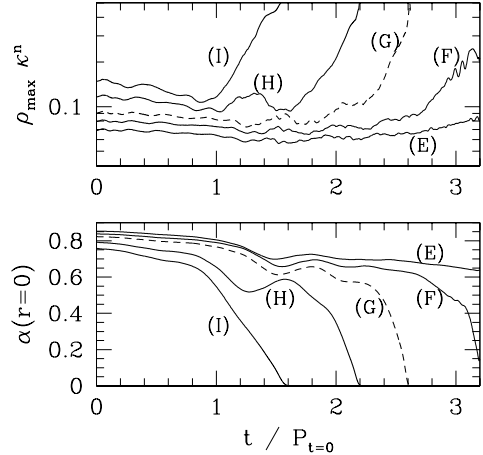}
\end{center}
\vspace{-8mm}
\caption{Evolution of the maximum values of $\rho$ and 
$\alpha$ at $r=0$ for models (E-1) (solid curve),
(F-1) (solid curve),(G-0) (dashed curve), 
(H-1) (solid curve) and (I-1) (solid curve). 
Note that the merger starts at $t \agt P_{t=0}$ for each model
and that apparent horizons are formed when $\alpha(r=0)$ becomes 
$\sim 0.04$. 
\label{alprho3}}
\end{figure}

\subsubsection{Dependence of merger on $\Gamma$}

The results of the simulation for $\Gamma=2$ 
are qualitatively the same as those for $\Gamma=2.25$, 
shown in Figs.~1--\ref{FIGALPRHO2}. 
As an example, we display the maximum value of $\rho$ and $\alpha$ at 
$r=0$ as a function of time for models (E-1), (F-1),
(G-0), (H-1) and (I-1) in Fig.~\ref{alprho3}. 
Black holes are formed for models (G)--(I),\footnote{
In Ref. \citen{bina}, we concluded that 
the product after merger for model (G) is a massive neutron star. 
In the present work, we perform a simulation again for a 
longer timescale than the previous one, and 
find that a black hole is eventually formed as a result of 
collapse of a transient massive object. Even for the less massive 
models (E) and (F), a black hole will eventually be formed, 
due to dissipation of the angular momentum by gravitational radiation
(see \S 4.2). Only the timescale for the formation of
the black hole depends on the models used.} and 
a neutron star is formed for model (E). 
For model (F), a black hole is likely formed at some 
$t \agt 3P_{t=0} \approx 800M_{\rm ADM0}$. 
However, it is difficult to determine the time of formation accurately, 
because numerical error accumulates for 
$t \agt 3P_{t=0}$, resulting in inaccurate computation.
We can be sure that at least 
the formation time would be $\agt 3P_{t=0}$, as in the case of model (A). 

Comparison of Figs.~\ref{FIGALPRHO} and \ref{alprho3} reveals that the 
time evolutions of the merger in models (E), (G), (H) and (I) for $\Gamma=2$ 
are similar to those in models (A), (B), (C) and (D) for $\Gamma=2.25$, 
respectively.  
A noteworthy point is that even if the value of
$R_{\rm mass}$ or $\comp$ are identical, 
the evolution process of the merger depends on $\Gamma$; 
e.g., the compactness of the neutron stars in model (A) is 
identical to that in model (G),
but the behavior of $\alpha$ at $r=0$ and $\rho_{\rm max}$ 
differ for the two. The minimum value of $R_{\rm~mass}$ 
for prompt formation of a black hole (i.e., for formation of 
a black hole without any transient object) is $\sim 1.5$ for 
$\Gamma=2.25$ and $\sim 1.6$ for $\Gamma=2$. Thus, 
for softer equations of state (i.e., for smaller $\Gamma$), 
this value is larger. 

In terms of the compactness $(M/R)_{\infty}$, the minimum value for 
prompt formation of a black hole is $\sim 0.16$ for $\Gamma=2.25$ and 
$\sim 0.14$ for $\Gamma=2$. Thus, for softer equations of state, 
this value is smaller. 
The compactness of a realistic neutron star of mass $1.4 M_{\odot}$ 
is likely in the range between 0.14 and 0.21, because 
the theory of neutron stars tells us that 
the radius is likely between 10 and 15km.\cite{ST}  
Thus, in a realistic merger, a black hole is likely 
formed soon after the onset of the merger for softer equations of state 
(with $\Gamma \alt 2$), while for stiffer equations of state 
(with $\Gamma \agt 2.25$), a transient object could be formed  
before forming a black hole in the
merger of low mass binary neutron stars.

\subsection{Gravitational waveforms}

\subsubsection{Character of waveforms and luminosity}

\begin{figure}[t]
\begin{center}
\epsfxsize=2.75in
\leavevmode
\epsffile{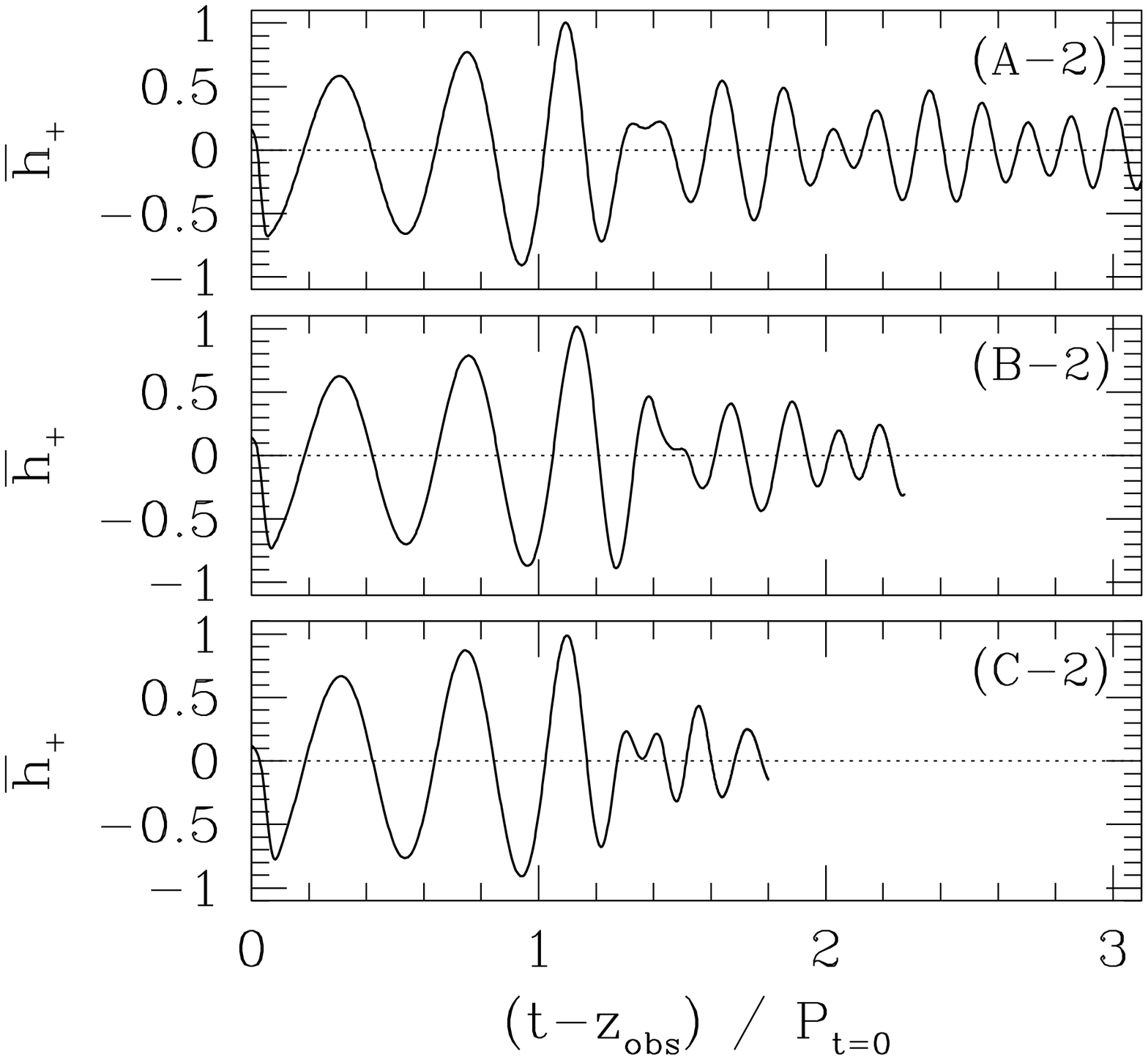}
\epsfxsize=2.75in
\leavevmode
\epsffile{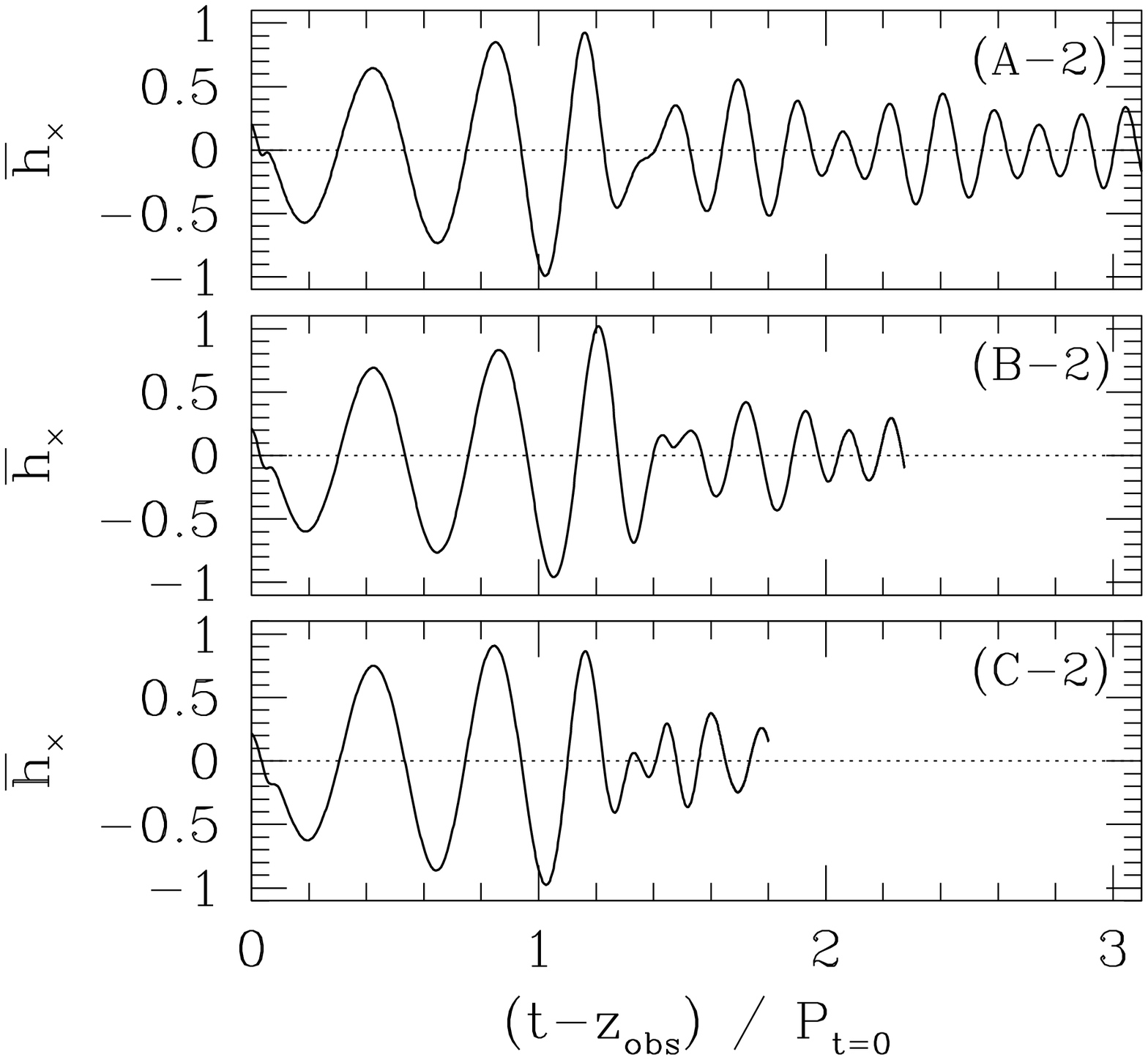}
\end{center}
\vspace{-7mm}
\caption{$\bar h_+$ and $\bar h_{\times}$ as functions of the retarded time 
for models (A-2), (B-2) and (C-2). 
\label{gw225}}
\end{figure}

In Fig.~\ref{gw225}, we plot $\bar h_+$ and $\bar h_{\times}$ as 
functions of retarded time $(t - z_{\rm obs})/P_{\rm t=0}$ 
for models (A-2), (B-2) and (C-2). 
(In the following, we always normalize the retarded time by $P_{t=0}$.) 
These results 
are obtained in a simulation with $(505,505,253)$ grid number. 
The waveforms are quite similar in these three cases for times satisfying
$t-z_{\rm obs}\alt P_{t=0}$, but they vary as the merger proceeds 
toward the final outcomes.

In the case of massive neutron star formation [model (A)],  
quasi-periodic gravitational waves of fairly large amplitude, which 
are excited due to nonaxisymmetric deformation and oscillation of the 
merged object, are emitted after the merger sets in. 
The wavelengths of these quasi-periodic waves are likely associated with 
several fundamental oscillation modes of the merged object, because 
the quasi-periodic oscillation is not a simple sine-cosine 
curve, but has a modulation. 
Since the radiation reaction timescale is much longer than 
the dynamical (rotational) timescale of a massive neutron star, 
the quasi-periodic waves will be emitted for many rotational cycles
(see below). 

Even for the case of black hole formation, 
quasi-periodic gravitational waves are excited 
due to the nonaxisymmetric oscillation of a transient 
merged object before collapsing to a black hole. 
Since the formation timescale of the black hole is different 
for models (B) and (C), depending on the initial 
compactness of the neutron stars, the duration of the emission of the 
quasi-periodic waves induced by the nonaxisymmetric oscillation of 
the merged objects is also different. Because 
the computation crashed soon after the formation of 
the apparent horizon, we cannot draw a definite conclusion from 
the present simulation 
with regard to gravitational waves in the last stage. However, 
we can expect that after the formation of 
black holes, QNMs of the black holes are  
excited and gravitational waves are eventually damped. 
To complete waveforms up to the ringing tail of QNMs, we 
need either to develop an implementation such as horizon excision 
\cite{excision} 
or to use the perturbative extraction technique recently developed 
by Baker, Campanelli and Lousto.\cite{BCL}

\begin{figure}[t]
\begin{center}
\epsfxsize=3.in
\leavevmode
\epsffile{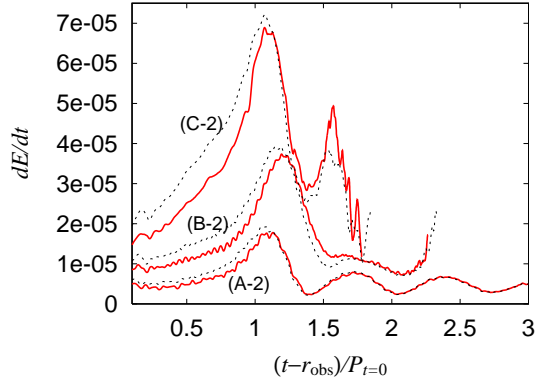}
\end{center}
\caption{Luminosity of gravitational waves as functions of the retarded 
time for models (A-2), (B-2) and (C-2). Solid and dashed curves represent 
the luminosity computed at $r_{\rm obs} \approx L$ and $0.77L$, 
respectively. $r_{\rm obs}$ denotes the radius at which the 
gauge invariant variables are extracted. 
For $t-r_{\rm obs} \alt P_{t=0}$, the wavelength of
gravitational waves is longer than $L$, but
for $t-r_{\rm obs} \agt 1.2P_{t=0}$, 
it becomes shorter than $L$ (see Fig.~9). 
The unit of luminosity is $c^5/G \approx 3.63 \times 10^{59}$~erg/sec. 
\label{dEdt}}
\end{figure}

The properties mentioned above are also reflected in the energy luminosity, 
as shown in Fig.~\ref{dEdt}. 
Here, we display the luminosity for $l=|m|=2$ modes 
as functions of the retarded time for (A-2), (B-2) and (C-2). 
We plot two luminosity curves that are extracted at 
$r_{\rm obs} \approx L$ (solid curves)
and $\approx 0.77L$ (dashed curves), respectively. 
It is found that during the early stage (i.e., for
$t-r_{\rm obs} \alt P_{t=0}$),
the luminosity extracted at $r_{\rm obs} \approx 0.77L$ is larger than
that at $r_{\rm obs} \approx L$. This is because the wavelength of 
gravitational waves is longer than $L$ during this stage, and hence 
gravitational waves are not accurately extracted from the metric. 
However, for $t - r_{\rm obs} \agt P_{t=0}$, the 
two curves agree fairly well, because the wavelength becomes 
shorter than $r_{\rm obs}$, and consequently accurate extraction of
gravitational waves becomes possible. 
We also note that it is difficult to obtain a smooth curve of $dE/dt$, 
because computing the energy luminosity requires taking 
time derivative of the gauge invariant quantity, which introduces 
a fairly large numerical noise.

\begin{table}[b]
\caption{
Total radiated energy and angular momentum in units of 
their initial values. We estimated the fluxes at 
$r_{\rm obs} \approx L$. If we estimated them at 
$r_{\rm obs} \approx 0.77L$, they are overestimated by $\sim 20\%$, 
because of inaccurate extraction of gravitational waves 
in the early stages during which the gravitational wavelength is longer than
$L$. 
}
\begin{center}
\begin{tabular}{|c|c|c|} \hline
Model & $\Delta E/M_{\rm ADM0}$ & $\Delta J/J_0$  \\ \hline\hline
(A-2) & 0.5\% & 8\% \\ \hline
(B-2) & 0.6\% & 10\% \\ \hline
(C-2) & 0.9\% & 12\% \\ \hline
(E-1) & 0.3\% & 6\% \\ \hline
(F-1) & 0.3\% & 7\% \\ \hline
(H-1) & 0.6\% & 10\% \\ \hline
\end{tabular}
\end{center}
\end{table}

We list the total radiated energy $\Delta E$ and 
angular momentum $\Delta J$ in Table III. 
A typical magnitude of $\Delta J/J_0$ is $\sim 10\%$, which is consistent 
with the decrease of the angular momentum $J$ computed by 
Eq. (\ref{eqj00}). (It was impossible to confirm 
the consistency of the change of $M_{\rm ADM}$ due to radiation reaction, 
because $M_{\rm ADM}$ varies by $\alt 10\%$ during simulations due to 
numerical error; see discussion in the Appendix.) 
It should be noted that $\Delta J/J_0$ is much larger than 
$\Delta E/M_{\rm ADM0}$, because the approximate relation 
$\Delta J \sim \Delta E / \Omega$ holds, and hence
\beq
{(\Delta J/J_0) \over (\Delta E/M_{\rm ADM0})} \sim 10 
\biggl( {1 \over q} \biggr)\biggl( {0.1 \over M_{\rm ADM0}\Omega} \biggr),
\eeq
where $\Omega$ here denotes a typical magnitude of the angular 
velocity. 

During the early stages ($t \alt P_{t=0}$), 
the luminosity of gravitational waves 
gradually increases, reaching a peak. 
The peak luminosity is approximately proportional to 
$\comp^5$, as expected from the quadrupole formula. 
However, after the peak is reached, 
the behavior of the curve
depends strongly on the outcome of the merger. 

For model (C-2), in which a black hole is formed in a short timescale 
at $t \sim 2P_{t=0}$ ($t-r_{\rm obs} \sim 1.7P_{t=0}$), 
a sharp peak is found before the black hole is formed. 
This peak is likely associated with an oscillation 
for a short-lived transient merged object. 
It is likely that 
peaks associated with QNMs appear after this peak, and 
the luminosity is eventually damped. 
For model (B-2), a long-lived transient object 
emits quasi-periodic gravitational waves for a certain duration. 
Consequently, the luminosity remains fairly large for a longer 
timescale than for model (C-2). 

In the simulations with models (B) and (C), the computation crashed 
soon after the formation of an apparent horizon, and hence it was not 
possible to observe gravitational waves 
associated with QNMs of the formed black hole. 
If it could be extracted, the peak associated with QNMs would 
appear at $(t-r_{\rm obs})/P_{t=0} \sim 2$ for model (C-2) and 
$\sim 2.5$ for model (B-2), before the luminosity is damped to zero. 

For the case of neutron star formation [model (A-2)], the evolution of 
the luminosity is very different from that for the other models after 
the first peak is reached. In this case, the luminosity 
oscillates with average amplitude $dE/dt \sim 5\times 10^{-6}$ 
after formation of the massive neutron star. 
We here estimate the approximate timescale for gravitational 
radiation reaction. 

The binding energy of the massive neutron star is approximately 
\beq
W \sim {M^2 \over R} = 0.2 M\biggl({5M \over R}\biggr),\label{eqW}
\eeq
where $M$ is the gravitational mass and $R$ is a 
typical radius of the massive neutron star $\sim 5M$. 
The ratio of the kinetic energy $T$ to $W$ is 
expected to be between 0.1 and 0.2, so that 
\beq
T \sim (0.02-0.04) M\biggl({5M \over R}\biggr). 
\eeq
Assuming that the energy luminosity remains 
$\sim 10^{-6}$, as indicated in Fig.~\ref{dEdt}, and that 
the emission timescale is approximately derived by dividing $T$ by $dE/dt$, 
we obtain 
\beq
t_{\rm GW} \sim {T \over dE/dt} 
\alt 0.1 {\rm sec} 
\biggl({M \over 3M_{\odot}}\biggr)\biggl({5M \over R}\biggr)
\biggl({5\times 10^{-6} \over dE/dt}\biggr). \label{tgw}
\eeq
Thus, within $\sim 0.1$ sec (or $\sim 50P_{t=0}$),
the massive neutron star will 
collapse into a black hole. 
As argued in Ref. \citen{BSS}, there are many other processes 
of dissipation and transport of angular momentum 
inside the massive neutron star, 
in addition to the emission of gravitational waves. 
However, the characteristic timescale for such processes is 
much longer than 1 sec. Thus, the emission timescale of 
gravitational waves would be shortest in this case. 
On the other hand, if $t_{\rm GW}$ were longer than 10 seconds, 
other processes, such as magnetic braking effect, could be more important 
than the effect of gravitational radiation.\cite{BSS} 

\begin{figure}[t]
\begin{center}
\epsfxsize=2.6in
\leavevmode
\epsffile{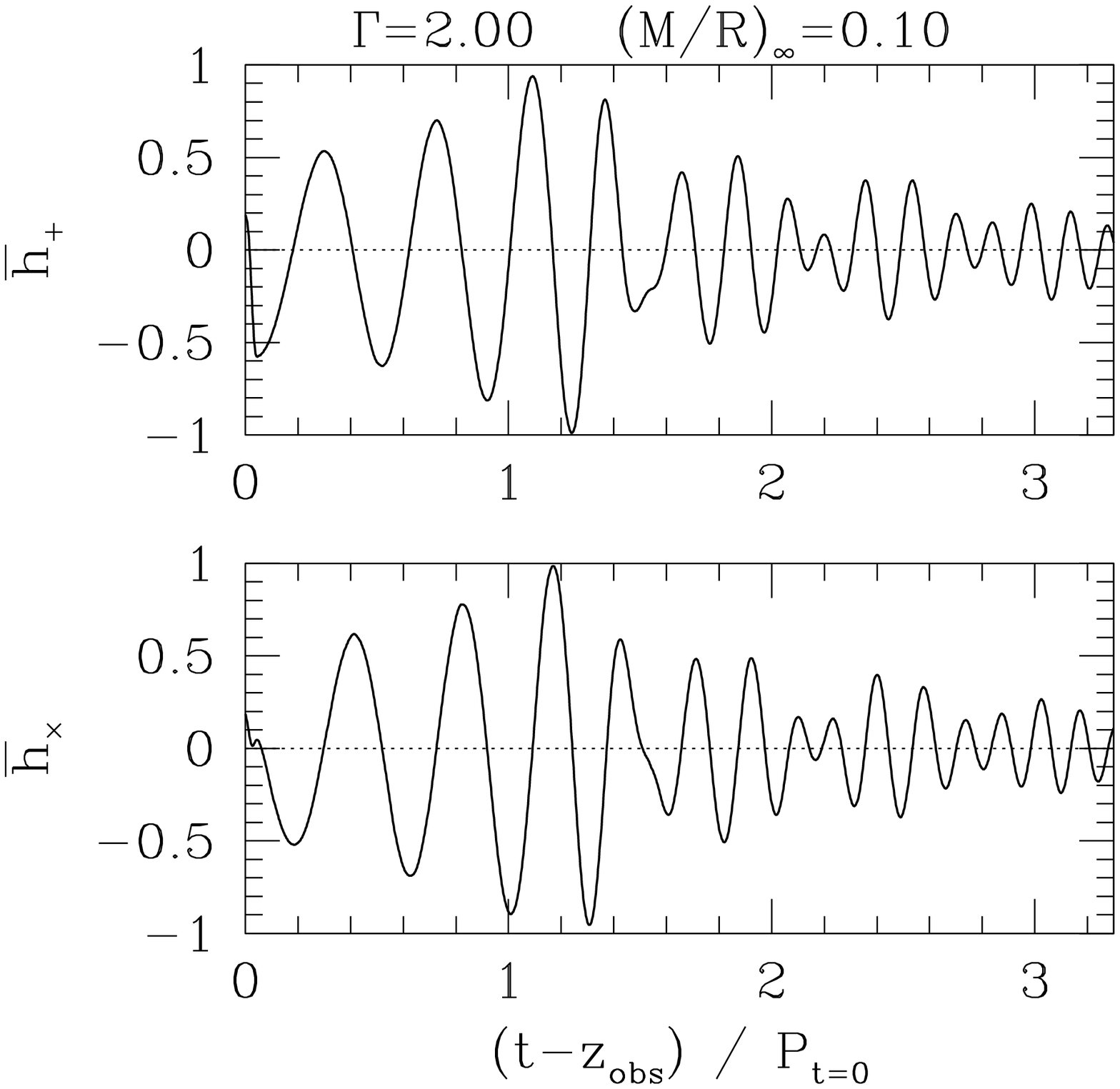}
\epsfxsize=2.6in
\leavevmode
\epsffile{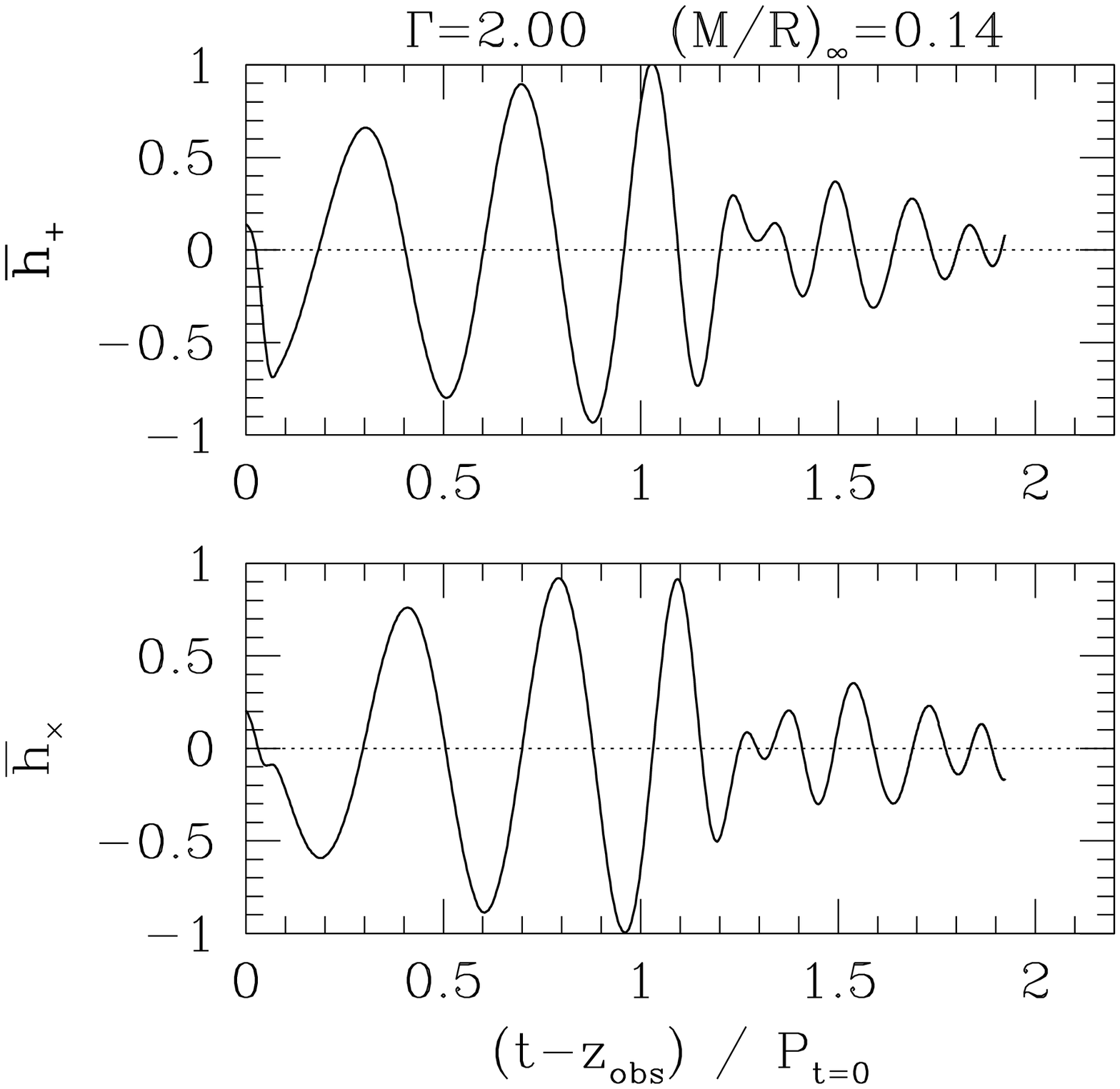}
\end{center}
\vspace{-7mm}
\caption{$\bar h_+$ and $\bar h_{\times}$ as functions of the retarded time 
for models (E-1) and (H-1). 
\label{gw20}}
\end{figure}

\begin{figure}[t]
\begin{center}
\epsfxsize=3.in
\leavevmode
\epsffile{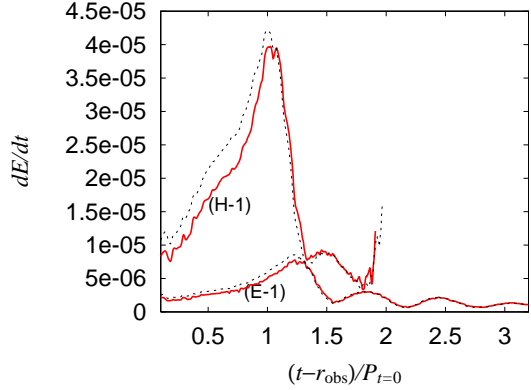}
\end{center}
\caption{
The same as Fig.~\ref{dEdt} but for models (E-1) and (H-1). 
\label{dEdt2}}
\end{figure}

All features mentioned above for $\Gamma=2.25$ are also found in 
gravitational waveforms for $\Gamma=2$. 
In Fig.~\ref{gw20}, we display $\bar h_+$ and $\bar h_{\times}$ for models 
(E-1) and (H-1). In Fig.~\ref{dEdt2}, we also display the luminosity 
for these models. 
For model (H-1), a black hole is formed, while 
for (E-1), a neutron star is the outcome. 
As in the $\Gamma=2.25$ case, 
this feature is well reflected in Figs.~\ref{gw20} and \ref{dEdt2}: 
Quasi-periodic oscillations are seen over a long duration 
for model (E-1), while for model (H-1), such oscillations 
are found only for a short timescale. 

For model (E-1), the luminosity remains $\sim 10^{-6}$
after formation of a massive neutron star.
Thus, the emission timescale of gravitational waves from
a formed massive neutron star is likely shorter than 1 sec, 
as shown in Eq. (\ref{tgw}).
This implies that, as in the case of model (A), 
it eventually collapses into a black hole, 
due to the angular momentum dissipation through gravitational waves. 

\subsubsection{Gravitational wave spectra}

\begin{figure}[t]
\begin{center}
\epsfxsize=2.7in
\leavevmode
\epsffile{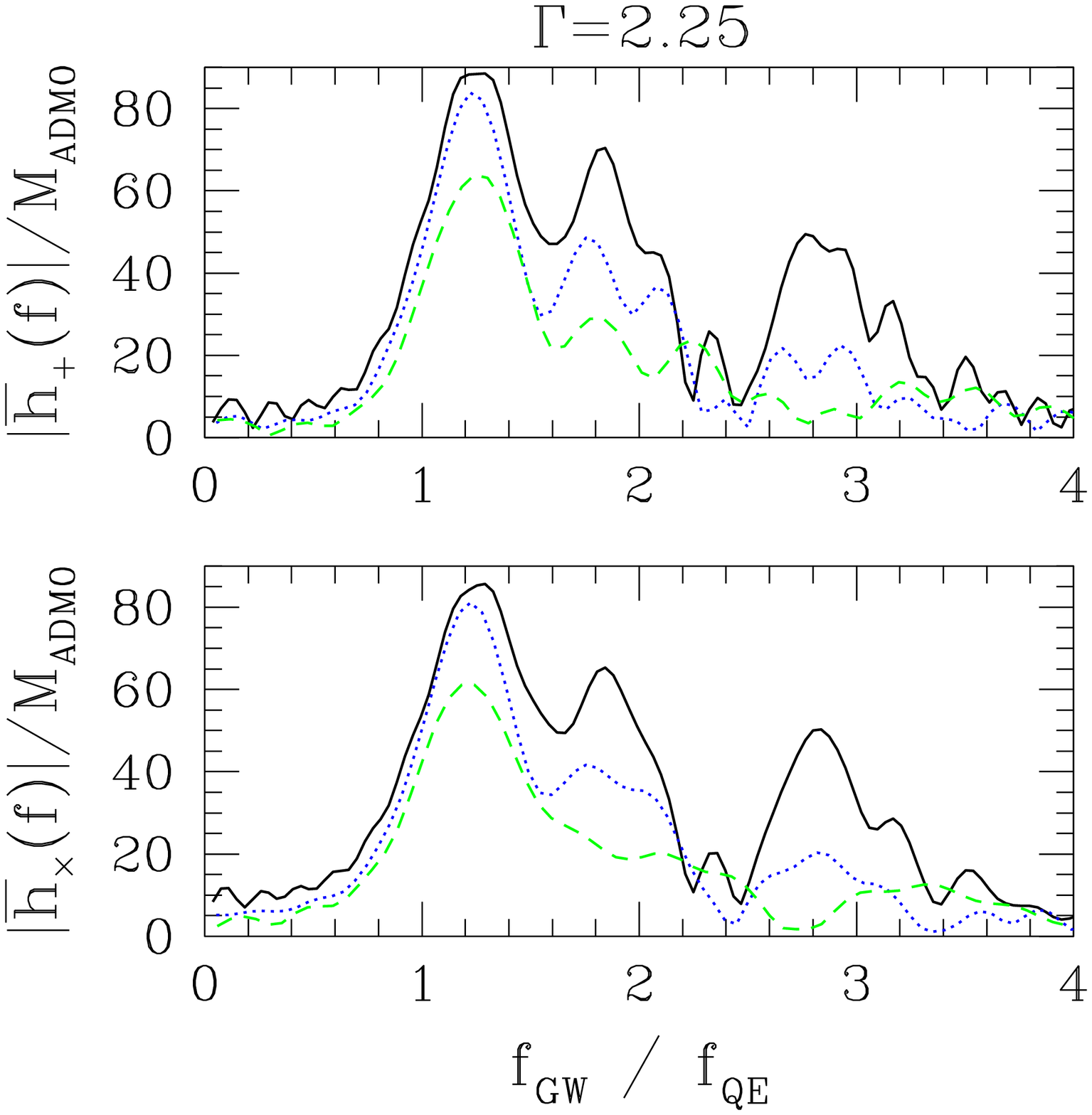}
\epsfxsize=2.7in
\leavevmode
\epsffile{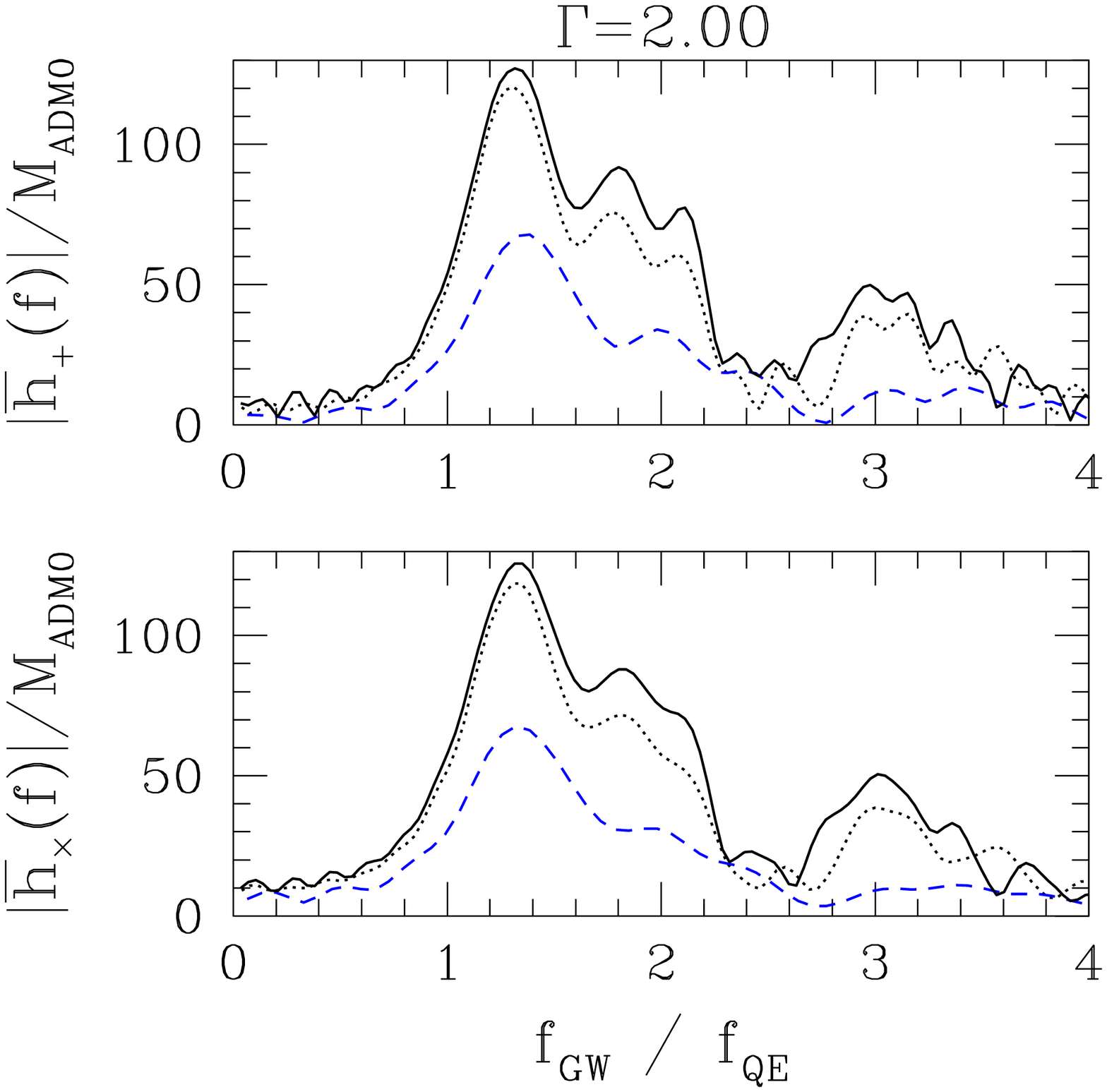}
\end{center}
\vspace{-7mm}
\caption{
Fourier spectra of $\bar h_+$ and $\bar h_{\times}$ 
for models (A-2) (solid curve), (B-2) (dotted curve) and 
(C-2) (dashed curve) [left], and  
for models (E-1) (solid curve), (F-1) (dotted curve)
and (H-1) (dashed curve) [right].
\label{spectrum}}
\end{figure}

In Fig.~\ref{spectrum}, spectra of gravitational 
waves are displayed for $\Gamma=2.25$
[models (A-2), (B-2) and (C-2) on the left] and for $\Gamma=2$ 
[models (F-1) and (H-1) on the right]. 
Since the computation crashed soon after formation of the black hole, 
the spectra of models (B-2), (C-2) and (H-1) are incomplete on the 
high frequency side [i.e., for $f_{\rm GW}/f_{\rm QE} \agt 4$, 
where $f_{\rm QE}$ is given in Eq. (\ref{fqe})]. 
In these cases, QNMs of black holes would be excited 
in the last stage of black hole formation. However, as mentioned 
in \S 1, the frequency would be very high 
($f_{\rm GW} > 5~{\rm kHz} \sim 5 f_{\rm QE}$). 
Here, we focus only on the spectra for quasi-periodic 
oscillations of merged objects for which 
$f_{\rm GW} \sim (2-3)~{\rm kHz}$, so that the incompleteness 
with regard to the spectra on the higher frequency side is not 
a problem. For the spectra of model (A), it should be kept in mind 
that the peak associated with the 
quasi-periodic oscillations is underestimated in Fig.~\ref{spectrum}, 
because quasi-periodic waves will be emitted perhaps to 
$t \sim 100 P_{t=0}$, as indicated in Eq.~(\ref{tgw}). 
For all models, we should also point out that in the real 
spectra of the coalescing binaries, 
$\bar h_{+,\times} \propto f_{\rm GW}^{-7/6}$ for 
$f_{\rm GW} \alt f_{\rm QE}$,\cite{CF}  
which is not taken into account in the present results.

{}From Fig.~\ref{spectrum}, we find two typical frequencies 
of the quasi-periodic oscillation for each $\Gamma$: 
\beqn
f_{\rm QPO} &\sim& \left\{
\begin{array}{ll}
\displaystyle 
1.8f_{\rm QE}
\approx 1.7~{\rm kHz}
\biggl({2.8M_{\odot} \over M_{\rm ADM0}}\biggr)
\biggl({C_0 \over 0.12}\biggr)^{3/2} & \\
\displaystyle 
2.8f_{\rm QE}
\approx 2.7~{\rm kHz}\biggl({2.8M_{\odot} \over M_{\rm ADM0}}\biggr)
\biggl({C_0 \over 0.12}\biggr)^{3/2} & 
\end{array}
\right. {\rm for}~\Gamma=2.25, \ \\
f_{\rm QPO} &\sim& \left\{
\begin{array}{ll}
\displaystyle 
1.8f_{\rm QE}
\approx 1.7~{\rm kHz}
\biggl({2.8M_{\odot} \over M_{\rm ADM0}}\biggr)
\biggl({C_0 \over 0.12}\biggr)^{3/2} & \\ 
\displaystyle 
3.0f_{\rm QE}
\approx 2.9~{\rm kHz}\biggl({2.8M_{\odot} \over M_{\rm ADM0}}\biggr)
\biggl({C_0 \over 0.12}\biggr)^{3/2} & 
\end{array}
\right. {\rm for}~\Gamma=2. \ 
\eeqn
This shows that the ratio of $f_{\rm QPO}$ to $f_{\rm QE}$ 
is not very sensitive to $\comp$. However,
this ratio for higher frequencies $\sim 3f_{\rm QE}$ 
depends on the stiffness (or $\Gamma$) of the equations of state. 
The numerical results indicate that this ratio 
is larger for softer equations of state
for a peak of higher frequency. (This fact agrees with 
the results of Newtonian and post Newtonian simulations.
\cite{C,FR,ORT}) As pointed out in Ref.~\citen{C}, 
from the frequencies of quasi-periodic oscillations, 
we could constrain the stiffness of equations of state. 

As explained above, the peaks of the Fourier spectra 
that are associated with the quasi-periodic oscillations 
are determined by its accumulated cycles. 
Hence, if we would not find a large peak, 
we could conclude that a black hole is formed in a short timescale 
after merger. This implies that even without detecting QNMs or other 
signals of the black hole, it is possible to obtain evidence for 
black hole formation. On the other hand, if we would find a 
large peak, we could conclude that a massive neutron star 
is formed, at least temporarily. The criterion for the formation of 
black holes depends on the compactness and stiffness of the 
equations of state of the merging neutron stars. 
Thus, this information is also useful for constraining the 
equations of state. 

{}From gravitational waves emitted in the inspiraling stage with 
post-Newtonian templates of waveforms,\cite{BIWW} 
the masses of the two neutron stars, and hence, the total mass, 
will be determined.\cite{CF}  As mentioned above, 
from the intensity of the peak of the 
spectra associated with quasi-periodic oscillations, 
we can infer the compactness (for a given stiffness of an 
equation of state). This combined observation could constrain 
the relation between the compactness and mass of neutron stars. 
In this way, it is likely possible to strongly 
constrain nuclear equations of state using observed quantities such as 
the intensity of the 
Fourier peak of quasi-periodic oscillation, the ratio of 
$f_{\rm QPO}$ to $f_{\rm QE}$, and the total mass of the system. 

Since $f_{\rm QPO}$ is rather high, it may be 
difficult to detect quasi-periodic oscillation 
by first generation, kilometer-size 
laser interferometers, such as LIGO I. 
However, resonant-mass detectors and/or specially designed 
advanced interferometers, such as LIGO II, may be capable 
in the future to detect such high frequency gravitational waves. 
These detectors will provide us a wide variety of 
information on neutron star physics. 

\subsubsection{Calibrations}

To assess the accuracy and robustness of our results as well as to investigate 
the influence of the approximate initial conditions on gravitational 
waveforms, we performed several test simulations. 

\begin{figure}[t]
\begin{center}
\epsfxsize=2.7in
\leavevmode
\epsffile{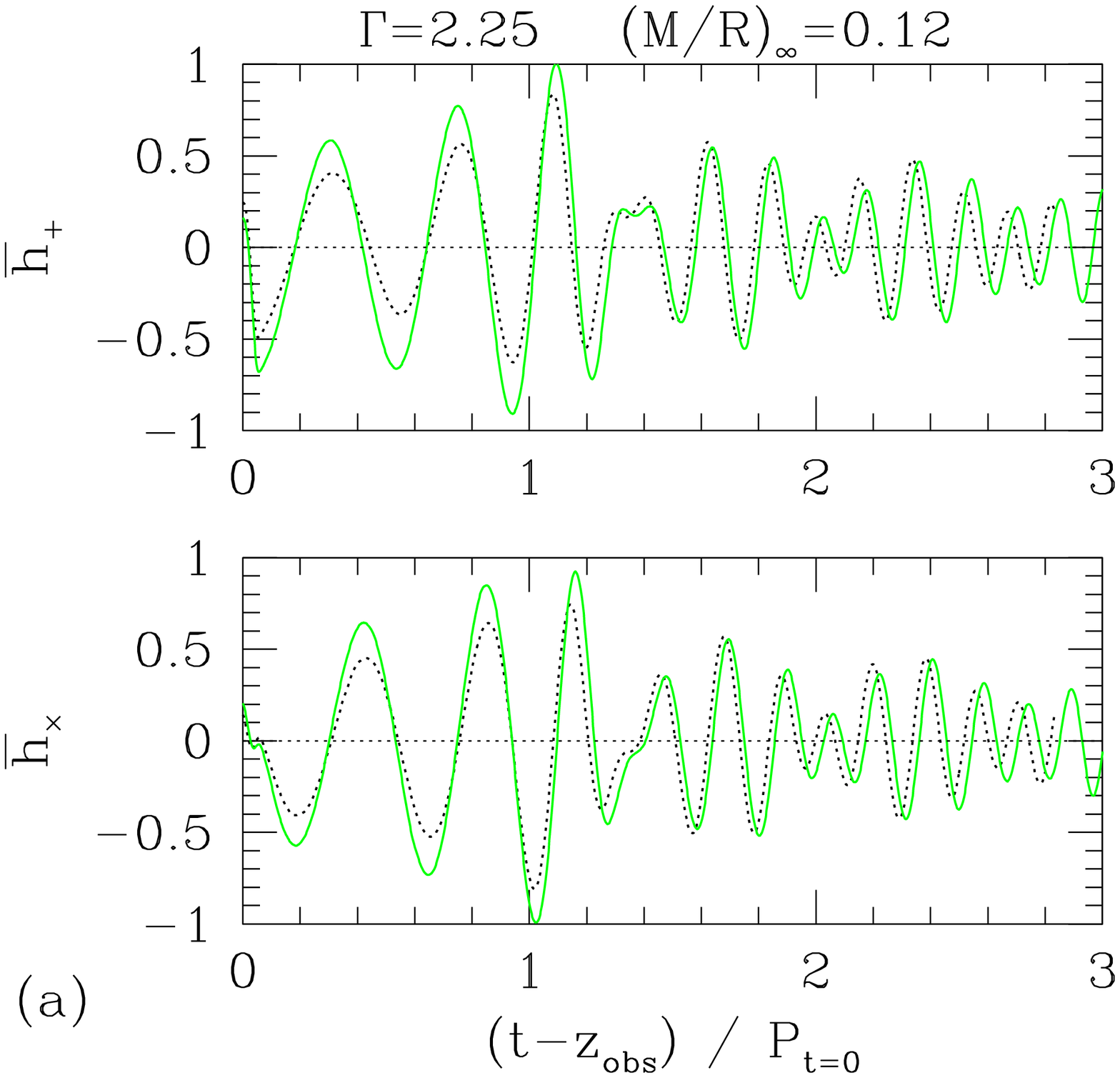}
\epsfxsize=2.7in
\leavevmode
\epsffile{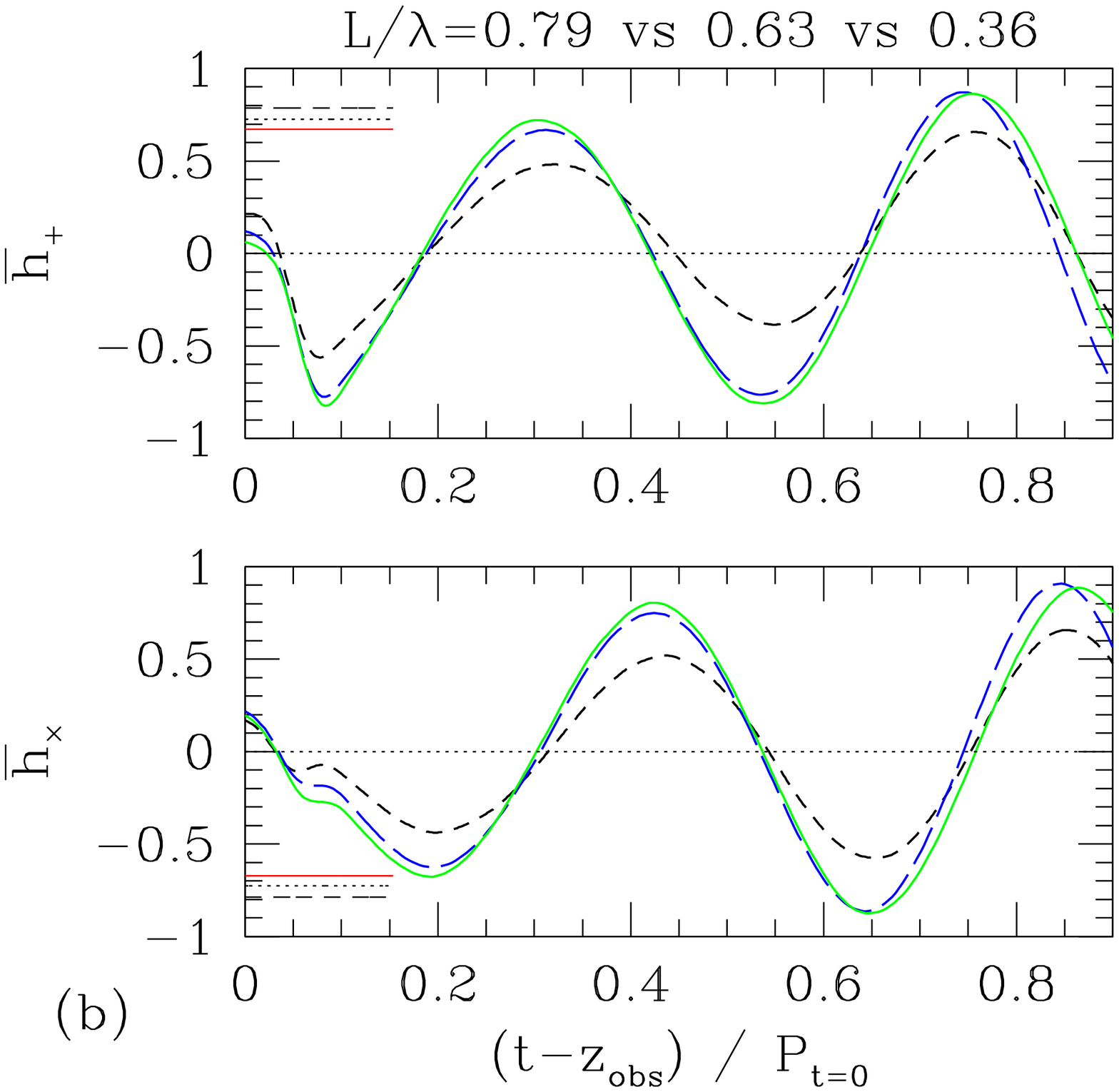}
\end{center}
\vspace{-7mm}
\caption{
(a): $\bar h_+$ and $\bar h_{\times}$ 
for models (A-2) (solid curve) and (A-0) (dashed curve). 
(b): $\bar h_+$ and $\bar h_{\times}$ 
for models (C-3) (solid curve), (C-2) (long dashed curve) 
and (C-0) (dashed curve). The horizontal solid, dotted and dashed lines 
drawn near the vertical axis denote the theoretical amplitudes 
at $t=0$ computed in the 2.5, second post-Newtonian and 
quadrupole approximations. 
\label{check}}
\end{figure}

As mentioned above, the numerical accuracy of the gravitational 
wave amplitude depends on the location of the outer boundaries $L$ 
and the grid spacing $\Delta x$.  It is indicated that (1) 
smaller $L(< \lambda)$ and larger $\Delta x$ result in 
the underestimation of $\bar h_{+,\times}$ and (2) 
smaller $L(\ll \lambda)$ also induces spurious modulation of
$\bar h_{+,\times}$ 
(by which $\bar h_{+,\times}$ deviate from zero systematically). 
However, we will show that with the typical grid size 
adopted in the present work ($L \agt 0.6\lambda$),
these numerical errors are sufficiently suppressed. 

First, we investigate the convergence of the wave amplitude 
resulting when the computational domain is enlarged. 
In Fig.~\ref{check} (a), we compare the merger waveforms for 
model (A) with (293,293,147) [(A-0)] and (505,505,253) [(A-2)] 
grid numbers. 
We note that $L \sim \lambda_{\rm QP}$ for the smaller grid number and 
$L \sim 2\lambda_{\rm QP}$ for the large grid number, where 
$\lambda_{\rm QP}$ here denotes the wavelength of 
quasi-periodic gravitational waves emitted by a formed 
massive neutron star. 
Figure~\ref{check} (a) shows that the amplitude of gravitational 
waves from quasi-periodic oscillations is not sensitive to $L$. 
This indicates that a value of $L$ that is
$\agt \lambda_{\rm QP}$ is sufficiently large for 
computing waveforms associated with quasi-periodic oscillations 
of a formed massive neutron star. On the other hand,
it is not clear if 
convergence is achieved for the wave amplitude for $t \alt P_{t=0}$. 

\begin{table}[t]
\caption{
Amplitudes of $\bar h_{+,\times}$ at $t=0$ 
for several levels of the post-Newtonian approximation. 
}
\begin{center}
\begin{tabular}{|c|c|c|c|c|c|} \hline
Model & quadrupole & 1PN & 1.5PN & 2PN & 2.5PN\\ \hline\hline
(A) & 0.755 & 0.611 & 0.738 & 0.705 & 0.681 \\ \hline
(B) & 0.766 & 0.594 & 0.760 & 0.712 & 0.676 \\ \hline
(C) & 0.785 & 0.578 & 0.793 & 0.727 & 0.671 \\ \hline
(E) & 0.735 & 0.620 & 0.712 & 0.690 & 0.675 \\ \hline
(H) & 0.759 & 0.590 & 0.752 & 0.705 & 0.670 \\ \hline
\end{tabular}
\end{center}
\end{table}

To evaluate the numerical error in the amplitude for $t \alt P_{t=0}$, 
we compare the waveforms for models (C-0), (C-2) and (C-3)
in Fig.~\ref{check} (b), in which $L/\lambda_{0}=0.36$, 0.63, and 0.79, 
respectively. In addition to the waveforms, the 
amplitudes predicted from the post-Newtonian theory 
are shown by horizontal lines. Here, 
the wave amplitude of the $|m|=2$ modes (sum of $l=2$ to $l=6$ modes) 
for a binary of equal point mass in circular orbits can be written as 
\beq
{M_{\rm tot} v^2 \over D}
\biggl[1 - {17  \over 8}v^2 + 2\pi v^3 
-{15917 \over 2880}v^4-{17  \over 4}\pi v^5 \biggr]
\eeq 
at the 2.5 post-Newtonian order,\cite{BIWW,B}
\footnote{In Ref. \citen{BIWW}, post-Newtonian 
waveforms are given up to second order. 
To derive 2.5 post-Newtonian waveforms, 
we need to use the 2.5 post-Newtonian luminosity derived in Ref. \citen{B}.} 
where $M_{\rm tot}$
is the sum of the gravitational masses of the two stars in isolation 
(hence $M_{\rm tot} \not= M_{\rm ADM0}$ for a finite separation, due to 
the existence of 
the binding energy between the two stars) and $v=(M_{\rm tot}\Omega_0)^{1/3}$. 
We list the amplitudes of $\bar h_{+,\times}$ for several levels of 
the post-Newtonian approximation in Table IV. It is seen 
that the amplitude converges to $\sim 0.7$ for each model. 
Figure~\ref{check} (b) indicates that 
for model (C-0), the amplitude is underestimated 
by 30--40 \%, but as $L$ increases, a convergence is 
achieved. The wave amplitudes for models (C-2) and (C-3) 
are nearly coincident and also 
agree with the 2.5 post-Newtonian amplitudes 
within $< 10\%$ error. Thus, with these computational domains 
with $L \agt 0.6 \lambda_{0}$,  
the amplitude seems to be computed within $10\%$ error, and hence 
we may assume that the error on the wave amplitude calculated for models 
(B-2), (C-2), and (H-1), 
shown in Figs.~\ref{gw225} and \ref{gw20}, is less than $10\%$.
This conclusion agrees with that of our recent numerical experiment 
on gravitational waveforms from inspiraling binary neutron stars 
in quasiequilibrium states.\cite{SU01} 

\begin{figure}[t]
\begin{center}
\epsfxsize=2.7in
\leavevmode
\epsffile{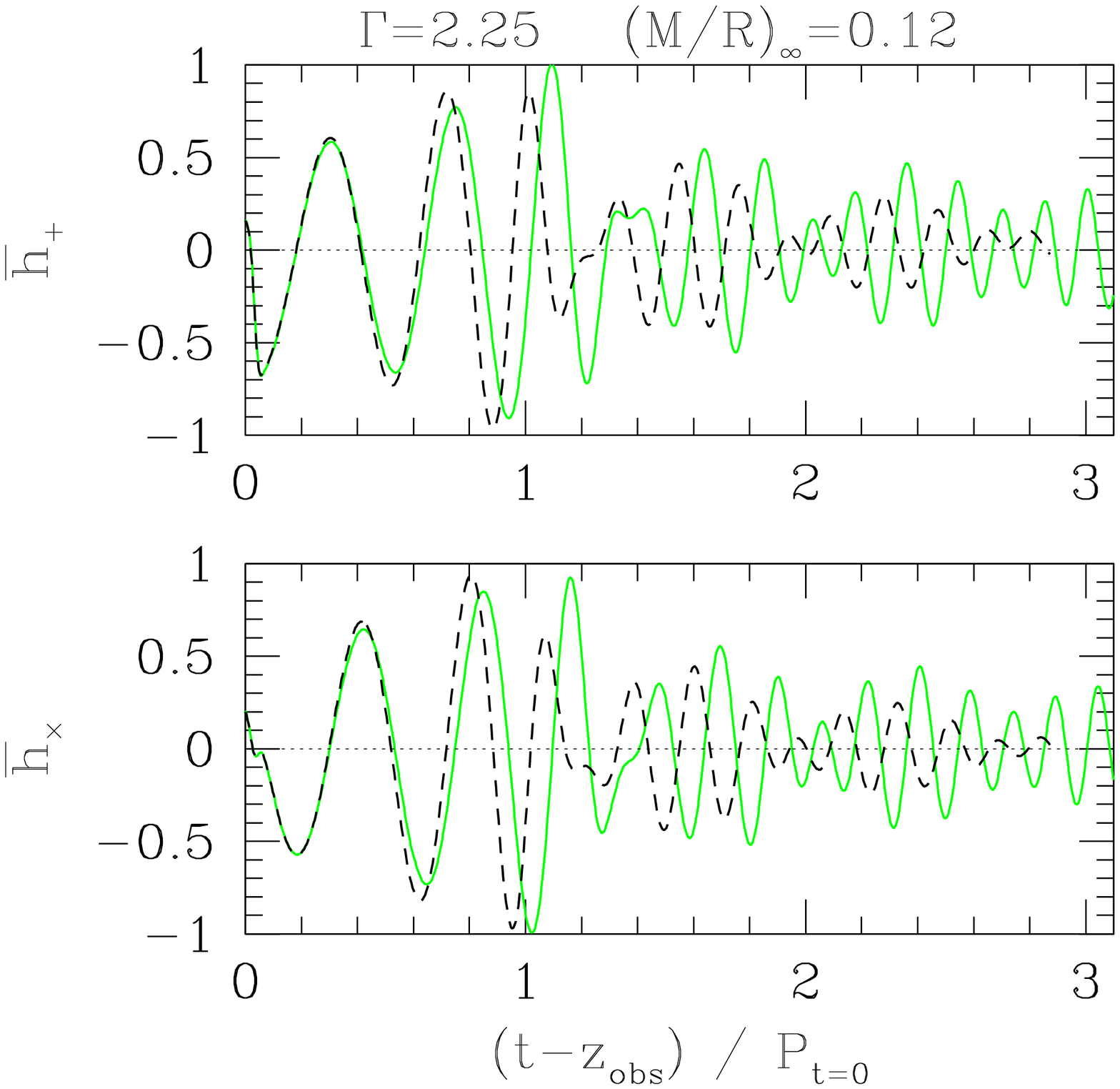}
\epsfxsize=2.7in
\leavevmode
\epsffile{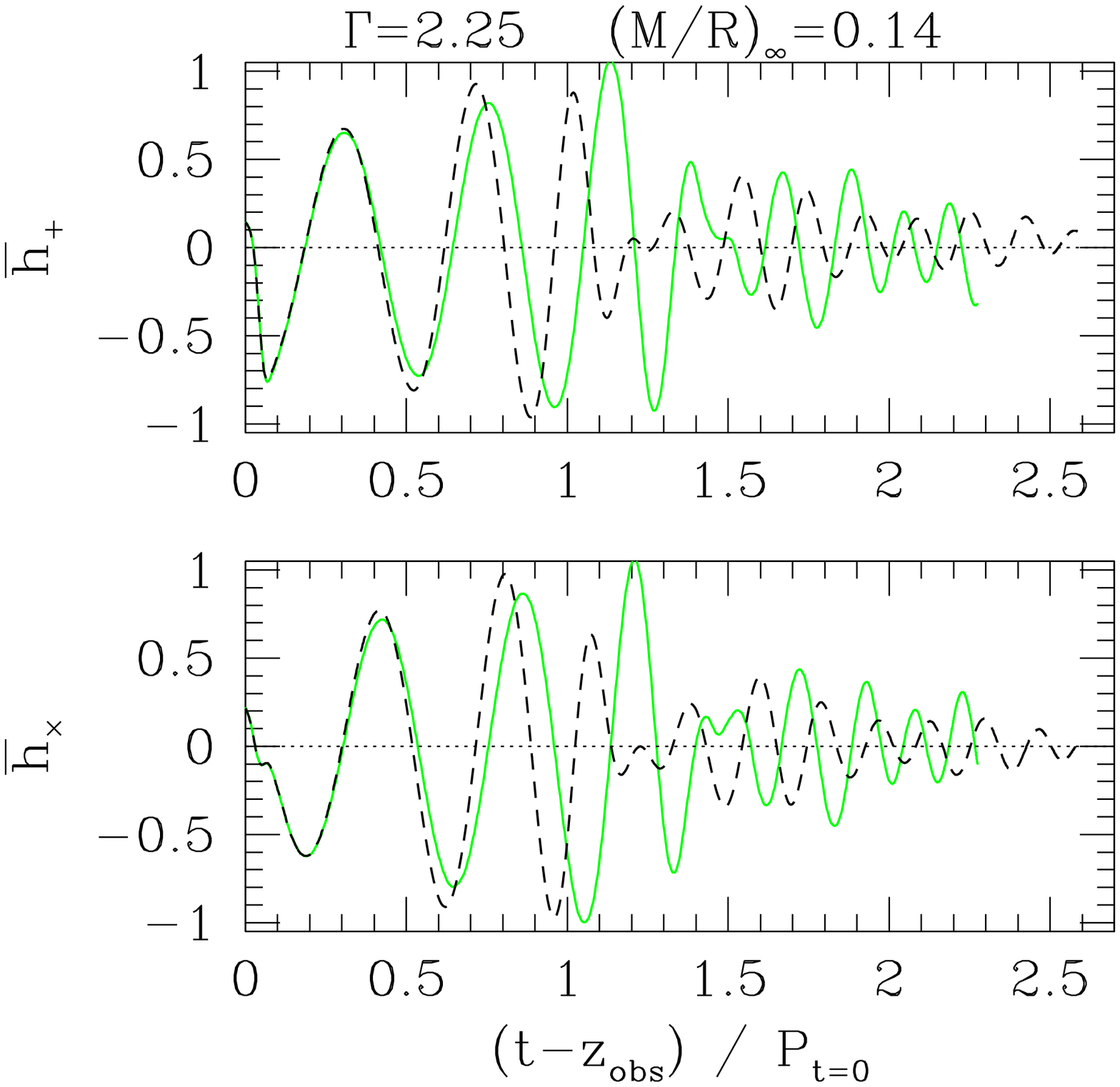}
\end{center}
\vspace{-5mm}
\caption{
Left: $\bar h_+$ and $\bar h_{\times}$ as functions of the retarded time 
for models (A-2) (solid curve) and (A-3) (dashed curve). 
Right: The same as (a) but for (B-2) (solid curve) and (B-3)
(dashed curve). 
\label{figconv}}
\end{figure}

Next, we investigate the effect of the grid resolution ($\Delta x$) on 
gravitational waveforms. 
In Fig.~\ref{figconv}, we show $\bar h_+$ and $\bar h_{\times}$ for 
models (A-2) and (A-3) and models (B-2) and (B-3) 
for comparison. At a glance, we find that 
the global properties in the waveforms are similar with regard to 
results for high and low resolution. However, two 
differences are also found : (1) 
since the merger starts earlier, due to larger numerical dissipation, 
merger waveforms start earlier for simulations with lower 
resolution; (2) the wave amplitude is underestimated for lower resolution.  
Property (2) can be understood in the following manner. 
In the simulation with lower resolution, 
the density gradient is less accurately computed. The 
wave amplitude becomes higher in association with 
the formation of higher density peaks. 
Thus, the lower resolution results in underestimation of the 
gravitational wave amplitude. [This property has been 
found in many Newtonian simulations using
Eulerian codes; e.g., Ref. \citen{RJ}.]

It is desirable to further investigate the convergence improving 
the resolution in order to draw definite conclusions about 
whether the convergence of the waveforms is fully achieved. However, it is
pragmatically difficult to do this with
the current computational resources, because
a simulation with a large grid size, e.g., (633, 633, 317), would take 
a very long computational time, perhaps $\agt 200$--250 CPU hours, 
which is about 1/3 of the computational time assigned to us.
In the future, it will be feasible to improve the resolution using better 
computational resources.  In addition, an adaptive mesh
refinement (AMR) technique for high density regions 
might be necessary to save computational time and memory.\cite{AMR}

\begin{figure}[t]
\begin{center}
\epsfxsize=2.7in
\leavevmode
\epsffile{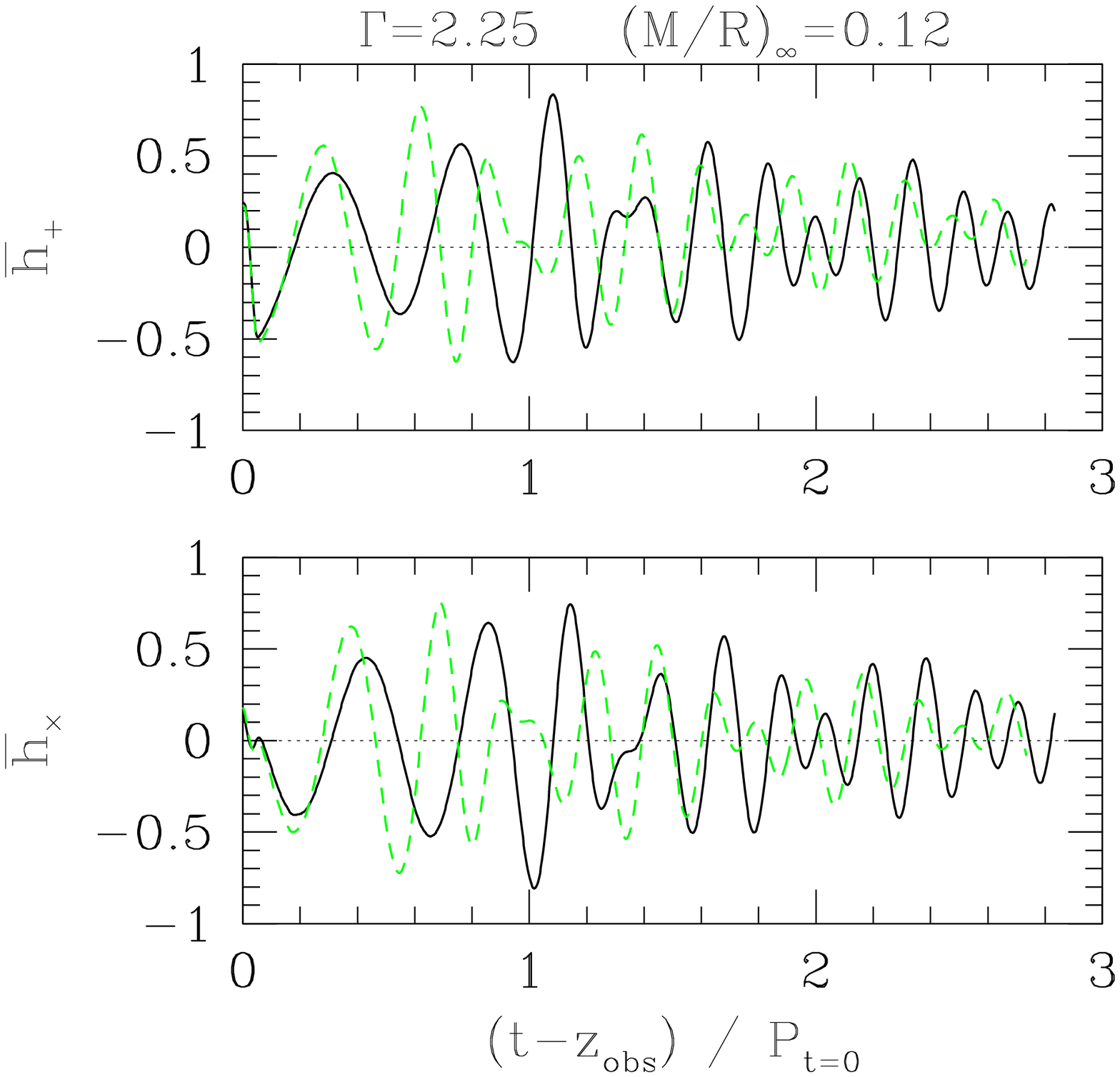}
\epsfxsize=2.7in
\leavevmode
\epsffile{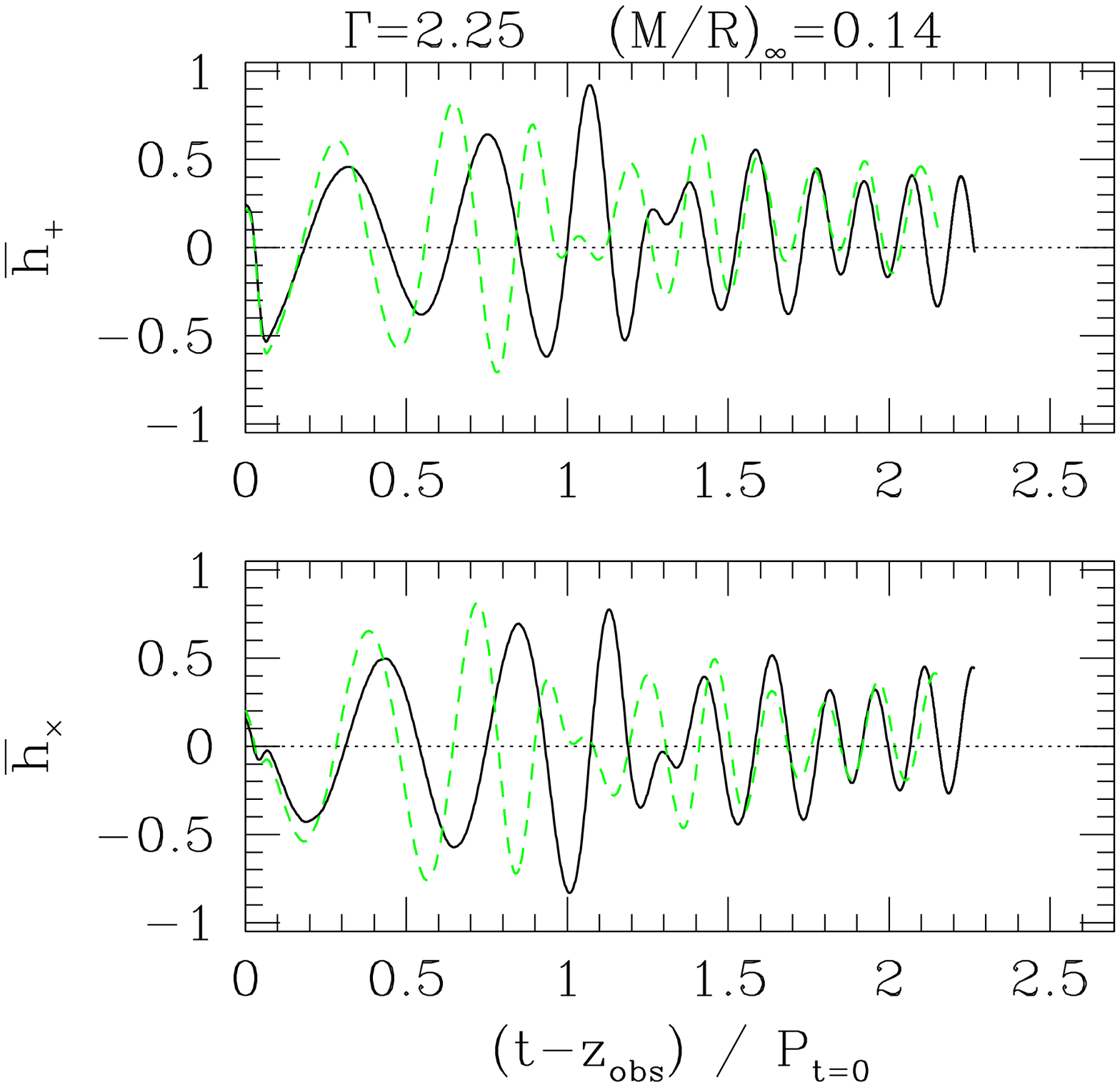}
\end{center}
\vspace{-7mm}
\caption{
Left: $\bar h_+$ and $\bar h_{\times}$ 
for models (A-0) (solid curve) and (A-1) (dashed curve). 
Right: The same as (a) but for models (B-0) (solid curve) and
(B-1) (dashed curve). The amplitude is underestimated by
30--40 \% in the early stages ($t \alt P_{t=0}$), because 
the computational domain in these simulations is too small 
(see Figs.~\ref{gw225} and \ref{check} for comparison). 
\label{approach}}
\end{figure}

Finally, we investigate the effect of an approaching velocity, which 
we ignored in our main simulations for $\Gamma=2.25$. 
In this test, we perform a simulation with a 
large approaching velocity as $v_{\rm app}=0.1 v_{\rm orb}$ for 
$\comp=0.12$ [model (A-1)], and 
for $\comp=0.14$ [model (B-1)] with (313,313,157) grid number. 
For model (B-1), we also decrease the reduction factor of 
the angular momentum by 1\% (see Table II). In Fig.~\ref{approach}, 
we show $\bar h_+$ and $\bar h_{\times}$ 
for these two cases. For comparison, we also plot the waveforms 
of models (A-0) and (B-0). 
Because of the approaching velocity, the waveforms before merger starts 
are very different for the two cases; the wavelength 
becomes short for models (A-1) and (B-1) soon after the
simulations start. 
However, the quasi-periodic waveforms after the onset of
the hydrodynamic interaction of two neutron stars 
are quite similar. This is because the 
quasi-periodic waves are associated with the fundamental 
oscillation modes of merged objects, which appear to depend 
weakly on the approaching velocity. 
Thus, we may conclude that the waveforms in the merger do not 
depend strongly on the approaching velocity, as 
long as $v_{\rm app}$ is not extremely large, i.e.,
as long as $\leq 0.1v_{\rm orb}$. 

We note that $\bar h_+$ in Fig.~\ref{approach} 
deviates from zero in the late stages for $t-z_{\rm obs}\agt 1.5P_{t=0}$ 
[in particular for model (B)]. This spurious effect 
results from the fact that the outer boundaries are too close to the 
strong field zone. Indeed, this numerical effect disappears when we 
increased the grid number to enlarge the computational domain 
(compare with Fig.~\ref{gw225}).

\section{Summary}

Using a new large scale supercomputer at NAOJ, 
we have performed fully GR simulations 
for the merger of equal mass binary neutron stars, 
particularly focusing on gravitational waveforms. 
Since the computational domain is much larger than that used in previous 
simulations, gravitational waveforms can be computed much more accurately 
than those reported in Refs.~\citen{bina} and \citen{binas}. 
In the following, we summarize the results obtained in this paper. 

The merger process and final products depend strongly on the 
compactness and the stiffness of the equations
of state for the merging neutron stars. 
For binaries of less compact neutron stars, massive 
neutron stars are formed, at least temporarily. 
In contrast, black holes are formed in the merger 
of sufficiently compact neutron stars on a dynamical timescale. 
However, the formation timescale of a black hole depends strongly on 
the compactness of the neutron stars. 
The criterion regarding compactness for prompt formation of 
a black hole depends also on the stiffness of equations of state. 
For $\Gamma=2.25$ and 2, threshold values are 
$\sim 0.16$ and $\sim 0.14$, respectively. 

The nature of the gravitational waveforms during the merger 
depend sensitively on the compactness (or mass) of the merging neutron stars. 
For mergers of less compact binaries, such as in models (A), (E) and (F), 
a transient massive neutron star is formed, and survives for the 
emission timescale of gravitational radiation, which is much 
longer than the dynamical timescale. Such massive neutron stars 
are highly nonaxisymmetric, so that quasi-periodic 
gravitational waves associated 
with nonaxisymmetric deformation are emitted. These 
quasi-periodic waves have typical frequencies of $\sim 2 - 3$ kHz, 
which yield Fourier peaks in the 
frequency domain of gravitational waves. 
The ratio of the peak frequency to the frequency at innermost orbits, 
$f_{\rm QPO}/f_{\rm QE}$, is not sensitive to the compactness, but 
depends on the stiffness of the equations of state. This 
indicates that its observation could constrain the stiffness of the 
equations of state.\cite{C}  It is also found that the luminosity of 
quasi-periodic gravitational waves is fairly large, so that 
a massive transient neutron star likely collapses into a black hole 
eventually,
due to the angular momentum dissipation through gravitational radiation. 

In models of slightly larger compactness, such as models (B) and 
(G), a transient massive object is also formed after merger sets in. 
This massive object is also highly nonaxisymmetric. 
Since the lifetime of such a transient massive 
object is fairly long, it 
causes characteristic peaks associated with 
the quasi-periodic oscillation in the Fourier domain of 
gravitational waves. However, for models with large compactness, such as 
models (C), (D), (H) and (I), the merged object collapses into a black hole 
on a dynamical timescale (i.e., on a few rotational periods of
the merged objects), and hence quasi-periodic gravitational waves 
are excited only on a short timescale. 

The intensity of the peaks in the Fourier spectra of gravitational waves 
associated with the nonaxisymmetric, quasi-periodic oscillations 
of the merged object 
depends on the lifetime of the transient massive object. 
If the transient object is long-lived, 
the peak becomes high, while 
if it collapses into a black hole in a few oscillation periods, 
the peaks are small. 
Therefore, from the intensity of the peak associated with 
the quasi-periodic oscillation, 
it may be possible to determine the object formed after the merger 
in future observations. 

To this time, we have performed simulations focusing on binaries 
of two identical neutron stars. Although all
the binary neutron stars observed so far 
are composed of neutron stars of two nearly identical masses,\cite{T} 
it seems that there is 
no fundamental reason for the production of such symmetric systems.
In the merger of two neutron stars of unequal mass, the merger process 
would be different from that for identical neutron stars. Since a 
less massive star is less compact, it is likely to be tidally disrupted 
before the dynamical instability of the orbital motion sets in. 
In this case, the tidally disrupted star could form accretion disks 
around the more massive companion. 
Sufficient accretion to the massive companion 
is likely to trigger gravitational collapse and the
formation of a black hole. As a result,
a system consisting of a black hole surrounded by accretion disks 
might be formed. Associated with the change of the merger process, 
gravitational waveforms would be modified.\cite{NO}
To survey possible outcomes 
in the merger of binary neutron stars, 
it is obviously necessary to perform simulations for two 
neutron stars of unequal mass. 

It is also desirable to improve the implementation of the initial conditions. 
In simulations carried out to this time,
we have used quasiequilibrium states of a conformally flat 
three-metric as the initial conditions for simplicity. 
The conformal flatness approximation becomes a source of 
a certain systematic error when attempting to obtain 
realistic quasiequilibrium states. As a result, 
this approximation introduces a systematic error on
the initial conditions and subsequent merger simulation. 
Since the magnitude of the ignored terms in the conformal flatness 
approximation seems to be small, we expect that this effect is not 
very serious. However, this conclusion is not entirely certain.
To rule out this problem, 
it is necessary to perform simulations using more realistic 
quasiequilibrium states of generic three geometries as 
initial conditions.\cite{US01}

In addition, there are technical issues requiring improvement. 
As discussed above and in the Appendix, black hole forming region does not 
have good resolution in our current computation.
Obviously, it is necessary to improve the resolution
around the black hole forming region. 
Since we have to prepare a large computational domain $L$,
which is at least equal to the wavelength of gravitational waves, using 
restricted computational speed and memory, 
it is desirable to develop numerical techniques such as
the AMR technique or nested grid technique to overcome this
problem. This is also an issue to be resolved in the future.

\vskip 6mm
\begin{center}
{\large\bf Acknowledgments}
\end{center}
\vskip 3mm

Numerical computations were performed on the FACOM VPP 5000 machines 
at the data processing center of NAOJ. 
This work was in part supported by a Grant-in-Aid (No. 13740143) of 
the Japanese Ministry of Education, Science, Sports, Culture, 
and Technology and also in part by NSF Grant PHY00-71044.

\appendix

\section{Accuracy}

We have monitored the degree of violation of the Hamiltonian
constraint, conservation of the baryon rest-mass,
ADM mass and angular momentum during numerical simulation. 

We have found that $f_{\psi}$
is less than 0.1 for a region in which $\rho$ is
larger than $\sim 10^{-3}\rho_{\rm max}$, where $\rho_{\rm max}$ 
is the maximum value of $\rho$.
For a less dense region, $f_{\psi}$ is
often $O(0.1)$, because low density regions are 
not well resolved in the finite differencing scheme we used for the
hydrodynamic equations. 
When the merged object starts collapsing,
$f_{\psi}$ increases to $O(0.1)$, even in a high density region.
This indicates that the resolution in the black hole forming
region is not very good either.  

The baryon rest-mass was found to be conserved within the truncation error
in all computations. This is because 
no mass is ejected outside the computational domain, which is wide
enough. 

\begin{figure}[t]
\begin{center}
\epsfxsize=2.5in
\leavevmode
\epsffile{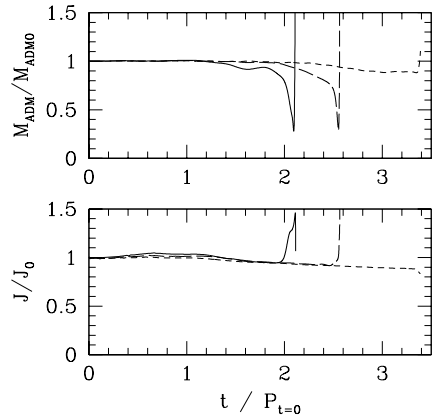}
\end{center}
\vspace{-7mm}
\caption{Evolution of $M_{\rm ADM}$ and $J$ as functions of
$t/P_{t=0}$ for models (A-2) (dashed curve), (B-2) (long-dashed curve)
and (C-2) (solid curve). An apparent horizon is formed at
$t/P_{t=0} \approx 2.5$ for (B-2) and  
$t/P_{t=0} \approx 2$ for (C-2). 
We do not add loss due to gravitational waves to $M_{\rm ADM}$ and
$J$. 
\label{error}}
\end{figure}

In the case of massive neutron star formation, the 
ADM mass is conserved within 10\% error throughout the simulation. 
At the last moment, at which excessive numerical error seems to accumulate,
the simulation crashes, and the error of the ADM mass suddenly diverges, 
as shown in Fig.~\ref{error}.
Such a crash took place typically at some time in the range
$t = 700$ -- $900M_{\rm ADM}$ 
(see \S 4). In the case of black hole formation, 
the ADM mass is conserved within 10\% error 
until the merged object starts collapsing into a black hole. The error 
systematically increases whenever the lapse 
at the center approaches to zero, and 
the error in the conservation of the ADM mass is typically 
$\sim 20\%$ at the time of the formation
of the apparent horizon (see Fig.~\ref{error}).
After this formation, the error rapidly increases 
to $O(1)$, and the computation crashes, at which time the error is
of order unity. The main source of the error in the ADM mass
appears to be the volume integral of $\tilde R_k^{~k}$, because it includes
the second spatial derivative of $\tilde \gamma_{ij}$, 
and hence loses accuracy easily whenever $\tilde \gamma_{ij}$ has
a steep gradient. 

The angular momentum is conserved within $\sim 5\%$ error 
throughout the simulation in the case of massive neutron star 
formation, as discussed in \S 4 (except for the last moment, 
at which excessive numerical error accumulates). 
For black hole formation cases, the angular momentum is conserved 
within $\sim 5\%$ error until the merged object starts collapsing.  
The error increases when the lapse 
at the center approaches to zero. 
The error for conservation in the angular momentum 
is typically $\sim 10\%$ at the time of the
formation of the apparent horizon. 
After the formation, the error increases to $O(1)$.

The divergence of the numerical error 
after formation of the apparent horizon is likely due either to
the lack of numerical resolution around black hole forming
region or to the inappropriate choice of the spatial gauge. 
However, at present it is not clear to us whether one of 
these is the actual source.


\begin{thebibliography}{99}

\bibitem{HT} R. A. Hulse and J. H. Taylor, Astrophys. J. {\bf 201} 
(1975), L55. \\
J. H. Taylor and J. M. Weisberg, Astrophys. J. {\bf 345} (1989), 434. 

\bibitem{BNST} V. Kalogera, R. Narayan, D. N. Spergel and J. H. Taylor, 
astro-ph/0012038: See also \\
E. S. Phinney, Astrophys. J. {\bf 380} (1991), L17: \\
R. Narayan, T. Piran and A. Shemi, Astrophys. J. {\bf 379} (1991), L17. 

\bibitem{KIP} For example, 
K. S. Thorne, in {\it Proceeding of Snowmass 
95 Summer Study on Particle and Nuclear Astrophysics and 
Cosmology}, eds. E. W. Kolb and R. Peccei (World Scientific, 
Singapore, 1995), p. 398, and references therein. \\
M. Ando et al. (the TAMA collaboration), 
Phys. Rev. Lett. {\bf 86} (2001), 3950. 

\bibitem{Leaver} E. W. Leaver, Proc. R. Soc. London {\bf A402} (1985), 285.\\
T. Nakamura, K. Oohara and Y. Kojima, Prog. Theor. Phys. Suppl. No.~
{\bf 90} (1987), 1. 

\bibitem{USE} K. Ury\=u, M. Shibata and Y. Eriguchi, Phys. Rev. D {\bf 62} 
(2000), 104015. 

\bibitem{RS} F. Rasio and S. L. Shapiro, Astrophys. J. {\bf 401} (1992), 226; 
{\it ibid} {\bf 432} (1994), 242. 

\bibitem{C} X. Zang, J. M. Centrella and S. L. W. McMillan, 
Phys. Rev. D {\bf 50} (1994), 6247; {\it ibid} {\bf 54} (1996), 7261. 

\bibitem{ONS} K. Oohara, T. Nakamura and M. Shibata, Prog. Theor. Phys. 
Suppl. No.~{\bf 128} (1997), 183. 

\bibitem{RJ} M. Ruffert, H.-Th. Janka and G. Sch\"afer, Astron. Astrophys. 
{\bf 311} (1996), 532. 

\bibitem{FR}  J. A. Faber and F. A. Rasio, Phys. Rev. 
D {\bf 62} (2000), 064012; {\it ibid} {\bf 63} (2001), 044012. 

\bibitem{AP} S. Ayal, T. Piran, R. Oechslin, M. B. Davies and 
S. Rosswog, Astrophys. J. {\bf 550} (2001), 846. 

\bibitem{ORT} R. Oechslin, S. Rosswog and F. Thielmann, gr-qc/0111005. 

\bibitem{bina} M. Shibata and K. Ury\=u, Phys. Rev. D {\bf 61} (2000), 064001.

\bibitem{binas} M. Shibata and K. Ury\=u, 
in {\em Proceedings of 
Gravitational Waves: A Challenge to Theoretical Astrophysics}, eds. 
V. Ferrari, J. C. Miller and L. Rezzolla 
(The Abdus Salam ICTP Publications, 2001), p. 137. \\
in {\em Proceedings of 20th Texas Symposium on Relativistic Astrophysics}, 
eds. J. C. Wheeler and H. Martel
(AIP conference proceedings {\bf 586}, 2001), p. 717 (astro-ph/0104409). 

\bibitem{Ale} A. Buonanno and Y. Chen, private communication. 

\bibitem{gr3d} M. Shibata, Phys. Rev. D {\bf 60} (1999), 104502.

\bibitem{other} 
K. Oohara and T. Nakamura, Prog. Theor. Phys. Suppl. No.~{\bf 136}
(1999), 270.\\
W.-M. Suen, Prog. Theor. Phys. Suppl. No.~{\bf 136} (1999), 251.\\
J. Font et al., gr-qc/0110047. 


\bibitem{SN} M. Shibata and T. Nakamura, Phys. Rev. D {\bf 52} (1995), 5428. 

\bibitem{gw3p2} M. Shibata, Prog. Theor. Phys. {\bf 101} (1999), 1199. 


\bibitem{ST} e.g., S. L. Shapiro and S. A. Teukolsky, {\em Black
Holes, White Dwarfs, and Neutron Stars}, Wiley Interscience
(New York, 1983). 

\bibitem{SY} L. Smarr and J. W. York, Phys. Rev. D {\bf 17} (1978), 1945. 

\bibitem{rot1} M. Shibata, T. W. Baumgarte, and S. L. Shapiro,
Phys. Rev. D {\bf 61} (2000), 044012.\\
M. Shibata, T. W. Baumgarte, and S. L. Shapiro,
Astrophys. J. {\bf 542} (2000), 453. 

\bibitem{moncrief} V. Moncrief, Ann. of Phys. {\bf 88} 
(1974), 323.

\bibitem{irre} M. Shibata, Phys. Rev. D {\bf 58} (1998), 024012.\\
S. A. Teukolsky, Astrophys. J. {\bf 504} (1998), 442. 
See also, \\
S. Bonazzola, E. Gourgoulhon and J.-A. Marck, Phys. 
Rev. D {\bf 56} (1997), 7740.\\
H. Asada, Phys. Rev. D {\bf 57} (1998), 7292. 

\bibitem{UE} K. Ury\=u and Y. Eriguchi, Phys. Rev. D {\bf 61} (2000), 124023.\\
See also Ref. \citen{GBM} for another method. 

\bibitem{GBM} E. Gourgoulhon et al., Phys. Rev. D {\bf 63} (2001), 064029. 

\bibitem{York} J. W. York, Jr., in {\it Sources of Gravitational Radiation}, 
ed. L. L. Smarr (Cambridge University Press, 1979), p. 83. 

\bibitem{WM} J. R. Wilson and G. J. Mathews, Phys. Rev. Lett. {\bf 75}
(1995), 4161.

\bibitem{SU01} M. Shibata and K. Ury\=u, Phys. Rev. D {\bf 64} (2001), 
104017. 

\bibitem{Bishop} N. Bishop et al., Phys. Rev. Lett. {\bf 76} (1996), 4303.\\
A. M. Abrahams et al., Phys. Rev. Lett. {\bf 80} (1998), 1812.\\
L. Rezzolla et al., Phys. Rev. D {\bf 59} (1999), 064001. 

\bibitem{AH} M. Shibata, Phys. Rev. D {\bf 55} (1997), 2002: 
M. Shibata and K. Ury\=u, Phys. Rev. D {\bf 62} (2000), 087501. 

\bibitem{BSS} T. W. Baumgarte, S. L. Shapiro, and M. Shibata, 
Astrophys. J. Lett. {\bf 528} (2000), L29. 

\bibitem{excision} For example, 
M. Alcubierre et al., Phys. Rev. D {\bf 64} (2001), 061501. 

\bibitem{BCL} J. Baker, M. Campanelli and C. Lousto, gr-qc/0104063.\\
J. Baker et al., Phys. Rev. Lett. {\bf 87} (2001), 121103, 
and references cited therein. 

\bibitem{BIWW} L. Blanchet, B. R. Iyer, C. M. Will and 
A. G. Wiseman, Class. Quant. Grav. {\bf 13} (1996), 575. 

\bibitem{B} L. Blanchet, Phys. Rev. D {\bf 51} (1995), 2559. 

\bibitem{CF} C. Cutler and E. E. Flanagan, Phys. Rev. D {\bf 49}
(1994), 2658.\\
E. Poisson and C. M. Will, {\it ibid} {\bf 52} (1995), 848. 

\bibitem{AMR} M. Berger and J. Oliger, J. Comp. Phys. {\bf 53} (1984), 484. 

\bibitem{T} E. S. Phinney and S. R. Kulkarni, Annu. Rev. Astron. 
Astrophys. {\bf 32} (1994), 591. 

\bibitem{NO} T. Nakamura and K. Oohara, Prog. Theor. Phys.
{\bf 86} (1991), 73. 

\bibitem{US01} K. Ury\=u and M. Shibata, in preparation. 

\end{thebibliography}
\end{document}